\documentclass[prodmode,acmtecs]{acmsmall} 

  \usepackage{amssymb}
  \usepackage{latexsym}
  \usepackage{amsfonts}
  \usepackage{amsmath}
  \usepackage{graphics}
  \usepackage{setspace}
  \usepackage{graphicx}
  \usepackage{epstopdf}
  \usepackage{color}
  \usepackage{float}
  \usepackage{wrapfig}
  \usepackage{color}
  \usepackage{rotating}
  \usepackage{array} 
  \usepackage{dblfloatfix}
  \usepackage{breakurl}
  \usepackage[table]{xcolor}
  \usepackage{multirow}

\usepackage[ruled]{algorithm2e}

\SetAlFnt{\small}
\SetAlCapFnt{\small}
\SetAlCapNameFnt{\small}
\SetAlCapHSkip{0pt}
\IncMargin{-\parindent}

\acmVolume{0}
\acmNumber{0}
\acmArticle{00}
\acmYear{2017}
\acmMonth{0}

\doi{0000001.0000001}

\issn{1234-56789}

\begin{document}

\markboth{C. Perera et al.}{Fog Computing  for Sustainable Smart Cities: A Survey}

\title{Fog Computing  for Sustainable Smart Cities: A Survey}
\author{
\textcolor{black}{CHARITH~PERERA}
\affil{The Open University}
\textcolor{black}{YONGRUI  QIN}
\affil{University of Huddersfield}
\textcolor{black}{JULIO C. ESTRELLA}
\affil{University of  Sao Paulo}
\textcolor{black}{STEPHAN REIFF-MARGANIEC}
\affil{University of Leicester}
ATHANASIOS V. VASILAKOS
\affil{Lulea University of Technology}}

\begin{abstract}
The Internet of Things (IoT) aims to connect billions of smart objects to the Internet, which can bring a promising future to smart cities. These objects are expected to generate large amounts of data and send the data to the cloud for further processing, specially for knowledge discovery, in order that appropriate actions can be taken. However, in reality sensing all possible data items captured by a smart object and then sending the complete captured data to the cloud is less useful. Further, such an approach would also lead to resource wastage (e.g. network, storage, etc.). The Fog (Edge) computing paradigm has been proposed to counterpart the weakness by pushing processes of knowledge discovery using data analytics to the edges. However, edge devices have limited computational capabilities. Due to inherited strengths and weaknesses, neither Cloud computing nor Fog computing paradigm addresses these challenges alone. Therefore, both paradigms need to work together in order to build an sustainable IoT infrastructure for smart cities. In this paper, we review existing approaches that have been proposed to tackle the challenges in the Fog computing domain. Specifically, we describe several inspiring use case scenarios of Fog computing, identify ten key characteristics and common features of Fog computing, and compare more than 30 existing research efforts in this domain. Based on our review, we further identify several major functionalities that ideal Fog computing platforms should support and a number of open challenges towards implementing them, so as to shed light on future research directions on realizing Fog computing for building sustainable smart cities.

\end{abstract}


%
%




\begin{bottomstuff}
Charith Perera's work is supported by European Research Council Advanced Grant 291652 (ASAP).
\end{bottomstuff}

\maketitle

\section{Introduction}
\label{sec:Introduction}

Over the last few years, the Internet of Things (IoT) has become a popular term that many different communities interpret in many different ways. The term IoT has been vaguely described in the literature and has very faded boundaries. From the functional point of view IoT is defined as \textit{``The Internet of Things allows people and things to be connected Anytime, Anyplace, with Anything and Anyone, ideally using Any path/network and Any service''} \cite{P006}. Most of the objects  (from living room couches to street lights to parking slots) around us are expected to be embedded with sensors. These sensors will have a wide range of sensing capabilities \cite{Perera2015a}. Then, the collected sensing data needs to be analyzed to derive knowledge in order to assist and accelerate decision making processes \cite{FC18}.

Data is the primary commodity in the data driven economy \cite{HAT1}. Surprisingly, large volumes of data can provide much more detailed insights into ourselves and our behaviors. Therefore, over the recent years \textit{Big Data} \cite{ZMP003} has become a buzz word in both industry and academia. Having said that, \textit{Big Data} is not a new concept or idea. In the early days,   \textit{Big Data} was produced by large tech companies such as Google, Yahoo, Microsoft, and large scientific research institution such as European Organization for Nuclear Research (CERN). Due to  advances in technology and economy of scale, computational hardware costs have significantly reduced. As a result, sensors are being integrated into all kinds of hardware devices. These sensors generate large volumes of data. Further, due to cheaper storage costs, organizations are storing data for much longer periods than before. However, \textit{Big Data} has created significant challenges in terms of processing and analyzing. As a result, \textit{Big Data} has become a buzz word in industry where different parties attempt to tackle these new challenges. More importantly, \textit{`Big Data'} does not have a clear definition. Typically, \textit{`Big Data'} is defined with the help of its characteristics, widely known as 3V's \cite{IBMBigData,ZMP003}: volume, variety, and velocity.

\begin{itemize}
\item \textbf{Volume:} \textit{``Volume relates to the size of the data such as terabytes (TB), petabytes (PB), zettabytes (ZB), etc''} \cite{ZMP003}.

\item \textbf{Variety:} \textit{``Variety means the types of data. In addition, different sources can produce big data such as sensors, devices, social networks, the web, mobile phones, etc. Therefore, data could be web logs, RFID sensor readings, unstructured social networking data, streamed video and audio, and so on''} \cite{ZMP003}.

\item \textbf{Velocity:} \textit{``This means how frequently the data is generated, for example, every millisecond, second, minute, hour, day, week, month, or year. Processing frequency may also differ with user requirements. Some data needs to be processed in real-time and some may only be processed when needed. Typically, we can identify three main categories: occasional, frequent, and real-time''} \cite{ZMP003}

\end{itemize}

Big data is usually discussed with respect to the cloud computing paradigm. Due to the fact that cloud computing theoretically provides infinite amounts of computational and storage resources, the typical ideology is to send all the data back to the cloud for processing. However, this is not always the case. There are downsides of heavily depending on cloud computing. For example, sending all the data collected all the time to cloud has high costs in terms of bandwidth, storage, latency, energy consumption (for communication) and so on. We discuss these issues later in this paper.

To address the weaknesses inherited by the cloud computing paradigm, a notion called \textit{`fog computing'} has been proposed. Fog computing encourages a shift of handling everything in the core of cloud computing to handling at the edges of a network. To be specific, instead of sending all the data collected to the cloud, fog computing suggests to process the data at leaf nodes or at the edges. This idea is also called \textit{`edge analytics'}. Local data processing helps to mitigate the weakness of cloud computing. In addition, fog computing also brings new advantages, such as greater context-awareness, real-time processing, lower bandwidth requirement, etc. We will discuss more advantages in detail later in this paper.

Our objectives in revisiting the literature related to fog computing are threefold: 1) to learn how the fog computing paradigm can be used to build sustainable Smart Cities on top of the IoT infrastructure, 2) to identify the most critical functionalities of an ideal fog computing platform, and 3) to highlight open challenges and to discuss future research directions with respect to developing fog platforms to support a wider ranges of use case scenarios.

This paper is organized into sections as follows: Section \ref{sec:Fog_Computing} introduces the fog computing paradigm from the IoT perspective by discussing its strengths, weaknesses and roles in the IoT era. Section \ref{sec:UseCase} presents a number of use cases that would benefit from the fog computing paradigm. In Section \ref{sec:Characteristics}, we identify ten major characteristics and common features of the fog computing paradigm with the support of use case scenarios presented in Section \ref{sec:UseCase}. Comparisons and in-depth analysis of existing research efforts are presented in Section \ref{sec:Comparison}. Based on our review, we discuss lessons leaned in Section \ref{sec:Lessons_Learned}. Then, Section \ref{sec:Lessons_Learned} discusses future research directions and challenges, before we conclude our review in Section \ref{sec:Conclusions}.

\subsection{Main Contributions}
In this section, we summarize the contributions and the novelty of this paper. There have been several surveys conducted in this field. We briefly introduce these surveys in chronological order. 

Bonomi et al. \cite{FC01} have discussed the role of fog computing in the internet of things domain. Yi et al. \cite{FC66}  have surveyed fog computing under seven themes, namely, fog networking, quality of service (QoS), interfacing and programming model, computation offloading, accounting, billing and monitoring, provisioning and resource management, security and privacy. Yi et al. \cite{FC77}  have surveyed fog computing  from the security point of view, where they have focused on number of security aspects such as trust and authentication, network security, secure data storage, secure and private data computation, privacy, access control and intrusion detection.

It is important to note that, none of these surveys have reviewed the fog computing domain from platform developers' and end users' perspectives. \textcolor{black}{Further, this survey is mainly focuses on connectivity and device configuration aspects of the fog computing paradigm.} Our objective is to identify major features that fog computing platforms need to support specially towards building sustainable sensing infrastructure for smart city applications. Towards this, we built a taxonomy (of common features) after conducting an extensive  literature review. This paper is  enriched by both real world use case scenarios and literature findings. We also compare different research efforts conducted in the past and present these in Table \ref{TbL:LiteratureReview}. Subsequently, we identified research trends and gaps based on the literature review. Finally, we highlighted research challenges and directions.

\section{Fog Computing: An Overview}
\label{sec:Fog_Computing}

Fog computing \cite{FogDef,FC01} is \textit{``an architecture that uses one or a collaborative multitude of end-user clients or near-user edge devices to carry out a substantial amount of storage (rather than stored primarily in cloud data centers), communication (rather than routed over the internet backbone), and control, configuration, measurement and management (rather than controlled primarily by network gateways such as those in the LTE (telecommunication) core)''} \cite{Chiang2015,FC12FC61}.

The fog computing paradigm does not necessarily stick to one architecture, instead it represents a notion that supports to push data analytics towards leaves (i.e., edge nodes) \cite{FC14}. The important message to understand is \textit{`towards leaf /edge node'}. Therefore, fog computing does not suggest that all analytics needs to be done on leaves, instead it promotes to  use leaf nodes as much as possible within realistic limitations. Leaf nodes also represent the closeness to the users. Further, it also suggests that multiple nodes close to the leaves would work in a cooperative manner to accomplish a task. Fog networking constitutes two planes, namely, 1) control plane and 2) data plane \cite{Chiang2015}. The responsibility of the control plane is to configure and manage the network (e.g., sync routing table information, efficient traffic control). The data plane is responsible for transferring data from the originator to the destination based on the routing information provided by the  control plane logic \cite{Chiang2015}. Chiang \cite{Chiang2015} has highlighted main differences between Cloud and Fog.

\begin{figure*}[b!]
 \includegraphics[scale=.70]{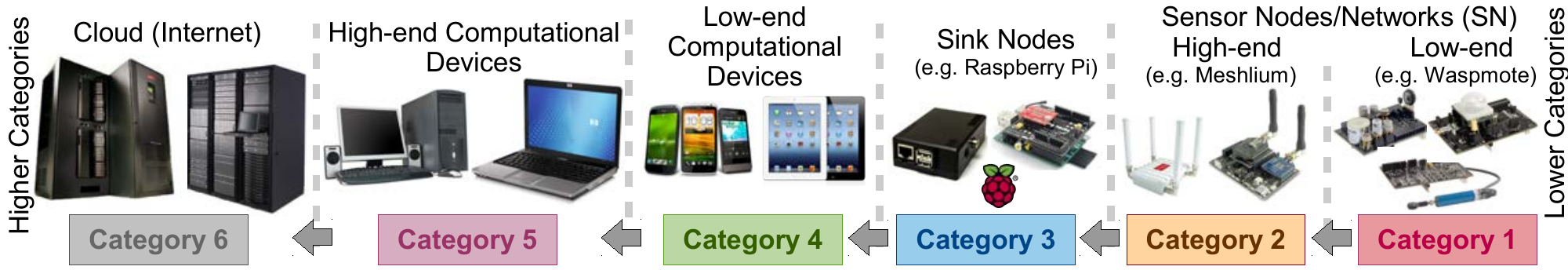}
 \caption{Categorizations of IoT devices based on their computational capabilities. }
 \label{Figure:Layered_Architecture}	
\end{figure*}

\begin{itemize}
\item Store most of the data at	or near the edge instead of sending all the data to the cloud all the time. \textcolor{black}{Features such as Mobility (Section \ref{sec:Mobility}), Context Discovery and Awareness (Section \ref{sec:Context_Discovery}), and Data Analytics (Section \ref{sec:DataAnalytics}) facilitate these characteristics.}

\item Use local networks (i.e., closer to the leaf nodes) to transfer data to processing nodes instead of using backbone	networks (e.g., Internet) to route the data.  \textcolor{black}{Features such as Multi-Protocol Support: Communication Level (Section \ref{Multi_Protocol_Support_Communication_Level}
), Multi-Protocol Support: Application Level ( Section \ref{Multi_Protocol_Support_Application_Level}
), Mobility (Section \ref{sec:Mobility}), Security and Privacy (Section \ref{sec:SecurityandPrivacy}
), and  Cloud Companion Support (Section \ref{sec:CloudCompanionSupport}) facilitate these characteristics.}

\item Process substantial	amount	of data at the edge devices (e.g. leaf nodes) instead of using cloud computing infrastructure.  \textcolor{black}{ Features such as Semantic Annotation (Section \ref{sec:SemanticAnnotation}), Data Analytics (Section \ref{sec:DataAnalytics}) and Context Discovery and Awareness (Section \ref{sec:Context_Discovery}) facilitate these characteristics.}

\item Edge device are self-governed, managed, and controlled instead of controlled cloud decision makers.  \textcolor{black}{Features such as Dynamic Discovery of Internet Objects (Section \ref{sec:Dynamic Discovery of Internet Objects}), Dynamic Configuration and Device Management (Section \ref{sec:DynamicConfigurationandDeviceManagement}), Data Analytics (Section \ref{sec:DataAnalytics}), Context Discovery and Awareness (Section \ref{sec:Context_Discovery}) facilitate these characteristics.}

\end{itemize}

Figure \ref{Figure:Layered_Architecture} illustrates how data moves from leaf nodes to the cloud backbone devices. The computational capabilities (e.g., processing, memory, communication) of each category is different. Right most device categories have less computational capabilities and cheaper in price. Left most  device categories have higher computational capabilities and  cost more. Even though there is no strict definition, fog computing refers to the devices belonging to categories 1 to 4. They have some capabilities to process data before sending it to the cloud.

It is important to note that, over time capabilities of these categories will increase dramatically. For example, a few years ago smart phone like devices had less than 1GB memory. However, today we have smart phones with 4GB or more memory. We can also observe the same pattern when considering different versions of Raspberry Pi Devices\footnote{https://www.raspberrypi.org/magpi/raspberry-pi-3-specs-benchmarks/}. The idea is that capabilities of devices of each  category will increase over time. However, the difference between categories will always remain due to form factor limitation and more importantly price points. Form factor limitation means that the amount of computational power that can be concentrated into a device always correlates to the size of the device. Some common characteristics of fog computing are \textit{``proximity to end-users and client objectives, dense geographical distribution and local resource pooling, latency reduction for quality of service (QoS) and edge analytics/stream mining, resulting in superior user-experience  and redundancy in case of failure''} \cite{FogDef,FC18}. 

One of the common goals of fog computing is to make \textit{`big data'} smaller. This is done through intelligent sensing (i.e. sensing only useful data based on the knowledge and intelligence available locally at a given edge device), filtering, and data analytics (e.g. averaging, knowledge discovery and throwing away raw data). In 2020, the annually captured data in total is foretasted to exceed 1,600 exabytes, or equivalently 1.6 zettabytes \cite{ABIResearch2015}. Fog computing  is expected to reduce the amount of data collected and needing to be processed to a manageable level.

\textcolor{black}{Further, data synchronization techniques from low end to high end will play a significant role in making the sensed data manageable, specially at the high ends of fog computing, where data is expected to be gathered and processed \cite{FC12FC61}. It is also critical to effectively push back information to the low ends whenever necessary. Typically, there are three schemes for data synchronization: Pull Scheme, Push Scheme, Push-Pull Scheme. The choice of data synchronization strategies will depend on the communication protocols in use and the requirements in the actual fog computing scenarios.
}

Another advantage of fog computing is lower latency. As shown in Figure \ref{Figure:Layered_Architecture}, it takes time for the data to move from sensing edge devices to the cloud computing backbone where data will be processed. However, the actions based on data analysis need to be taken fast, sometimes in real-time, in order for such actions (or data) to be useful. The fog computing vision fits well in scenarios where data will be processed within edge devices. Most of the smart home event detection scenarios typically need to support quick responses (e.g., if a motion in the living room is detected during day time while occupants are not there, then send an alert to the occupants including an image of the living room). Some enterprise-level IoT applications may require real-time or near real-time physical interactions with the connected assets to avoid any machinery failures. 

Fog computing can also ensure higher availability. Connecting to the cloud is less reliable due to a variety of different connectivity issues \cite{FC02}. This reduces the dependency on cloud infrastructures and enables edge devices to function uninterruptedly for a reasonable time period and perform analytics even if the connection to the cloud is lost. Madsen et al. \cite{FC10} have highlighted the importance of developing techniques to improve the reliability of fog computing. Having intelligent edge devices can also increase overall security of an IoT solution by encrypting data at the source. Further, edge devices can process raw data and produce less privacy and security sensitive secondary context knowledge \cite{FC02}, so the data communication and transfer face a lesser security threat. Further, edge devices that can fully process data  locally can eliminate the communication based security threats completely, which is a further advantage in the access to context knowledge. As a result of being close to users or physical fields, fog based IoT applications can gather more knowledge about the local setting.

Considering cost, fog computing will cost more than deploying dumb devices at the edge and thus it would increase the total cost of a fog-based IoT solution. However, in the long term such additional costs will become justified due to potential savings (i.e., reduced connectivity costs and extended life cycles of battery-operated devices, and more importantly potential savings in highly dynamic and analytically complex settings).

The above mentioned advantages, that fog computing brings, are traditional weaknesses of cloud computing. Therefore, it is clear that fog computing can mitigate weaknesses of the cloud computing paradigm. However, fog computing has its own weaknesses as well. Edge devices are less computationally capable (e.g., less processing power, less storage, and less memory). This means that edge devices can only handle smaller amount of data and provide less processing capacity. Further, edge devices have less knowledge about the big picture. Therefore, some types of processing (i.e., data analysis) that require global knowledge may not be able to perform. In addition to the resource limitations, edge devices may have other limitations, such as energy. Some edge devices may run on batteries or alternative energy sources such as solar power. Therefore, certain types of data processing  that require significant amounts of energy may not be suitable to run on edge devices.

Smart home solutions have traditionally used fog computing approaches over the last decade \cite{Perera2015a,PereraIEEEAccess}. However, to realize the real potentials of fog computing, edge analytics needs to be incorporated with IoT solutions in enterprise-level applications such as smart cities, manufacturing, healthcare, agriculture transportation, etc. In the next section we introduce several different and promising scenarios where fog computing has a potentially important role to play. These use cases have motivated us to explore and investigate challenges related to fog computing.

\section{Use Case Scenarios}
\label{sec:UseCase}

In this section, we present four different use case scenarios. Our objective is to use these scenarios  to extract major functional requirements and characteristics of fog computing platforms. The use cases presented in this section are on 1) Smart Agriculture, 2) Smart Transportation, 3) Smart Healthcare, and 4) Smart Waste Management. These generic use case scenarios were extracted, with minor alterations, from two of our previous publications \cite{Perera2014,Perera2016}. We refer to the scenarios throughout this paper and have used them to extract major functionalities of fog computing platforms.

\subsection{Smart Agriculture}
\label{sec:SmartAgriculture}

\begin{figure}[t!]
 \centering
 \includegraphics[scale=.51]{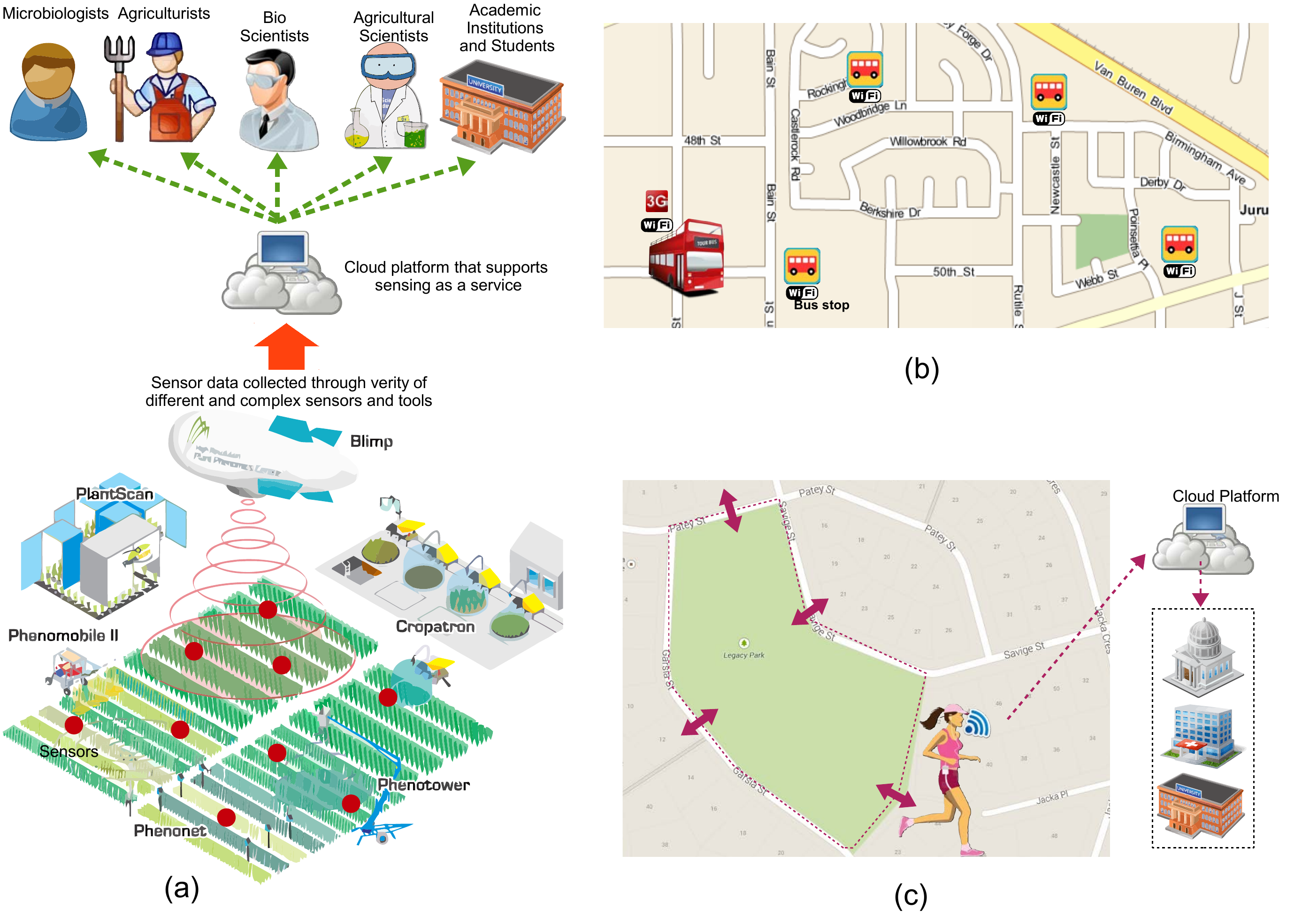}
 \caption{Fog Computing based Use Case Scenarios: (a) Different types of agricultural field vehicles can be fitted with sensing devices in order to monitor the growth of  seeds and plants, (b) Sensors are deployed in buses in a Smart City environment and the data is expected to be collected based on context information and conditions provided by the data consumer, (c) Wearable sensors can be used to monitor movements of the  general public who use public spaces such as parks for exercising and recreational activities. So the authorities can plan further development and upgrades of the infrastructure.}
 \label{Figure:UseCase2}	
\end{figure}

\textit{``Agriculture is an importation part of smart cities as it contributes to the food supply-chain that facilitates a large number of communities concentrated into cities. This smart agriculture case study is based on two real world projects: \textit{Phenonet} \cite{P412} and \textit{OpenIoT} \cite{P377}. In this scenario, the general public is not directly involved as in the smart home domain. To be specific, in Figure \ref{Figure:UseCase2}(a), we illustrate how the sensing works in the smart agriculture domain. In the following, we demonstrate the applicability of fog computing towards agricultural research through describing the \textit{Phenonet} project in detail. \textit{Phenonet} has a network of sensors collecting information over a field of experimental crops. Researchers at the High Resolution Plant Phenomics Centre are testing a network of smart sensor nodes that are able to monitor plant growth, performance information and climate conditions. Experimental plants are monitored using different types of sensing devices and techniques. First, sensor stations are used to monitor the plants in the field. Second, air balloons called \textit{`blimps'} are used to sense the field from sky. Third, field vehicles such as \textit{Phenomobiles} are used to capture data on plant growth. Fog computing can play a significant role in performing the above mentioned sensing tasks more efficiently and adaptively. Ideally, fog gateway devices can be used in all three types of sensing approaches mentioned above to make the sensing process more efficient''} \cite{Perera2014}. \textcolor{black}{As shown in Figure \ref{Figure:UseCase2}(a), stationary sensor stations, \textit{Phenomobiles}, and \textit{`blimps'} act as edge devices. They combines both sensors and gateway devices. However, in some situations, sensor stations only have sensors and gateway devices fit into \textit{Phenomobiles} act as the gateway. Average the data over different time frames in order to reduce data storage and  communication are the most common data aggregation techniques performed on the gateway devices. Further, data analysis  techniques are also used measure quality of service parameters.}

\textit{``Context information plays a critical role in efficient sensing in smart agriculture related applications. The objective of collecting sensor data is to understand plant growth better by applying fusing and reasoning techniques. In order to accomplish this task, sensor data needs to be collected in a timely and location-sensitive manner. Each sensor needs to be configured by considering context information. In some situations, severe frosts and heat events can have a devastating effect on crops. Flowering time is critical for cereal crops and a frost event could damage the flowering mechanism of the plant. However, the ideal sampling rate could vary depending on both the season of the year and the time of day. For example, on one hand, a higher sampling rate is necessary during winter and nighttime. In contrast, lower sampling would be sufficient during summer and daytime. On the other hand, some reasoning approaches may require multiple sensor data readings. One example for this is that, a frost event can be detected by fusing air temperature, soil temperature, and humidity data. However, if the air temperature sensor stops sensing due to malfunction, it would become useless to continue sensing humidity. The reason is that frost events cannot be detected without temperature. In such circumstances, configuring the humidity sensor to sleep is ideal until the temperature sensor is replaced and starts sensing again. Such intelligent (re-)configuration can save energy by eliminating ineffectual sensing and network communication''} \cite{Perera2014}.

\subsection{Smart Transportation}

\textit{``John, a researcher at the Department of the Environment, is interested in measuring and monitoring air pollution in cities. John's team has deployed sensor kits in buses. Each of these sensor kits consists of multiple sensors and a communication device with both WiFi and 3G capabilities. The team has developed an application that processes data collected by these sensor kits. This application consists of a number of different algorithms that analyze and visualize air pollution in a city. However, according to the way that the  algorithms are written, John only needs to collect data when a bus is moving. Sensor data that is captured while the bus is stopped at a bus stop, or in traffic, does not add any value. Therefore, John would like to collect sensor data only when the bus is moving. Further, John does not need real-time data in most of the occasions. Therefore,  it is sufficient to push the  sensor data to the cloud when the bus reaches a bus stop. The communication devices fitted in the bus will connect to the bus stop's WiFi and push the data collected since the last bus stop to the cloud, as illustrated in Figure \ref{Figure:UseCase2}(b). Meanwhile, John is also interested to receive sensor data in real-time when raining. Therefore, when raining, the communication devices need to use 3G to upload the sensor data to the cloud. However, they still need to adhere to the first rule  that says \textit{sense only when moving}''} \cite{Perera2016}. \textcolor{black}{In this scenario, sensor kits on  buses act as ICOs. The gateway devices are fitted to the bus stations.}

In this scenario both sensor kits attached to the bus as well as to the bus stop  act as fog devices to make the sensing process more efficient. Fog computing technologies are widely used to build connected vehicle based applications \cite{FC01,Datta2015}. \textcolor{black}{ Datta et al. \cite{Datta2015}  have proposed an architecture for connected vehicles with Road Side Units (RSU) and M2M Gateways. RSUs are access points where  vehicles can connect themselves to M2M gateways. M2M data analytics and vehicle  discovery are the core features of this architecture.}

%



\subsection{Smart Health and Well-Being}

\textit{``Michael is working for the Department of Public Health and Well-Being. He has been asked to develop a plan to improve the public health  in cities by improving the infrastructure that supports  exercise and recreational activities (e.g., parks and the paths that supports jogging, cycling, and venues for bar exercise). Michael developed a wearable light-weight sensor kit that can monitor a variety of different parameters, such as air quality, sound and movement. Further, Michael has recruited volunteers who are willing to wear those sensor kits when  exercising. A sensor kit can collect data and push to the volunteer's smart phone. A smart phone application pushes data to the cloud once it gets connected to the Internet. Firstly, Michael  only needs to collect data when a volunteer enters  the park areas as illustrated in Figure \ref{Figure:UseCase2}(c). Further, Michael only needs to perform sensing only when the volunteers are moving (e.g., walking, running, and cycling). Meanwhile, Michael has noticed that there are a large amount of people coming to the park during the weekend. In order to reduce the burden to the volunteers, Michael only needs to collect data from a maximum of 30 sensors  kits (i.e., volunteers) despite the actual number of volunteers visiting the park at weekends''} \cite{Perera2016}. \textcolor{black}{In this scenario, the sensor kits are ICOs and the mobile devices are the gateways.}

In this scenario smart phones act as the fog devices to make the sensing process more efficient. In line with this use case, Zao et al. \cite{FC11} have used fog computing to develop brain computer interaction application. In their application, EEG data is pre-processed on the fog reducing latency and data communication.

\subsection{Smart Waste Management}

\begin{figure}[b!]
 \centering
 \includegraphics[scale=.55]{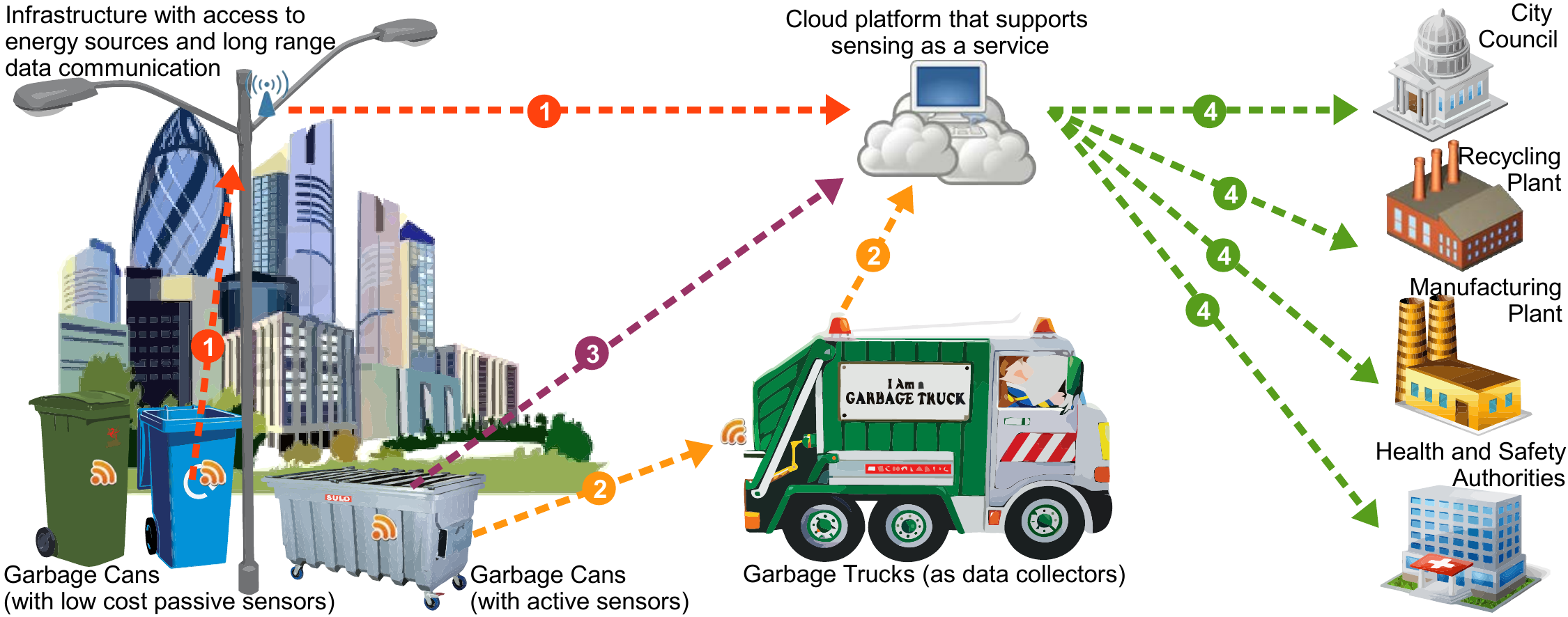}
 \caption{Efficient waste management in Smart Cities}
 \label{Figure:UseCase3}	
\end{figure}

\textit{``Waste management is one of the toughest challenges that modern cities have to deal with. Waste management consists of different processes such as collection, transport, processing, disposal, managing, and monitoring of waste materials. These processes cost a significant amount of money, time, and labor. Optimizing waste management processes help to save money that can be used to address other challenges faced by smart cities. Figure \ref{Figure:UseCase3} illustrates how the sensing   works in the waste management domain. In a modern smart city, there are several parties who are interested in waste management (e.g., city council, recycling companies, manufacturing plants, and authorities related to health and safety). For example, a city council may use sensor data to develop optimized garbage collection strategies, so they can save fuel cost of garbage trucks. Additionally, recycling companies can use sensor data to predict and track the amount of  waste coming into their plants for further processing so that they can optimize internal processes. Finally, health and safety authorities can monitor and supervise the waste management process without spending a substantial amount of money for manual monitoring inspections''} \cite{Perera2014}.

\textit{``In order to perform waste management, different types of sensors need to be deployed in different places such as garbage cans and trucks. These sensors need to detect various kinds of information, including the amount of garbage, types of  garbage, and so on. As we have depicted in Figure \ref{Figure:UseCase3}, direct and indirect communication strategies can be used to collect and communicate sensor data to the cloud. Sensors with energy harvesting capabilities are important in this domain \cite{P633}. As represented in step (1) in Figure \ref{Figure:UseCase3}, low powered and low capable sensors \cite{P634} can be used to sense and data can be uploaded to the cloud with the help of nearby infrastructure (e.g., through communication devices attached to street lights or similar infrastructures that have access to rich energy sources and communication capabilities). In addition, when long range communication is not available, data can be uploaded to the cloud with the help of auto-mobiles, as depicted in step (2) in Figure \ref{Figure:UseCase3}, such as garbage trucks, city council vehicles and buses that operate in the near areas. Furthermore, both active and passive sensors can be used to sense the environment \cite{P017}. Direct communication can be done via technologies such as 3G,  which makes this approach less dependent on third parties (as depicted in (3) in Figure \ref{Figure:UseCase3})''} \cite{Perera2014}. \textcolor{black}{In this scenario, waste bins are fitted with ICOs and street lights and garbage trucks act as gateways.}

In this scenario, we highlight the ability of fog computing platforms in collecting data from sensors with  short range capabilities and opportunistically uploading data to the cloud after aggregation. \textcolor{black}{For example, sensors may record how the garbage bins get filled over time. However, in order to reduce data communication (also energy usage), they may only send the data point that is sufficient to plot a graph in acceptable level of accuracy.} 

The above scenarios highlight some major capabilities of the fog computing paradigm. We will revisit these scenarios in the next section, where we identify individual features of fog computing.

\subsection{Smart Water Management}

\begin{figure}[h]
 \centering
 \includegraphics[scale=0.35]{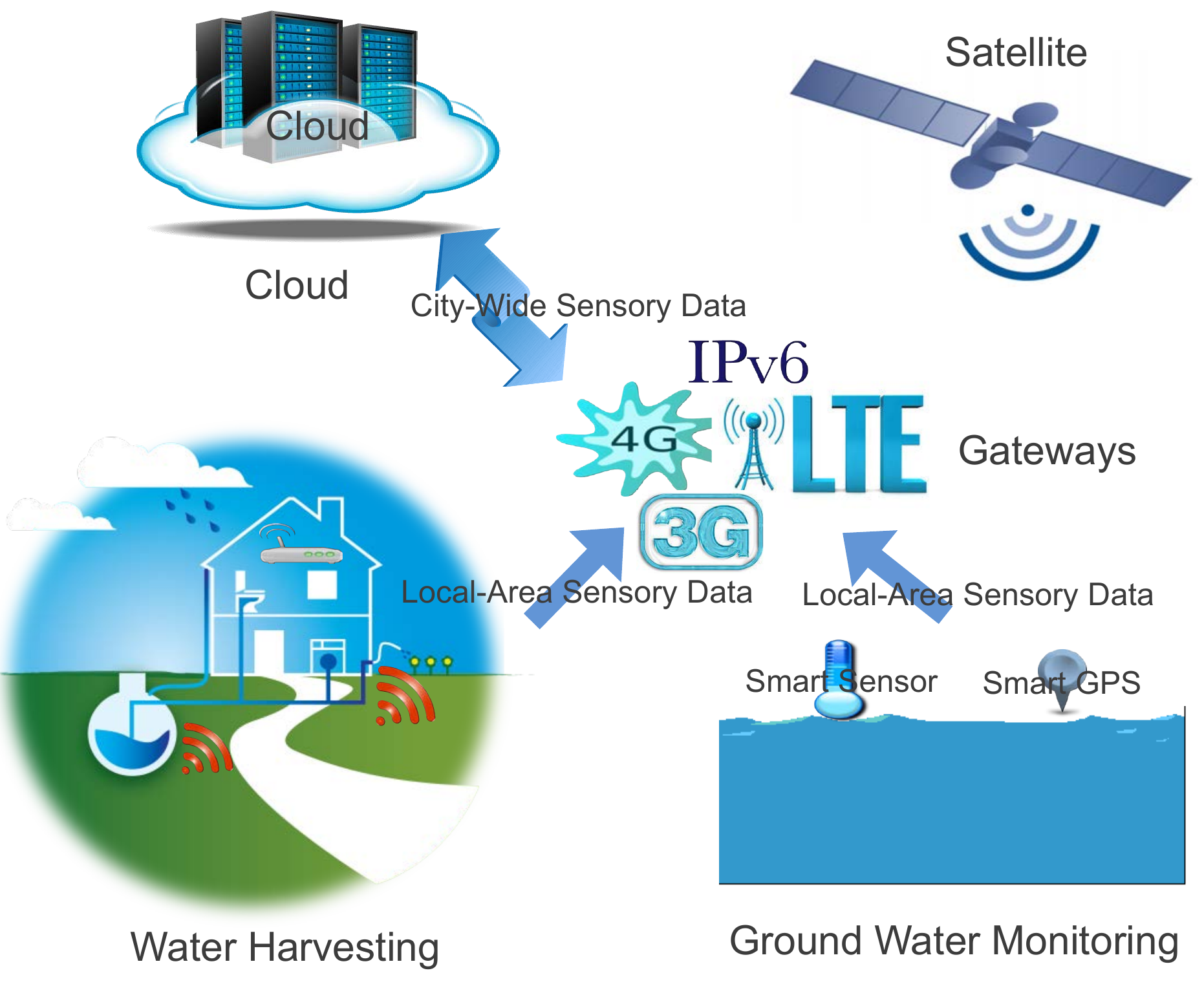}
 \caption{Smart Water Management.}
 \label{Figure:SmartWater}
\vspace{-8pt}
\end{figure}

\textcolor{black}{The water system of a city is one of the most important aspects for building future smart cities. Effective and energy-efficient transportation and use of water, and cost-effective and environment-friendly treatment of wastewater are critical in smart water management at city-scales. A smart water system is expected to help monitor city-wide water consumption, transportation, prediction of future water use, and so on. For example, Figure \ref{Figure:SmartWater} describes the needs of water harvesting and ground water monitoring, which will rely on the support from Fog/Cloud computing infrastructure, such as wireless sensors, smart meters, GPS devices, Fog gateways, Cloud platforms, IPv6 technology and long-distance communication protocols like 3G, 4G, LTE, etc. On top of all these useful aspects of the city water network, the smart water system is expected to analyze collected information and generate actionable data for management, reduction of water losses and other improvement of the city water system.}

\subsection{Smart Greenhouse Gases Control}

\begin{figure}[b!]
 \centering
 \includegraphics[scale=0.40]{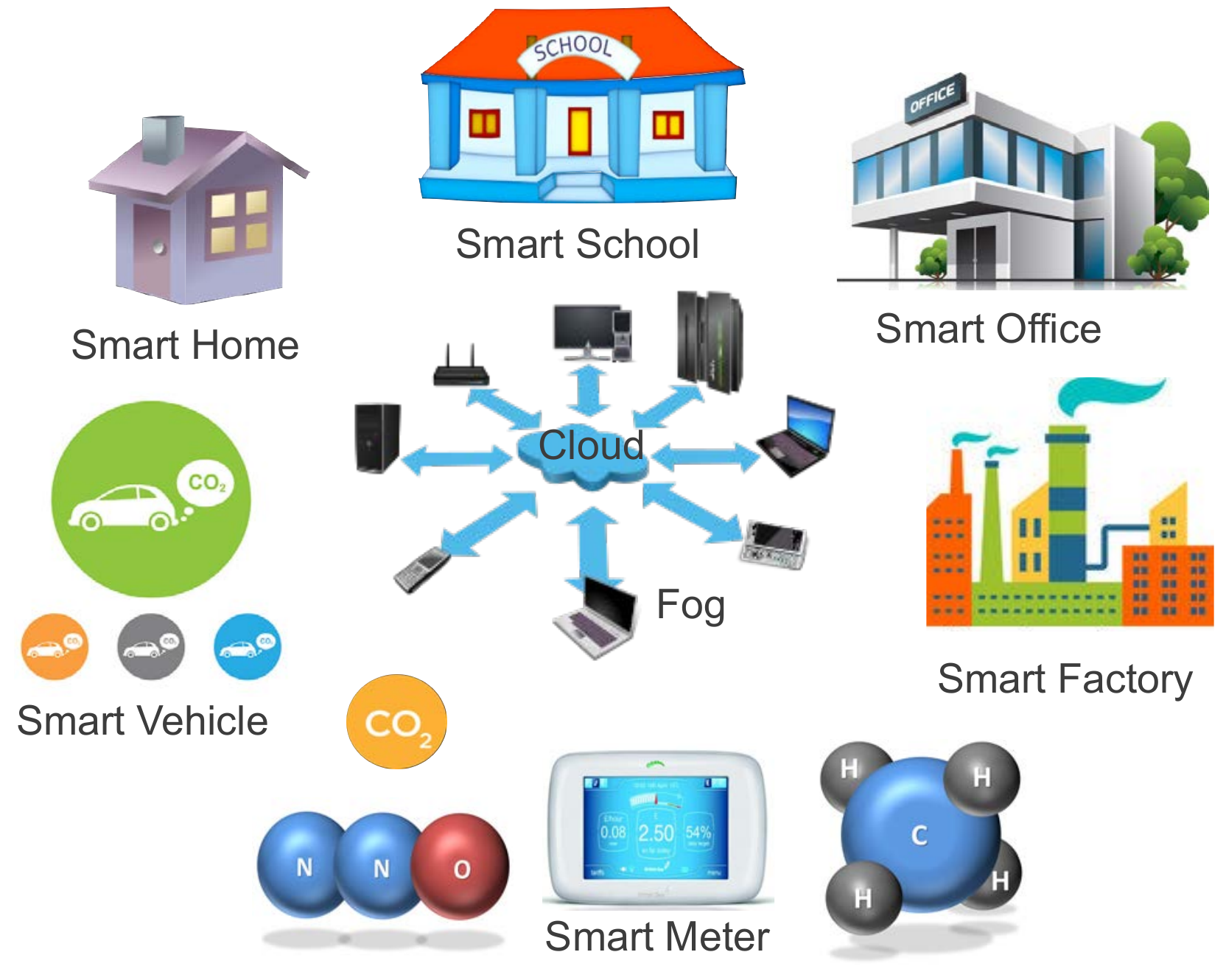}
 \caption{Smart Greenhouse Gases Control.}
 \label{Figure:SmartGreenhouseGases}
\vspace{-8pt}
\end{figure}

\textcolor{black}{Greenhouse gases control requires collective efforts. Governments can take actions and personal lifestyles matter too. Smart greenhouse gases control systems in future smart cities can help to collect critical information for governments to make better decisions on policies in order to effectively reduce greenhouse gases. Such smart systems can also work with city habitats to guide them on making their own efforts in helping reduce greenhouse gases. Figure \ref{Figure:SmartGreenhouseGases} shows an example of  a future greenhouse gases control system. Smart monitoring can cover homes, schools, offices, factories, vehicles, and so on. Individuals and workers can be reminded instantly based the smart monitoring results for taking simple and quick actions to help reduce greenhouse gases, specially with the help of Fog devices in the Fog smart computing paradigm. Governments can make timely decisions and policies on top of the smart monitoring results of greenhouses in city-wide environments and infrastructure.}

\subsection{Smart Power Grid}

\begin{figure}[h]
 \centering
 \includegraphics[scale=0.40]{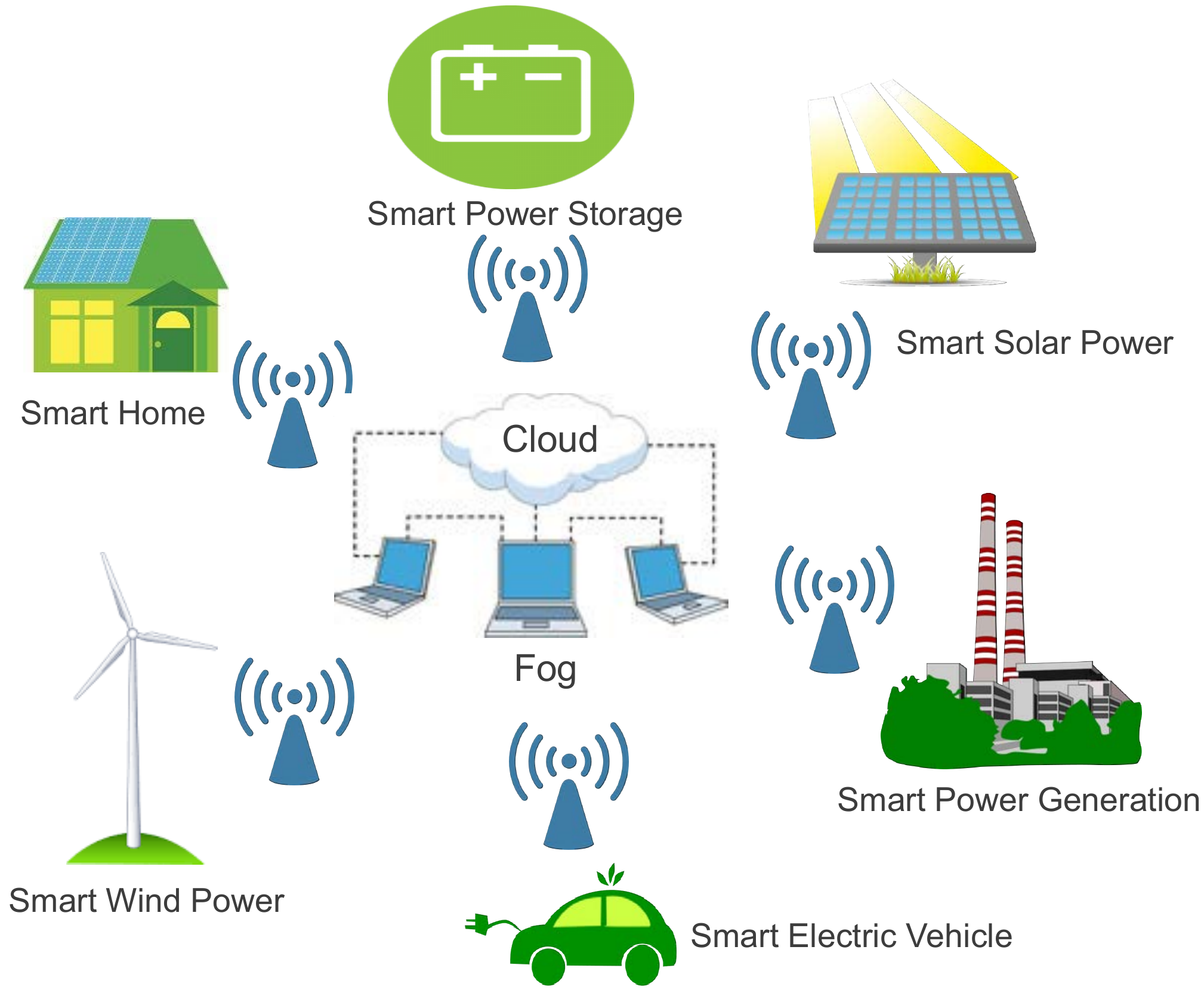}
 \caption{Smart Power Grid Control.}
 \label{Figure:SmartPowerGrid}
\vspace{-8pt}
\end{figure}

\textcolor{black}{In future smart cities, smart power grids will be critical in ensuring reliability, availability, and efficiency in city-wide electricity management. Figure \ref{Figure:SmartPowerGrid} demonstrates an example future smart grid system, where Fog/Cloud computing can play a significant role. A successful smart grid system will be able to help improve transmission efficiency of electricity, react and restore timely after power disturbances, reduce operation and management costs, better integrate renewable energy systems, effectively save electricity for future usage, and so on. It will also be critical in building better electricity networks to help bring down electricity bills and balance the whole electricity system. In addition, the smart power grid system should  monitor power generation, power demands and help make storage decisions. In terms of security, a smarter grid will also add resiliency to large-scale electric power systems so as to help governments react promptly to emergencies or natural disasters, e.g., severe storms, earthquakes, large solar flares, and even terrorist attacks, etc.}

\subsection{Smart Retail Store Automation}

\begin{figure}[h]
 \centering
 \includegraphics[scale=0.35]{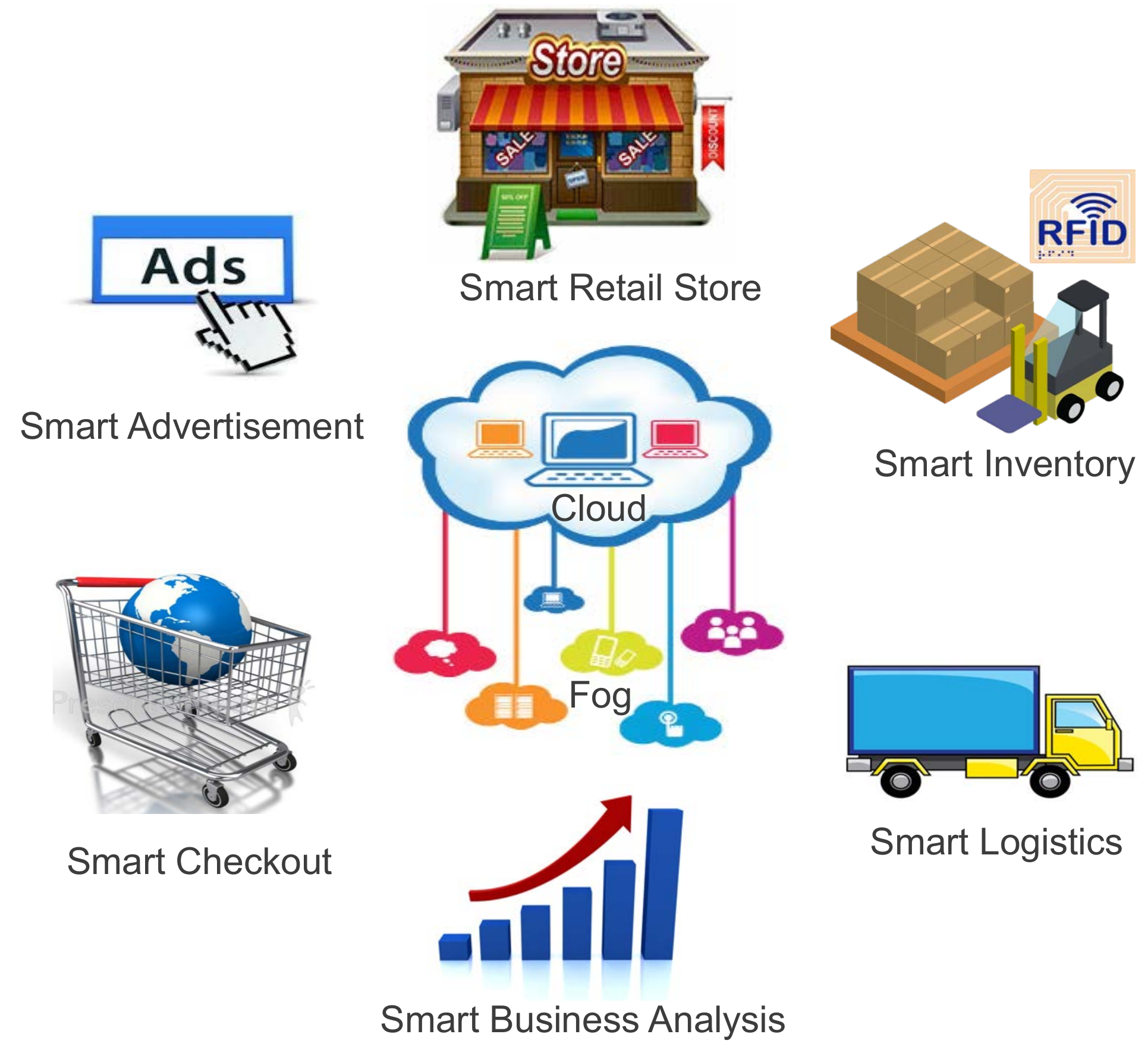}
 \caption{Smart Retail Automation.}
 \label{Figure:automatedretail}
\vspace{-8pt}
\end{figure}

\textcolor{black}{
In future smart cities, shopping experience can be improved dramatically with smart retail store automation. As shown in Figure \ref{Figure:automatedretail}, with the adoption of Fog/Cloud computing technologies, the whole cycle of retail store can be automated with better machine intelligence. For example, a smart retail store physically can support automated scanning, express self-shopping, high-/medium-volume checkout, etc. Virtually, a smart retail store system can help advertise to the best social networks/communities/groups to attract best interests from customers, keep track of sales of assorted products, and hence manage inventory and products ordering effectively and automatically. Cost-effective transportation will also be recommended. Such retail store automation systems will be expected to reduce operation and management costs, react promptly to demands from customers.}

\section{Common Features in Fog Computing}
\label{sec:Characteristics}
In this section, we identify the major   features that are useful in supporting different types of use cases in smart cities. We extract them by both analyzing use case scenarios and reviewing relevant literature. Note that, there are already a number of different academic and industrial research projects implementing some of the features we discuss in this section. More details about those projects will be provided in the next section.  

\begin{figure}[b!]
 \centering
 \vspace{-0.43cm}
 \includegraphics[scale=.41]{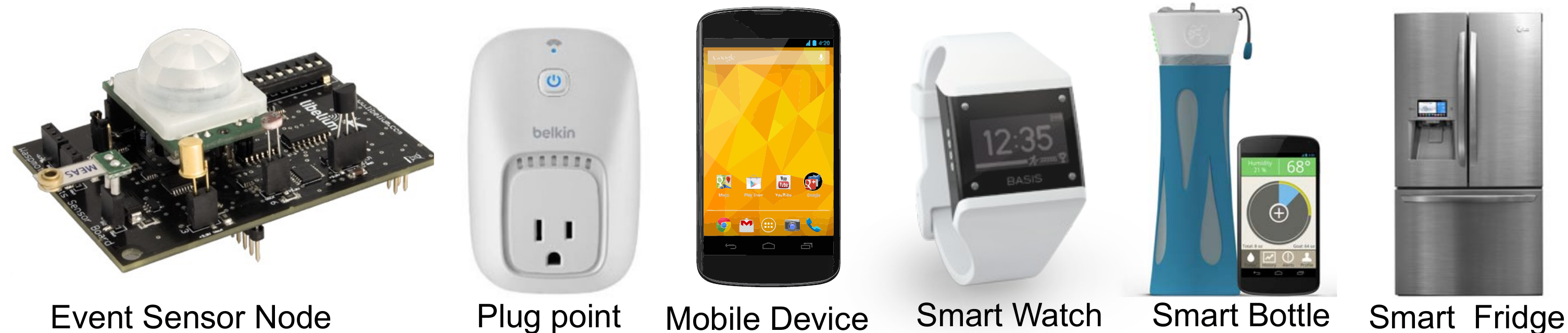}
 \caption{Sample Internet connected objects. ICOs can be in any form as shown}
 \label{Figure:ICO}	
\end{figure}

Let us first introduce what are \textit{edge devices / leaf nodes}. Previously, we introduced a spectrum of devices, in Figure \ref{Figure:Layered_Architecture}, that are expected to find use in IoT solutions. In this paper, we use the term \textit{`edge devices'} or \textit{`leaf nodes'} to refer to any of the devices in category 1,2,3, and 4. As shown in Figure \ref{Figure:Layered_Architecture}, these devices have less computational capabilities compared to cloud composing infrastructures. Figure \ref{Figure:ICO} illustrates several different types of edge devices. These devices could be raw sensors (e.g., Libelium sensors \cite{P416}), smart plugs, mobile devices, smart watches, smart bottles, or smart fridges, just to mention a few. Each of these devices can become a leaf node in an IoT application and perform edge analytics. Any data analysis task performed within an edge device (or leaf node) can be identified as edge analytics. For example, without pushing energy consumption data every second to the cloud, a smart plug itself may perform data processing and avoid redundant data. A smart plug may only send data when there is a fluctuation of the energy consumption. This way the smart plug reduces the total data communication by acting as a fog device. In other words, both communication bandwidth and the total amount of data sent to the cloud over a certain period can be reduced. \textcolor{black}{Internet Engineering Task Force (IETF) 7228\footnote{https://tools.ietf.org/html/rfc7228} has proposed a terminology for constrained-node networks. They aim to standardise small devices   with severe constraints on power, memory, and processing resources.}

\textcolor{black}{IETF defines constrained nodes / devices as \textit{``A node where some of the characteristics that are      otherwise pretty much taken for granted for Internet nodes at the       time of writing are not attainable, often due to cost constraints       and/or physical constraints on characteristics such as size,       weight, and available power and energy.  The tight limits on       power, memory, and processing resources lead to hard upper bounds       on state, code space, and processing cycles, making optimization       of energy and network bandwidth usage a dominating consideration       in all design requirements.  Also, some services such as full connectivity and broadcast/multicast may be lacking''}}

\textcolor{black}{IETF categorise device based on number of criteria: 1) availability of memory, 2) availability of energy, 3) availability of communication. Under each criteria, they have identified few different classes as follows:}

\begin{enumerate}
\item Availability of memory
	\begin{itemize}
	\item C0: $<<$ 10 KiB  data size (e.g., RAM), $<<$  100 KiB code size (e.g., Flash)
	\item C1: appx. 10 KiB data size (e.g., RAM), ~ 100 KiB  code size (e.g., Flash)
	\item C2: appx. 50 KiB data size (e.g., RAM),  ~ 250 KiB code size (e.g., Flash)	
	\end{itemize}
	
\item Availability of energy
	\begin{itemize}
	\item E0: Event energy-limited (e.g., Event-based harvesting )
	\item E1: Period energy-limited  (e.g., Battery that is periodically recharged or replaced)
	\item E2: Lifetime energy-limited (e.g., Non-replaceable primary battery)
	\item E9: No direct quantitative limitations to available energy (e.g, Mains-powered)
	\end{itemize}	
	
\item Availability of communication
	\begin{itemize}
	\item P0: Normally-off (Reattach when required)
	\item P1: Low-power (Appears connected, perhaps with high)
	\item P9: Always-on (Always connected)
	\end{itemize}
	
\end{enumerate}

\textcolor{black}{However, for our discussion in this paper, the exact hardware specifications are not that important. In this paper, we only wanted to highlight the differences of device categories broadly. Typical, Micro-controllers without an OS can be identified as ICOs. Gateway devices usually run an Operating system.}

\textcolor{black}{Over the next ten sub sections, we will introduce the most common features that need to be facilitated by an ideal fog computing platform for IoT applications. We extracted these features through extensive literature and use case scenario analysis. In each section, we present representative set of approach  proposed by researchers in the past. Further details can be obtained though the references presented in Table III.}

\subsection{Dynamic Discovery of Internet Objects}
\label{sec:Dynamic Discovery of Internet Objects} 
\begin{figure}[h!]
 \centering
 \includegraphics[scale=0.40]{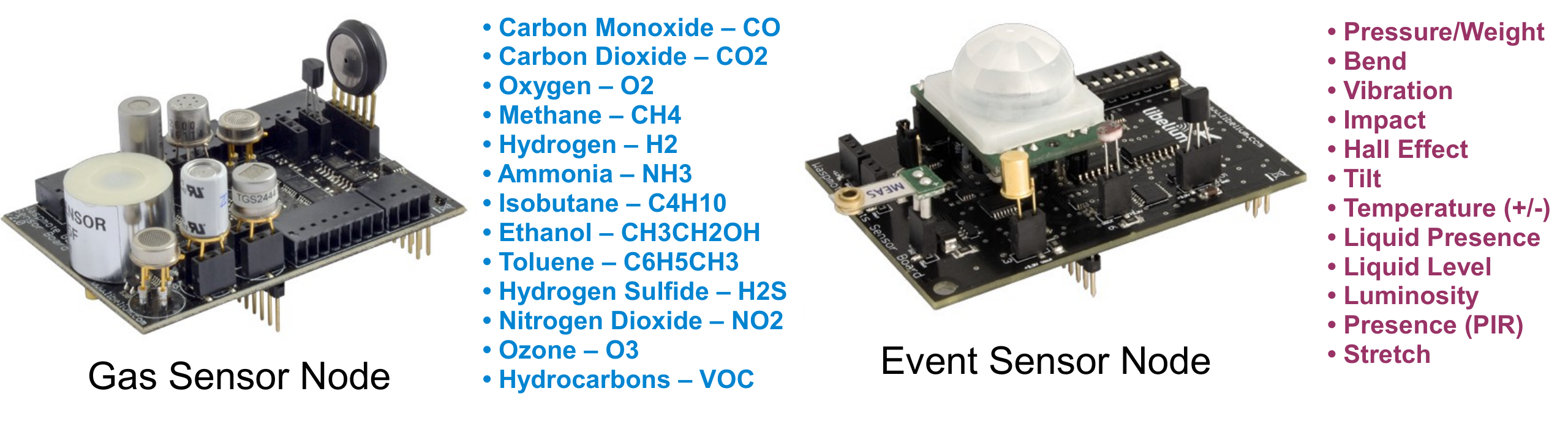}
 \caption{Heterogeneity in terms of sensing/measurement capabilities of sensor nodes}
 \label{Figure:Heterogeneity}
\end{figure}

\begin{figure}[h]
 \centering
 \includegraphics[scale=0.35]{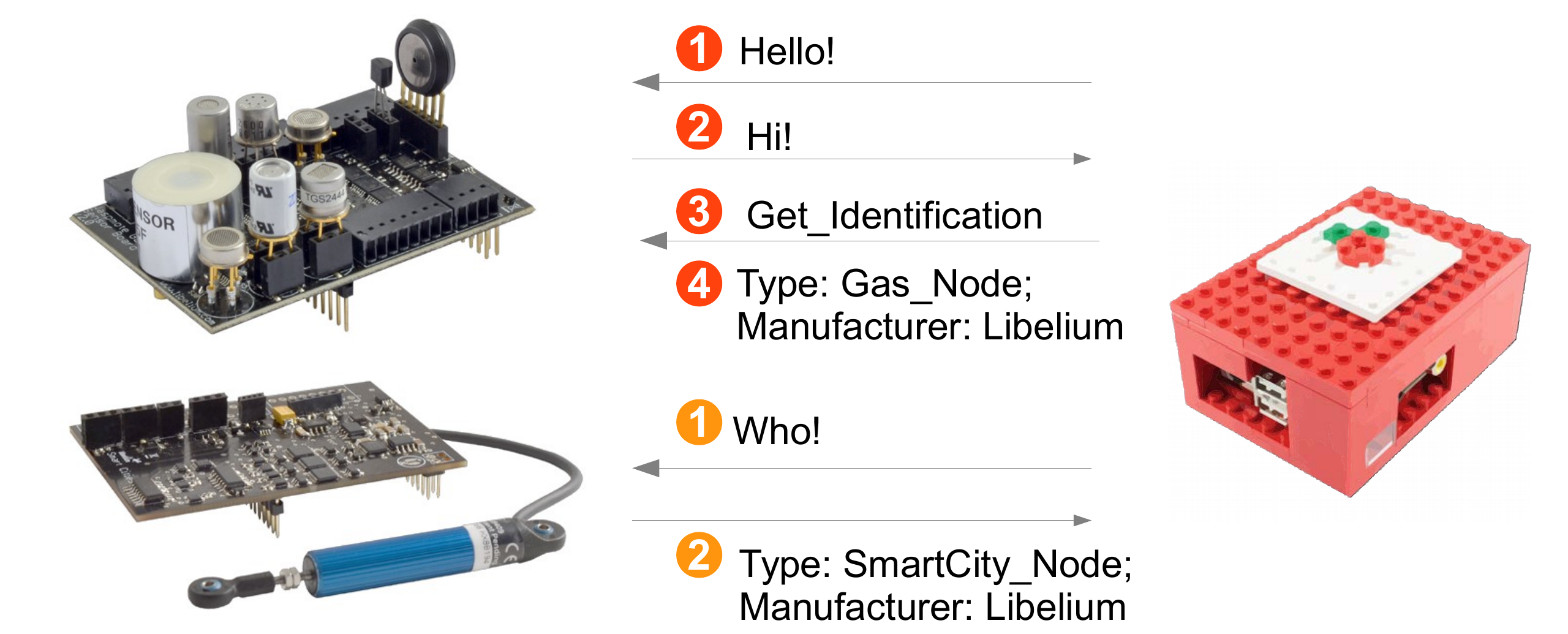}
 \caption{Heterogeneity in terms of communication  sequences.}
 \label{Figure:Communication_Sequence}
\end{figure}

Discovery in IoT has different meanings. Some discoveries happen in the cloud where a user may try to find an Internet Connected Object\footnote{We use both terms, \textit{`Internet Connected Objects'}, \textit{`objects'} and \textit{`things'} interchangeably to give the same meaning as they are frequently used in IoT related documentation. Some other terms used by the research community are \textit{`smart objects'}, \textit{`devices'}, \textit{`nodes'} \cite{FC03}.} (ICO) from a data store or a management system. Figure \ref{Figure:ICO} illustrates some ICOs. In this section, we use the term discovery to refer to the activity of connecting an ICO to a gateway device \cite{FC12FC61} in order to build a fog network. Some  major challenges in discovery include heterogeneity, security, and dynamicity. Heterogeneity is in multiple levels from communication protocols, application protocols, discovery sequences, sensing and capabilities, and so on. For example, each ICO may have different sensing capabilities as shown in Figure \ref{Figure:Heterogeneity}. Further, each sensor may use different types of application level discovery sequences and security measurements, which are hard to standardize \cite{PereraC011}. An example is shown in Figure \ref{Figure:Communication_Sequence}. 

\textcolor{black}{Alongside sensing data from environments, reaction to environmental changes is critical. To this end, actuators can be used to put something into automatic action and can work with assorted sensors. In Fog computing and Internet of Things, actuators will play an important role, where machine actions can be taken automatically without human intervention. This is important specially in the situations where human beings will be at risk or instant actions are required, and so on. 
}

\textcolor{black}{Actuators can be categorized into linear actuators and non-linear actuators, depending on the direction of motion generated by the actuator. These actuators can also be categorized into types of actuators based on the source of energy for the motion, including mechanical actuators, hydraulic actuators, pneumatic actuators, piezoelectric actuators, etc. Take linear actuators as an example, some common linear actuators include \cite{diffOfActuators}: Pneumatic linear actuators rely on pressure from an external compressor or manual pump to produce required motion. Hydraulic linear actuators are similar to pneumatic actuators, but they gain force from an incompressible liquid from a pump. An electric linear actuator relies on electrical energy to drive motion. Different types of actuators have their own advantages and disadvantages in a number of aspects, e.g., pressure limitation, temperature scope, accuracy, reaction time, maintenance and programming cost, etc \cite{diffOfActuators}. In microelectromechanical systems (MEMS), microactuators are fundamental and are designed based on different actuation principles, such as shape-memory alloys, electrostatic, electrothermal, piezoelectric, pneumatic and electromagnetic etc \cite{rpme/JiaX2013}.
}

On the other hand, nanoscale technology will develop devices as small as one to a few hundred nanometers ($10^{-9}$ meters) \cite{iotAtNanoScale}. 
\textcolor{black}{Dynamic discovery and configuration is critical in nanoscale sensor domain as the number small devices are growing rapidly and they are getting deployed everywhere. These nano sensors need to be discovered and configured continuously as they tend to be unreliable. They also tend to move from one place to another}. It is reported that the U.S. National Nanotechnology Initiative requested \$1.5 billion in federal funding in the year of 2016, which was also roughly the average amount of annual funding awarded to the U.S. National Nanotechnology Initiative in the last 15 years. Nowadays, there are some significant advances in the development of nanosensors \cite{iotGoesNano}. For example, some nanosensors to date created by the tools of synthetic biology can be used to modify single-celled organisms, such as bacteria. Other nanosensors made from non-biological materials include carbon nanotubes, which can both sense and signal, acting as wireless nanoantennas.

\textcolor{black}{There are many potential applications of sensors at nanoscale (nanosensors). For example, in-body networks can help monitor real-time blood, sickness and breath tests. Hence, intrabody nano-networks for healthcare applications can provide real-time health information and work status, where personal health can be monitored anywhere and anytime \cite{nanoThings}. Further, the spread of viruses and diseases can be monitored in public locations and real-time actions can be arranged. In the future, billions of nanosensors will help harvest huge amounts of real-time information of our cities, homes, offices, factories, and even our bodies. Fog computing and Cloud computing technologies can then provide much more detailed, inexpensive and up-to-date pictures of our living environments, leading to the Internet of Nano-Things. 
}

\begin{figure}[h!]
 \centering
 \includegraphics[scale=.30]{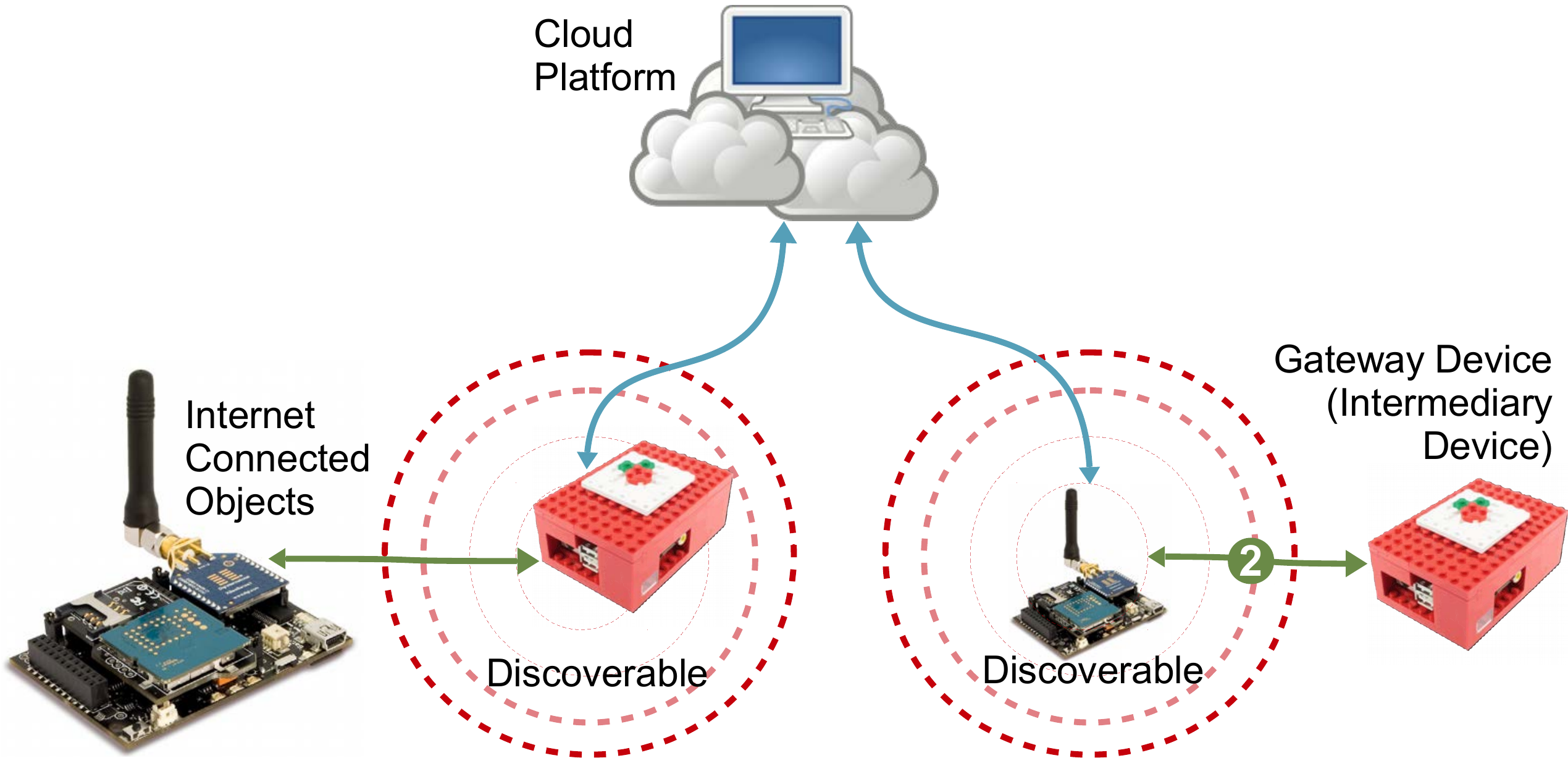}
 \caption{ICO detection and discovery}
 \label{Figure:ICO_Discovery}	
\end{figure}

The ICO detection and discovery may happen in two different ways as shown in Figure \ref{Figure:ICO_Discovery}. One method is that ICOs may continuously search for fog gateway devices while gateway devices are passively waiting to be connected with ICOs. In this method, gateway devices become discoverable. In the other approach, fog gateway devices may continuously search for ICOs around them. In this method ICOs need to be discoverable. Typically, both methods may use  low range communication protocols. We will briefly introduce several different types of discovery and communication protocols proposed in the IoT domain in Section \ref{Multi_Protocol_Support_Application_Level}.

\subsection{Dynamic Configuration and Device Management}
\label{sec:DynamicConfigurationandDeviceManagement}
\begin{figure}[t!]
 \centering
 \includegraphics[scale=0.40]{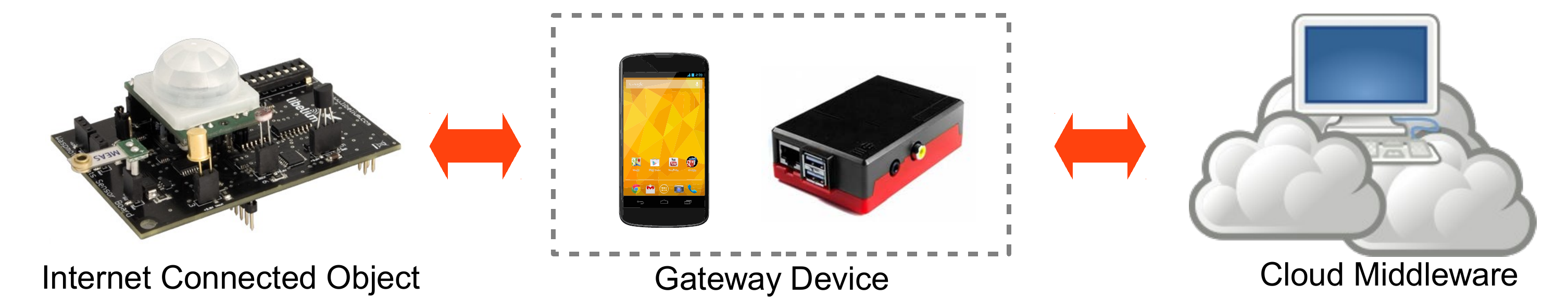}
 \caption{Connectivity between ICOs and Cloud  platforms need to be established and configured.}
 \label{Figure:Configuration}	
\end{figure}

In IoT, connection establishment and configuration need to be occurred in two different places. Firstly, ICOs need to be connected to gateway devices. Secondly, fog gateways need to connect to IoT cloud platforms. Above two approaches are illustrated in Figure \ref{Figure:Configuration}.  It is important to note that industrial IoT cloud platforms such as IBM Blumix and Microsoft Azure IoT distinctly identify sensing devices and fog gateways as two separate categories and provide specialized configuration options unique to each of the categories.

Let us walk through the process of connection establishment and configuration in an IoT application. First, ICOs need to connect to its corresponding fog gateway. To do this, both parties need to negotiate and agree to a  common protocol and message sequence \cite{Perera2014Book}. Next, ICOs need to introduce them to fog gateways by describing themselves (e.g., sensor ID, manufacture details, sensor types). Then, the fog gateways can use such information to configure themselves  accordingly (e.g., prepare to accept data in a certain way) \cite{Perera2014Book}. Further more, ICOs also need to configure their   scheduling calendar, communication frequency,  data acquisition methods and sampling rate  \cite{Perera2014Book}. In the next step, fog gateways need to send the data they collected from ICOs to the IoT cloud platforms. This allows the IoT clod platform to prepare itself to accept incoming data and send commands back to the ICOs. We identify this process as registration. As a result of the registration process, the cloud IoT platform gets to know about the data availability through each fog gateway. Additionally, the registration process provides several pieces of information to the cloud about fog gateways and ICOs (e.g., location, energy source and level). The cloud IoT platforms can use such information to develop energy and privacy aware data collection plans \cite{PereraC011}. As we will discuss later, such efficient data collection plans are vital towards building sustainable smart cities. \textcolor{black}{Typical, each ICO may have support one  protocol and message sequence that can be used for configuration \cite{Perera2014Book}. Gateways will be required to intelligently find out which message sequence need to be used to configure a given ICO. The ICO description languages such as Hypercat \cite{Hypercat} can be used to identify how to configure each ICO when a given ICO met a gateway for the first time.}

In a typical  pervasive computing environment, only few ICOs are used in an application (e.g., smart home, smart office). However, the IoT paradigm aims to connect billions of sensors to the Internet. Therefore, establishing connectivity between ICOs and IoT applications becomes a challenge that developers  need to face more frequently. It is inefficient and cumbersome to connect ICOs to IoT applications manually through hard wired coding \cite{Perera2014Book}. It is essential to develop automated (or at least semi-automated)  techniques to support dynamic configuration of ICOs. Towards this, one of the challenges is to develop ICO description techniques so the IoT application can find more information about each ICO and how to connect to them (e.g., sensors'  capabilities,  data structures they produce, hardware/driver  level configuration  details). Over the last few years, there has been few different approaches that aims to tackle this challenge of describing ICOs. Sensor ontologies \cite{P103}, Transducer Electronic  Data Sheet (TEDS)  \cite{P258}, Sensor Device Definitions \cite{ZMP002}, and Sensor Markup Languages (SensorML)   \cite{P256} to name a few. Further, there are recent application level protocols and frameworks proposed to automate the ICO configuration process and make it scalable. Some examples are AllJoyn, Iotivity \cite{IoTivity}, HyperCat \cite{Hypercat}. We will briefly introduce these protocols and frameworks in Section \ref{Multi_Protocol_Support_Application_Level}.

\subsubsection*{Based on Responsibility}

\begin{figure}[h]
 \centering
 \includegraphics[scale=0.34]{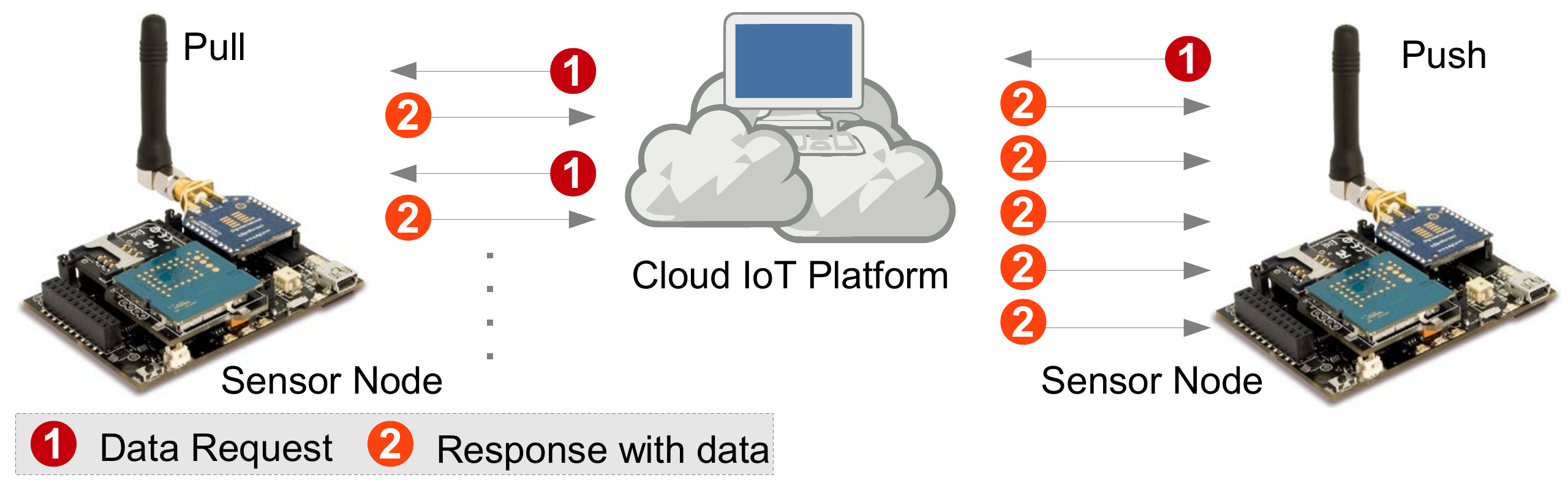}
 \caption{Both pull (left side) and push (right side) techniques can be used retrieve data from ICOs}
 \label{Figure:Push_and_Pull}
\end{figure}

As shown in Figure \ref{Figure:Push_and_Pull}, data can be captured by sensors using two main approaches \cite{P334}: push and pull. A detailed comparison is presented in Table V in \cite{ZMP007}.


\begin{itemize}

 \item Pull: The fog gateways  retrieve data from ICOs using this techniques. Further, IoT cloud platforms may also use pull techniques to acquire data from fog gateways.

 \item Push:  The ICOs may use this techniques to send data to fog gateway. Similarly, the fog gateway may also use this techniques to push data to the IoT cloud platform.
 
 The physical or virtual sensor  pushes data to the software component that is responsible to acquiring sensor data periodically  or instantly. However, periodical or instant pushing can be employed to facilitate a publish and subscribe model.

\end{itemize}

\subsubsection*{Based on Frequency}

More broadly, data is captured by sensors using two main approaches: instant events and interval  events. A detailed comparison is presented in Table VI in \cite{ZMP007}.


\begin{itemize}
 \item Instant (also known as threshold violation): This method sends data to a certain destination when an event get triggered. An event can be triggered by the sensors themselves (e.g., temperature reach a certain number, motion sensor detect a motion). This approach supports both push and pull techniques.  However, ICO can only use the push method where it sends data when an event occurs. On the other hand, a fog gateways may use pull approach to query it's ICOs when an event occurs.
 
 \item Interval (also known as periodically):  This method sends data to a certain destination periodically. That means that the event is triggered by  \textit{time}. For example, an ICO may be configured to sense and send data every 20 seconds. On the other hand,  a fog gateway may be configured to pull data from ICOs periodically every 20 seconds. We identify data sensing date as \textit{`sampling rate'}. This approach supports both push and pull techniques.  

\end{itemize}

\begin{figure}[h!]
 \centering
 \includegraphics[scale=0.35]{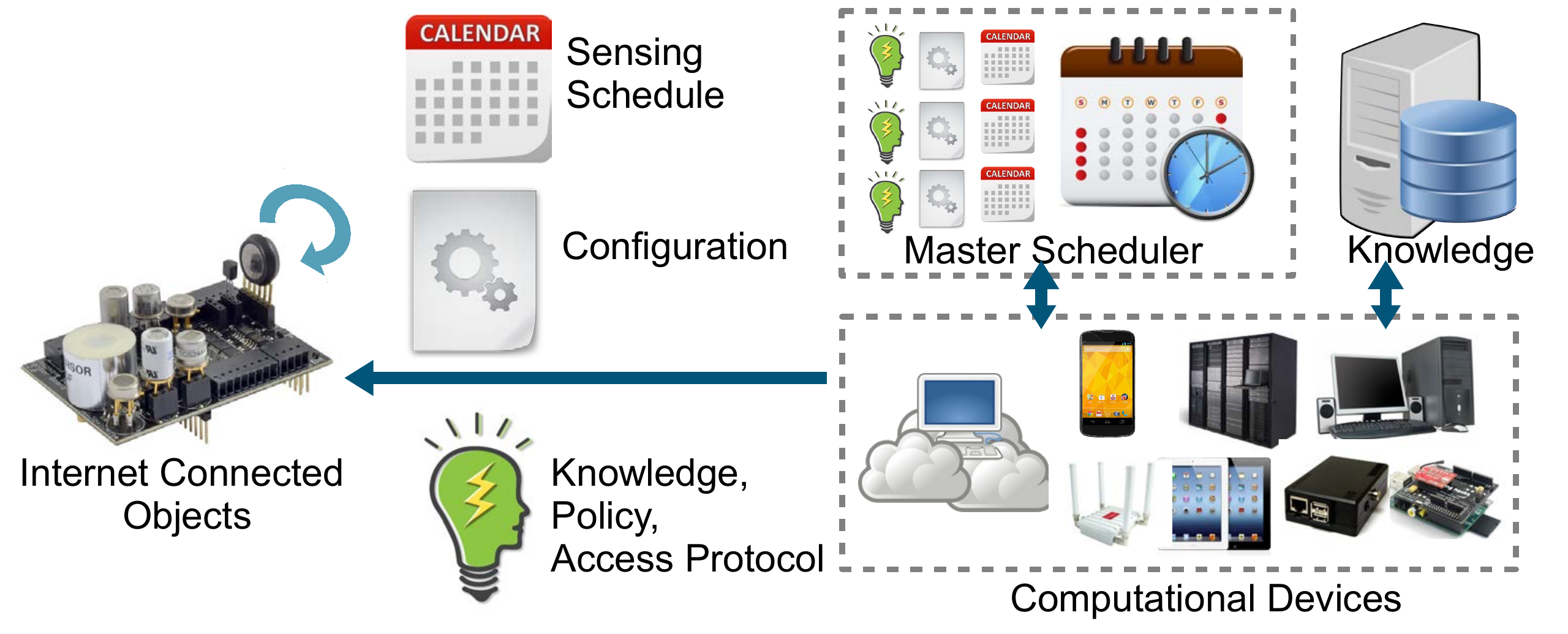}
 \caption{Dynamic configuration ICOs and fog gateways based on context}
 \label{Figure:Reconfiguration}
\end{figure}

So far we have discussed several different parameters that need to be configured in both ICOs and fog gateways. Such configurations are dynamically performed based on the application requirements. Sampling rate and communication frequency are the two most important parameters, because they make direct impact on the lifetime of ICOs due to energy constraints. This is a very important factor in building sustainable smart cities. Typically, both ICOs and fog gateways do not have the full picture of the sensing application. Therefore, ideally, the responsibility of designing an efficient sensing plan needs to be  passed-on to the cloud IoT platforms. Firstly, the cloud IoT platforms needs to establish an overall sensing strategy by considering the overall requirements of the smart city application. Secondly, it needs to delegate the sensing responsibility to individual fog gateways and ICOs by dividing the overall sensing plan into mini schedules. These mini sensing schedules and configuration details then need to be pushed into both gateways and ICOs as shown in Figure \ref{Figure:Reconfiguration}. However, it is important to note that overall sensing requirements of a smart city application may change over time at runtime. Therefore, the above mentioned process may need to take place repeatedly to maintain the overall efficiency of the application over time \cite{FC09}. \textcolor{black}{Both ICOs and fog gateways are typically unreliable and prone to breakdowns and malfunction that will require replacements. Cloud companions can be efficiently used to keep track of both ICOs and gateways. Ideally, an image  (or a configuration clone) needs to be store in the cloud so that in case of a breakdown, new devices can be quickly deployed and configured using the cloud images instead of conducting discovery and configuration processes which take time and energy.}

\subsection{Multi-Protocol Support: Communication Level}
\label{Multi_Protocol_Support_Communication_Level}

\begin{figure*}[t!]
 \centering
 \includegraphics[scale=0.45]{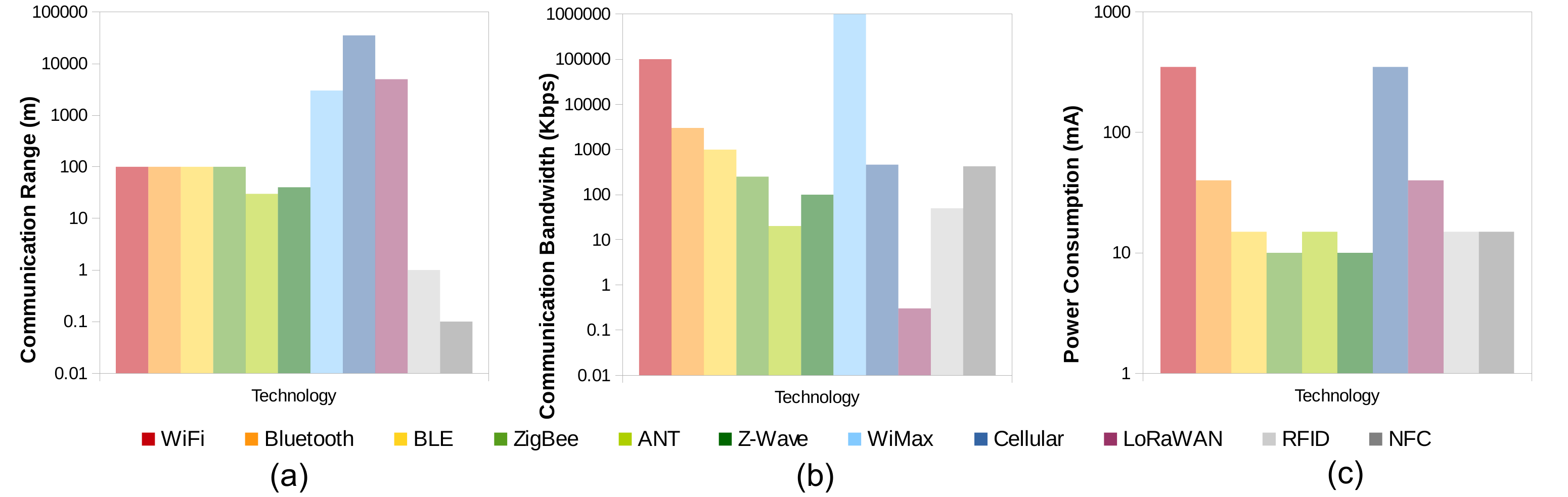}
 \caption{(a) Communication Ranges; (b) Communication Bandwidth; (c) Power Consumption. It is important to note that the above measurements can vary greatly based on both external factors, e.g., how the measurements are taken, and how the hardware devices are being built. Further, some protocols have a number of different variations that have different capabilities with regards to the above aspects. To simplify the presentation, we restrict our summary to the most popular version of the protocols. Our intention is to give a broader look at how these different protocols fit in the IoT space. Data Sources: \cite{Halperin2010,Gomez2013,Bluetooth2005,Mohandas1060,Dementyev2013,Texas2014,Nuaymi2007}}
 \label{Figure:ComparisonCommunicationProtocols}
\end{figure*}

Even though our intention is not to survey different communication protocols, we briefly introduce some major communication protocols used in smart city applications and especially, in the fog computing domain. It is important to note that none of these protocols is superior to another. This is simply because all of them are superior in some aspects and weak in other aspects. Ideally, each protocol is designed to support different use case scenarios efficiently than other alternatives. In the fog computing domain, specially from the gateways point of view, it is important to support different types of communication protocols so that a wider range of edge nodes or ICOs can be connected to these gateways. In Figure \ref{Figure:ComparisonCommunicationProtocols}, we present a comparison of IoT communication protocols from three different perspectives, namely, (a) communication range, (b) bandwidth, and (c) power consumption.

\begin{figure}[h!]
 \centering
 \includegraphics[scale=0.60]{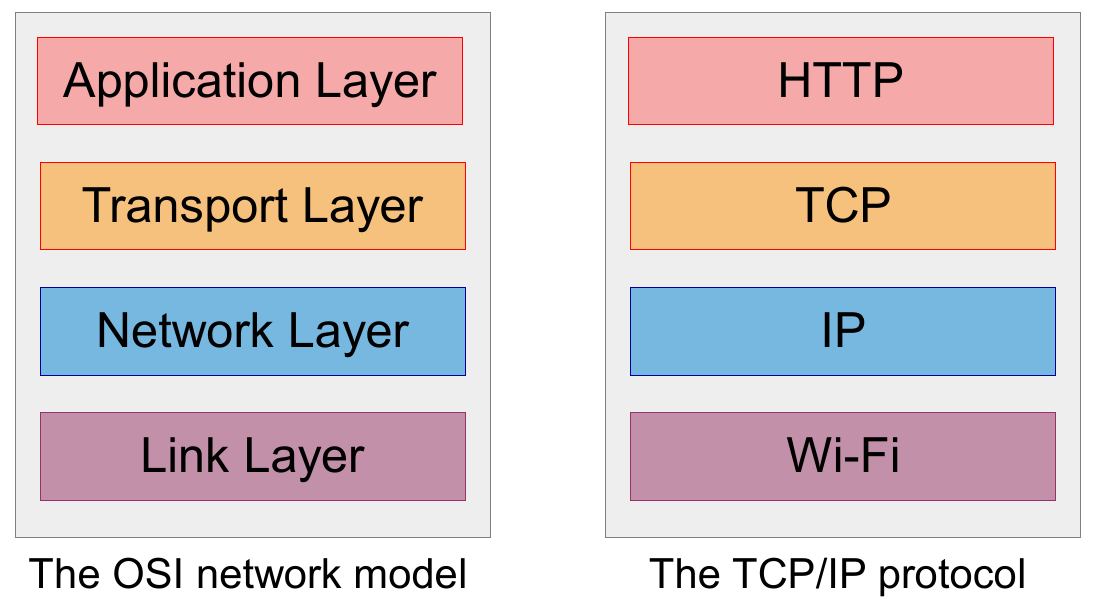}
 \caption{Simplified Version of OSI Model and TCP/IP/HTTP Stack}
 \label{Figure:OSIModel}
\end{figure}

Figure \ref{Figure:OSIModel} presents a simplified version of Open Systems Interconnection model (OSI model) \cite{Texas2014}. In parallel, we present an example using TCP/IP stack. Before we get into the discussion of different communication protocols, it is useful to understand where each protocol fits in. The link layer is responsible for converting radio signals to bits and vise versa. The network layer provides addressing capabilities that are being used to route data through  networks while the transport layer manages the communication between two application endpoints. Finally, the application layer is responsible for data formatting.

\begin{figure}[h!]
 \centering
 \includegraphics[scale=0.60]{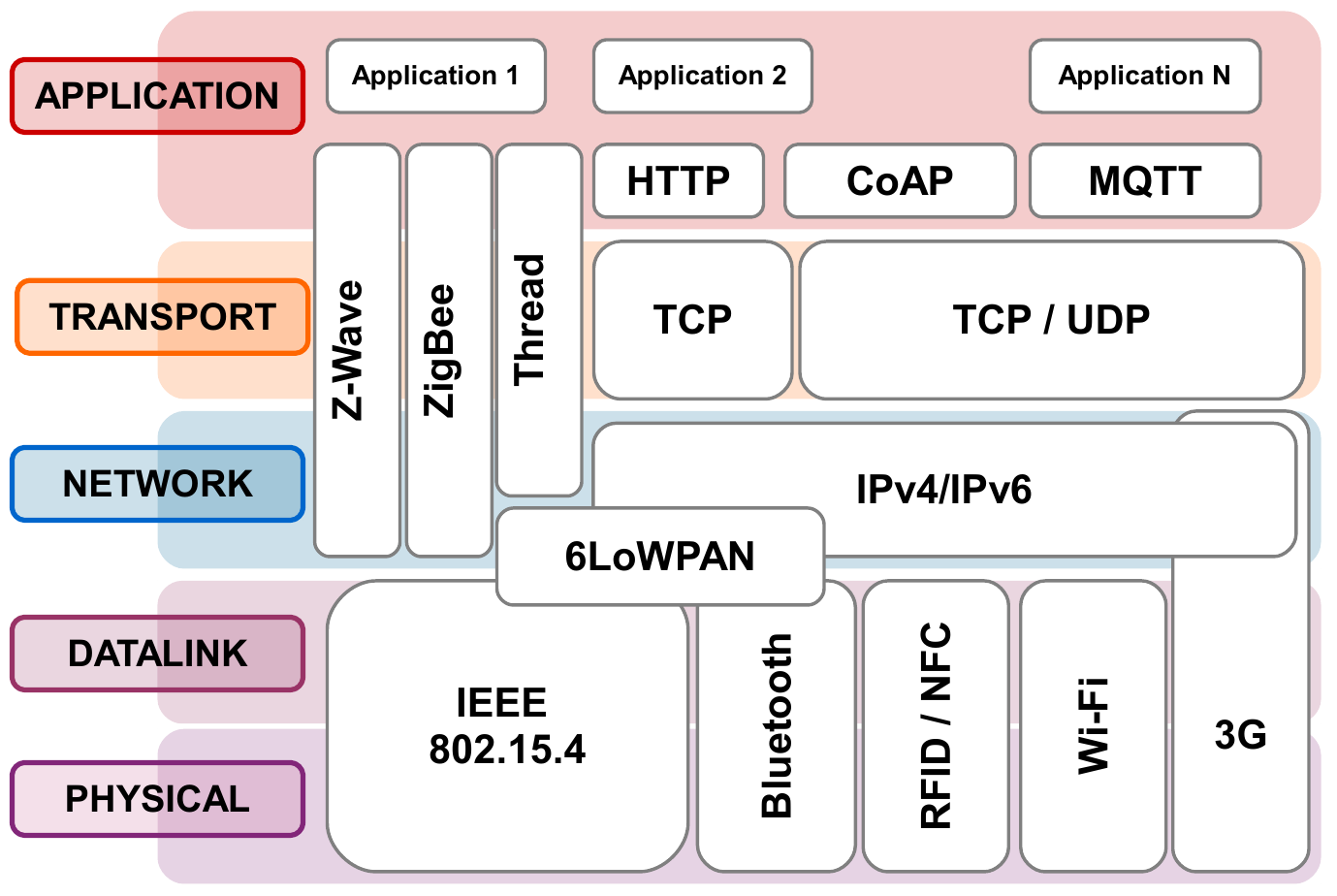}
 \caption{Overview of Communication Protocols with respect to OSI model \cite{Texas2014}}
 \label{Figure:Communication_Protocols_Comparison}
\end{figure}

In Figure \ref{Figure:Communication_Protocols_Comparison},  we illustrate how different communication protocols fit in with respect to the OSI model \cite{Texas2014}. Some of the protocols are designed to communicate over short distances and are more suitable to conduct communication between ICOs and fog gateways. Other protocols are typically suitable to communicate between gateway devices and the cloud  infrastructure. As a result, in fog computing,  gateways should be able to deal with different types of communication protocols, mostly in parallel as illustrated in Figure \ref{Figure:Configuration}.

As we mentioned earlier, each communication protocol has its own strengths and weaknesses \cite{Texas2014}. For example, one issue in TCP/IP based communication is that it is a fairly complex protocol. TCP/IP is larger in size and requires comparatively considerable amount of computational resources and memory. Such hardware requirements are sometimes hard to facilitate by ICOs in IoT due to miniature sizes and cost related issues. The complexity also leads to larger data packets and more power required to send and receive packets. To address this issue, different parties have developed a variety of different protocols as discussed below. However, today hardware is becoming cheaper and smaller than ever before. This movement encourages  adopting TCP/IP even for resource constrained ICOs. \textcolor{black}{The network topologies supported by each protocol is important as they  impact performance and cost of  deployment. They helps to determine which protocol to  use in which parts of the fog network in a given IoT application scenario.}

\textbf{Wi-Fi:} This is one of the most commonly used wireless local area network (WLAN)technology \cite{Yeh2003}. Wi-Fi can be used to connect different types of nodes (e.g., ICOs) to a network so they can communicate with each other. It is developed by Wi-Fi Alliance \cite{Henry2002}. Substantial amounts of IoT devices are expected to connect to the Internet via Wi-Fi \cite{TexasInstruments}. However, most of the time, Wi-Fi is used by high end devices such as computers, tablets, mobile phones, gaming devices printers, cameras, and more recently smart refrigerators and microwaves. These high end devices are connected to a WLAN using Wi-Fi through as access point (or hotspot). Typically, 20 meters (66 feet) coverage can be provided by a single hotspot. However, coverage may be extended  in out doors due to lack of obstacles. Typically, up to 250 devices can be connected to each Wi-Fi access point in parallel \cite{Texas2014}.  Typically, an IoT device can be connected to the Internet using Wi-Fi over a year using two AA alkaline batteries by using efficient power management (e.g., fast on-off time, long sleeps). Wi-Fi is an implementation of the standard IEEE 802.11.  Wi-Fi Protected Access (WPA), Wired Equivalent Privacy (WEP), and WPA with  Advanced Encryption Standard (WAP2) have been introduced over time where the latter ones are better then the previous \cite{Lashkari2009}. However, non of these methods are considered fully secure.  Virtual Private Networks  (VPN) \cite{Harmening2013} and secure Hypertext Transfer Protocol over Transport Layer Security (HTTPS) are recommended to use to secure the Wi-Fi based communication. Wi-Fi mostly uses the star network topology to connect devices.

\textbf{Bluetooth:}  This is also one of the most common wireless communication technologies developed to enable wireless personal area networks (WPANs) \cite{Zimmerman1996}. Bluetooth is deigned to support data communication over short distances. This standard is developed by Bluetooth Special Interest Group and standardized as IEEE 802.15.1 \cite{Bluetooth2005}. Bluetooth has three variations in terms of device classes where each class supports different communication ranges at the cost of power consumption\footnote{Device class 1 consumes 100 mA and intended transmission range is 100m. Device class 2 consumes 2.5 mA and intended transmission range is around  10m. Device class 3 consumes 1 mA and intended transmission range is less than  10m}. In IoT context, Bluetooth is used to enable communication between low-powered sensors and fog gateways. Bluetooth can only connect 8 devices at a time using a star network topology \cite{Texas2014}. The time taken to connect two Bluetooth devices is less than 6 seconds.

\textbf{Bluetooth Low Energy (BLE) / Bluetooth Smart:} This is a variation of traditional Bluetooth specifically designed to operate under low energy consumption \cite{Heydon2012}. It can communicate within 100m at 1 Mbps using less than 10mA. Further, in contrast to Bluetooth,  it does not have limitations on number of devices to be connected. BLE is widely used in developing beacons that are being used for indoor localizations and related IoT applications \cite{Tommy2016}. The time taken to connect two blueetooth devices is less than 0.006 seconds \cite{Mohandas1060}.

\textbf{Zigbee / ZigBee PRO:}
ZigBee  is a protocol suite that covers network, transport, and application layers. ZigBee physical and link layer support is built on IEEE 802.15.4 specification.It is one of the most popular, low-cost, low-throughput, and low-power wireless mesh networking standards \cite{Texas2014}. In mesh networks, data moves from one node to another in different directions until it reaches the destination. ZigBee can be configured to operate with very long sleep intervals. It can also be configured to have  very low duty cycles. \cite{Alliance2008}. ZigBee can connect 65,536 devices to a network. ZigBee PRO is an enhanced version of ZigBee. In order to connect ZigBee devices to Internet, application level gateway is required. This particular gateway needs to connect to the   ZigBee network as a node and it also needs connect to the Internet using either Ethernet or Wi-Fi. Additionally, the gateway also needs to support TCP/IP communication so it can transfer data from Zigbee network to the Internet.

\textbf{ANT:} This is an proprietary protocol designed for low bit-rate and low power sensor networks. Conceptually, ANT\footnote{https://www.thisisant.com/developer/ant-plus/ant-antplus-defined} is similar to both ZigBee and BLE \cite{Dementyev2013}. ANT defines a wireless communication protocol stack. In an ANT network, nodes can play dual roles where they can be either masters or slaves. As a result, each node can  transmit, receive or do both in order to enable routing data between nodes within the network \cite{Buratti2009}. One of the primary application of ANT is for fitness and sport related IoT devices. ANT can have  long sleep periods during which power is consumed in micro amps. ANT can connect 65,533  devices to a network. ANT supports point-to-point, star, tree, and mesh topologies.

\textbf{Z-Wave:} This is a  reliable, low-latency wireless communication certification specifically designed to facilitate machine to machine communication in order to support smart home applications \cite{ZWave2015}. The technology is designed to run on low powered battery operated devices devices. Each Z-Wave network can have  up to 232 nodes per controller node.  Controllers nodes and slave devices (nodes) are the two types of nodes that each Z-Wave network has. Z-Wave operates on mesh network topologies.

\textbf{6LowPAN:} 6LoWPAN is an acronym of \textit{``IPv6 over Low power, low throughput  Wireless Personal Area Networks''} \cite{Shelby2009}. This protocol aims at enabling IP based communication for small low powered devices with limited processing capabilities. IoT is one of the main application domains for which 6LoWPAN has been designed. The 6LoWPAN standard only defines an efficient adaptation layer between the network layer and the data link layer as shown in Figure \ref{Figure:ComparisonCommunicationProtocols}. 6LoWPAN needs an connectivity to the Internet in some way (e.g., Ethernet or Wi-Fi gateway). 6LoWPAN is built on the 802.15.4. Therefore, it has the advantages such as \textit{``mesh network topology, large network size, reliable communication and low power consumption''} \cite{Ma2008} as well as advantages of IP-based communication.

\textbf{Thread:} Thread is a network protocol that is IPv6-based,  royalty-free  protocol for smart home devices to communicate over a network \cite{ThreadGroup2015}. It is designed to compete with Z-Wave and Zigbee. Thread runs over 6LoWPAN, which in turn uses the IEEE 802.15.4 wireless protocol. As a result, Thread runs on mesh network topologies and can handle up to 250 nodes.

\textbf{WiMax:} Worldwide Interoperability for Microwave Access  is a set of standards  for wireless communications  \cite{Nuaymi2007}.  Roughly, WiMax can be considered as a a standardized wireless version of Ethernet. It focuses  on enabling  broadband access to users. WiMax is an interoperable implementation of the IEEE 802.16. Compared to Wi-Fi, WiMax provides higher speeds (1 Gbps) over greater distances (3 Km) and for a greater number of users. WiMax is a competitor of Long-Term Evolution (LTE).

\textbf{Cellular (GSM / HSPA):} \textcolor{black}{High Speed Packet Access (HSPA) protocol enables long range support for IoT applications.} It sends data over the exiting cellular networks. Data communications will cost in terms of both money and power. Cellular can communicate over 35km using GSM and 200km over HSPA. Cellular also has the capability to transfer data at different speeds based on the technology such as \textit{``35-170kbps (GPRS), 120-384kbps (EDGE), 384Kbps-2Mbps (UMTS), 600kbps-10Mbps (HSPA), 3-10Mbps (LTE)''} \cite{RSComponents}.

\textbf{LPWAN (LoRaWAN):} Long Power Wide Area Network (LPWAN) \cite{LinkLabs2016} is a low powered, low cost, low bit rate protocol design for two way secure communication in the IoT domain. LPWAN allows  battery-operated sensors to communicate over long distances. \textcolor{black}{LPWAN gateways deployed in a building or tower can connect to sensors more than miles away (2-5km (urban environment), 15km (suburban environment)). It can also penetrate through water, underground, or in basements (up to 50 meters).} \textit{Star-of-stars} topology is being typically used by LPWAN networks. Moreover, LPWAN uses gateways to bridge the communication between Internet and leaf nodes \cite{Vangelista2015}. Data can be communicated at 0.3-50 kbps \cite{LinkLabs2016}. Some competing LPWAN protocols proposed are ``Sigfox (30-50km rural environments, 3-10km urban environments, up to 10-1000bps), Neul (10km range, up to 100kbps)'' \cite{RSComponents}, and Weightless \cite{LinkLabs2016}.

\textbf{RFID:} Radio-frequency identification \cite{Want2006} uses  electromagnetic fields to transfer data. The tags contain electronically stored information. RFID systems can be categorized based on the types of tags and readers. Passive Reader Active Tag (PRAT) comprises a passive reader and active tags where the reader receives radio signals from active tags. Active Reader Passive Tag (ARPT)  comprises an active reader and passive tags. The active reader transmits the interrogator signals and read the data stored in the passive tag.  Active Reader Active Tag (ARAT) has an active reader and active tags. Active RFID can perform communication over 100m  and Passive RFID is normally limited within around 100cm. RFID tags come in different sizes and look like stickers, cards, and so on. RFID tags are widely used in IoT applications specially to enrich non-electronic dump objects. RFIDs are commonly used for tasks such as tracking (persons, animals, goods), management of asserts,  contact-less payment, travel cards (e.g. buses, trains, tubes), and automated toll collection.

\textbf{NFC:} Near field communication \cite{Coskun2013}  is a communication protocol that enables communication between devices within 10cm. Similar competing technologies are  bar codes and LF passive RFID tags. NFC supports a peer to peer communication model. NFC devices can work in three modes, namely, NFC card emulation (for smart card based payments), NFC reader/writer (for smart tag based smart posters), and NFC peer-to-peer (for machine to machine communication). Typically, NFC tags are passive data stores that can contain between 96 and 8,192 bytes. NFCs are read-only in normal use. NFC standards cover communication protocols and data exchange formats. These standards are generally based on existing RFID standards. NFC is not limited to tag based communication. NFC is also widely used to enable communication between two smart devices, such as between smart phones and washing machines \cite{Want2011}.

\begin{table*}[t!]
\label{tbl:HTTPMQTT}
\centering
\tbl{Performance Comparison Between HTTP and MQTT \cite{Karasiewicz2013}}{
\footnotesize
\begin{tabular}{llllll}
\hline
\multicolumn{2}{|c|}{Characterisitcs}                                                                      & \multicolumn{2}{c|}{3G}                                        & \multicolumn{2}{c|}{Wi-Fi}                                     \\ \hline
\multicolumn{1}{|l|}{}                                   & \multicolumn{1}{l|}{}                           & \multicolumn{1}{l|}{HTTP}     & \multicolumn{1}{l|}{MQTT}      & \multicolumn{1}{l|}{HTTP}     & \multicolumn{1}{l|}{MQTT}      \\ \hline
\multicolumn{1}{|l|}{\multirow{4}{*}{Received Messages}} & \multicolumn{1}{l|}{Messages / Hour}            & \multicolumn{1}{l|}{1,708}    & \multicolumn{1}{l|}{160,278}   & \multicolumn{1}{l|}{3,628}    & \multicolumn{1}{l|}{263,314}   \\ \cline{2-6} 
\multicolumn{1}{|l|}{}                                   & \multicolumn{1}{l|}{Percent Battery / Hour}     & \multicolumn{1}{l|}{18.43\%}  & \multicolumn{1}{l|}{16.13\%}   & \multicolumn{1}{l|}{3.45\%}   & \multicolumn{1}{l|}{4.23\%}    \\ \cline{2-6} 
\multicolumn{1}{|l|}{}                                   & \multicolumn{1}{l|}{Percent Battery / Messages} & \multicolumn{1}{l|}{0.01709}  & \multicolumn{1}{l|}{0.00010}   & \multicolumn{1}{l|}{0.00095}  & \multicolumn{1}{l|}{0.00002}   \\ \cline{2-6} 
\multicolumn{1}{|l|}{}                                   & \multicolumn{1}{l|}{Messages Received}          & \multicolumn{1}{l|}{240/1024} & \multicolumn{1}{l|}{1024/1024} & \multicolumn{1}{l|}{524/1024} & \multicolumn{1}{l|}{1024/1024} \\ \hline
\multicolumn{1}{|l|}{\multirow{3}{*}{Send Messages}}     & \multicolumn{1}{l|}{Messages / Hour}            & \multicolumn{1}{l|}{1,926}    & \multicolumn{1}{l|}{21,685}    & \multicolumn{1}{l|}{5,229}    & \multicolumn{1}{l|}{23,184}    \\ \cline{2-6} 
\multicolumn{1}{|l|}{}                                   & \multicolumn{1}{l|}{Percent Battery / Hour}     & \multicolumn{1}{l|}{18.796\%} & \multicolumn{1}{l|}{17.806\%}  & \multicolumn{1}{l|}{5.446\%}  & \multicolumn{1}{l|}{3.66\%}    \\ \cline{2-6} 
\multicolumn{1}{|l|}{}                                   & \multicolumn{1}{l|}{Percent Battery / Messages} & \multicolumn{1}{l|}{0.00975}  & \multicolumn{1}{l|}{0.00082}   & \multicolumn{1}{l|}{0.00104}  & \multicolumn{1}{l|}{0.00016}   \\ \hline
\multicolumn{6}{l}{\footnotesize The tests were done by sending and receiving 1024 messages of 1 byte each.}                                                                                                                                              
\end{tabular}}
\end{table*}

\subsection{Multi-Protocol Support: Application Level}
\label{Multi_Protocol_Support_Application_Level}

In addition to the network communication level protocols discussed earlier, a number of application level protocols heva been proposed and developed to support different types of applications. Primarily, in smart city applications, there are three common models of communication, namely,  Device-to-Device (D2D), Device-to-Server (D2S), and Server-to-Server (S2S) \cite{Buyya2016}. At the moment, smart city domain does not have a single widely accepted protocol supporting its wide range of requirements. Some of the requirements are \textit{``1) broadcasting information from one to many, 2) listening for events whenever they may happen, 3) distributing small packets of data in huge volumes, 4) pushing information over unreliable networks, 5) high sensitivity to volume (cost) of data being transmitted, 6) high sensitivity to power consumption (battery-powered devices), 7) high sensitivity to responsiveness (i.e., near real-time delivery of information), 8) security and privacy, and 9) scalability''} \cite{Karasiewicz2013}. From here on, we review several popular application level protocols used by smart city applications.

\textbf{HTTP:} Hypertext Transfer Protocol (HTTP) \cite{Fielding1999} is an application layer protocol designed for distributed hypertext documents and media. Hypertext is structured text. HTTP is a  D2S protocol where the device makes requests and the server responds by returning requester hypertext. In Table I, we present a performance evaluation between HTTP and MQTT protocol \cite{Stanfordclark2008,Karasiewicz2013}. MQTT protocol, which we discuss later in this section, is a lightweight protocol specifically designed for IoT needs. MQTT is one of the potential candidates to be used in smart city applications. According to the results, MQTT outperforms HTTP where HTTP uses more  battery, is less reliable, and a lot slower than MQTT. The main reason to present this table is to show that some of the widely used application protocols in the Internet domain are not suitable for the IoT domain.

\textbf{XMPP:} Extensible Messaging and Presence Protocol (XMPP) \cite{SaintAndre2011} is a message oriented protocol designed for (near) real-time communication. It uses  XML to format and model its data. This protocol is widely used for multi-party communication (e.g., voice and video communication) \cite{Schneider2013}. XMPP is originally designed and developed by Jabber open-source community. XMPP  is an open standard approved by IETF \cite{SaintAndre2011}. In order to support secure communication, secure authentication (SASL) and encryption (TLS) have been built into the core XMPP. XMPP is scalable even for 100K nodes. XMPP has much higher overhead compared to MQTT.

\textbf{MQTT:} Message Queue Telemetry Transport (MQTT) \cite{Stanfordclark2008} is  a D2S protocol which focuses on telemetry, or remote monitoring. MQTT designed to acquire data from large number of sources (e.g., ICOs) and communicate them to IoT sensing infrastructure (e.g., IoT cloud platforms). MQTT is initially developed by IBM \cite{Lampkin2012} for satellite communications with oil-field equipment \cite{Hunkeler2008}.  It is important to note that MQTT is an OASIS open-standard \cite{OASIS2014}. MQTT is a lightweight protocol. Its header packet size is 2 bytes. Further, client libraries also have low foot print (e.g., C\# M2Mqtt library sized at 30KB). In IoT context, MQTT is typically used to collect data from a large number of devices that need to be monitored through the cloud. MQTT is not designed to be fast or real-time. A \textit{hub-and-spoke}  is the commonly used architectural pattern when using MQTT \cite{Schneider2013}. MQTT is  designed to  operate using a `publish/subscribe' model. As a result, it requires a message broker to manage and route messages between connected ICOs \cite{Thangavel2014}. Through the `publish/subscribe' broker, MQTT enables efficient many-to-many communication. MQTT uses  TCP as it is  reliable, ordered and error-checked \cite{Thangavel2014}. MQTT uses Secure Sockets Layer (SSL) and Transport Layer Security(TLS)  \cite{Thomas2000} for security by having encrypted payload \cite{Singh2015}. Having said that, MQTT is a mature and stable protocol compared to CoAP \cite{Shelby2014}. MQTT is a binary  protocol that performs asynchronous communication.

\textbf{CoAP:} Constrained Application Protocol (CoAP) \cite{Shelby2014} is an open standard application level protocol specifically designed to enable communication between resource constrained devices \cite{Bormann2012}.  IETF's \cite{Shelby2014} sample node specification is 8-bit micro-controller with very less amounts of memory that runs on  IPv6 over Low-Power Wireless Personal Area Networks   (6LoWPANs) with throughput of 10s of kbps.  CoAP header packet size is 4 bytes \cite{Shelby2014}. CoAP is a client-server protocol where it supports  one-to-one (machine-to-machine) communication. However, CoAP also supports multi-cast communication \cite{Rahman2014}. CoAP is HTTP-like protocol specially similar to how RESTful communication works. Interoperability with HTTP / RESTful is supported through proxies e.g., simple middleware services) \cite{Castellani2011}. Resource discovery support is built into CoAP. CoAP uses UDP in transport layer  which is less reliable than TCP. As a result of being run over UDP (i.e., connectionless datagrams), CoAP's  has faster wake-up and transmit cycles. Further, CoAP also has smaller packets with less overheads.  SSL/TLS \cite{Thomas2000} based security is not possible in CoAP as it runs on UDP, not on TCP. However, CoAP uses Datagram Transport Layer Security (DTLS) \cite{Raza2012} to provide security similar to TLS using encryption based security techniques \cite{Raza2013}.

\textbf{AMQP:}  Advanced Message Queuing Protocol (AMQP) \cite{Vinoski2006} is an application layer, message oriented protocol  that follows open standards. AMQP  is an OASIS standard. Further, AMQP is secure and reliable protocol that supports queuing and routing. AMQP can work in both point-to-point and publish-and-subscribe settings. AMQP has a packet size of 60 bytes and security is based on SSL/TLS \cite{Thomas2000}. AMQP aims not to lose messages. It uses TCP to assure reliability. AMQP  is mostly used in business messaging where it was initially being developed to support banking applications at JPMorgan Chase in London, UK \cite{OHara2007}. In the IoT applications, AMQP is useful for the control plane or server-based analysis functions \cite{Schneider2013}. AMQP is similar to MQTT from the functional perspective. However, MQTT is designed as a low-overhead, simple to implement protocol that can be run on  many small, relatively dumb devices in order to send small messages on low-bandwidth networks. In contrast, AMQP is a very comprehensive protocol designed to support a large number of message queuing based use cases. AMQP uses a buffer-orientated approach enabling high-performance at the cost of larger foot print. For example, MQTT does not support transactions whereas  AMQP does support \cite{Team2013}.

\textbf{IoTivity:} IoTivity \cite{IoTivity} is not necessary a protocol, but an open source  framework that aims to act as a middleware where it offers four primary functionalities, namely, 1) devices and resources discovery in proximity and remotely, 2) data transmission based on messaging and streaming models, 3) data management through collection, storage and analysis of data from various resources, and  4) device management by configuring, provisioning and diagnostics of devices. IoTivity is a project hosted by Linux Foundation and sponsored by Open Interconnect Consortium \cite{IoTivity}. The objective is to design  an extensible and robust architecture  for smart and thin ICOs, specially for IoT applications. It aims at enabling seamless device-to-device connectivity. IoTivity is primarily targeting device with  platforms such as  Ubuntu (Linux), Android, Arduino, and Tizen.  It is expected to support multiple protocols where it currently supports CoAP and MQTT. Therefore, CoAP supports CoAP service discovery.  IoTivity is currently using UDP for transports and JSON payload is supported using CBOR serialization (cbor.io). IoTivity is a RESTful API and supports point-to-point and point-to-multipoint network topologies. IoTivity uses JSON to model data and RAML (raml.org) to model interactions. It also provides security via DTLS \cite{Raza2012} link-layer using technologies such as ECC, AES, X509 \cite{Delfs2007}.

\textbf{HyperCat:} HyperCat \cite{Hypercat} is a design description framework that allows to expose IoT assets over the web. It is built on top of web standards, e.g., HTTPS, REST/HATEOAS, JSON. HyperCat provides standard mechanisms to semantically annotate its resources and related APIs. HyperCat is designed to be an open standard. It is also a JSON-based hypermedia catalogue format with very low footprint. HyperCat is capable of making URIs available and discoverable to outside applications or services. In HyperCat catalogues, there are not any limitations on how many URIs can be exposed. Further, unlimited RDF-like triple statements can be used to describe each URI (i.e. resource). At minimum, developers need to provide an URL so the clients can access the catalogue using an HTTP GET request. HyperCat does not intend to standardize how IoT resources are accessed, instead it provides a standard mechanism to describe how each IoT resource can be accessed. That means HyperCat provides a common standard to describe APIs and how they can be used by an external entity (e.g., applications). In this way, developers can develop their IoT solutions or devices in anyway they like and use Hypercat specification to describe how to interact with their solutions using a common specification. HyperCat is a Web framework so HTTPS and other Web security mechanisms are  supported.

\textbf{AllJoyn:} AllJoyn \cite{AllJoyn} is an open source software framework that allows devices to communicate with each other in close proximity. At a high level, both IoTivity \cite{IoTivity} and AllJoyn \cite{AllJoyn} aim to achieve same goals, namely, standardization of wired protocols, standardization of schema definitions, standardization of data models, transport and OS independence, collaborative development, open source and freely available, and proximal discovery. However, proximity is an important feature in AllJoyn than IoTivity. AllJoyn is a full stack that consists its own application protocol and several other features. This framework offers an easy way for IoT devices to advertise their services and capabilities. Nearby devices can be connected to each other and discover each other's capabilities. AllJoyn goes further in terms of functionalities it offers and use cases it supports in comparison to application level data exchange protocols such as HyperCat \cite{Hypercat}. AllJoyn's aims to provide solutions to common IoT challenges as presented in \cite{AllJoyn}. AllJoyn currently supports Android, Arduino, iOS, Linux, OS X, and Windows. AllJoyn uses a binary payload with DBUS\footnote{https://dbus.freedesktop.org/doc/dbus-specification.html} serialization. It is designed to run on mesh of stars network topology. AllJoyn uses XML to define schema. AllJoyn follows a publish/subscribe interaction model complemented by Remote Method Invocations. AllJoyn provides application level security via technologies such as ECC, AES, X509 \cite{Delfs2007}.

\textcolor{black}{In the following Table II, we summarise the application protocols using fog computing related characteristics. Column 2 denotes on which part of the fog architecture, the given protocol will be mostly used (Device-to-Device, Device-to-Server). We categorised the application level protocols based on their latency in column 3. In column 4, we denote which category of devices will use a given protocols for their communication (Refer Figure \ref{Figure:Layered_Architecture}). At the beginning of Section \ref{Multi_Protocol_Support_Application_Level}, we presented nine requirements that an ideal IoT application level protocol should support. We map those requirements to each protocol in column 5.}

\begin{table}[h!]
\begin{center}

\renewcommand{\arraystretch}{1.3}
\tbl{Comparision of IoT Application Level Protocols}{
\begin{tabular}{p{3.3cm} l l l l}
Protocol & Usage & Latency & Supported Device Category & Supported Requirements \\ 
\hline
HTTP(S) & D2S, S2S & High & 3,4,5,6 & 8,9 \\ 
XMPP & D2S & Low & 4,5,6 & 1,2,4,8,9 \\ 
MQTT & D2S & Low & 1,2,3,4,5,6 & 2,3,4,5,6,7,8,9 \\ 
CoAP & D2D, D2S & Low & 1,2,3,4,5,6 & 3,5,6,7,8,9 \\ 
AMQP & S2S & High & 4,5,6 & 2,4,8,9 \\ 
IoTivity & D2D & High & 1,2,3,4 & 8,9 \\ 
HyperCat & D2D & High & 1,2,3,4 & 8,9 \\ 
AllJoyn & D2D & High & 1,2,3,4 & 1,8,9 \\

\multicolumn{5}{p{14cm}}{\vspace{2pt} \textbf{In \textit{Supported Requirements}: }1) broadcasting information from one to many, 
2) listening for events whenever they may happen, 
3) distributing small packets of data in huge volumes, 
4) pushing information over unreliable networks, 
5) high sensitivity to volume (cost) of data being transmitted, 
6) high sensitivity to power consumption (battery-powered devices), 
7) high sensitivity to responsiveness (i.e., near real-time delivery of information), 
8) security and privacy, and 
9) scalability}  
\end{tabular}}
\end{center}
\label{TbL:Comparison}
\end{table}

\subsection{Mobility}
\label{sec:Mobility}
Mobility is an important aspect in the fog computing domain, specially in many smart city applications \cite{FC14}. We have discussed a number of smart city applications that will be benefited by mobility in Section \ref{sec:UseCase}. In some applications, ICOs at the edge do not require to maintain communication with gateway devices all the time. Due to the fact that gateway devices have more sophisticated capabilities, they are comparatively expensive. Therefore it would be a waste to deploy unnecessary gateway devices \cite{Chirila2016}. Mobility allows small number of gateway devices to manage a large number of edge ICOs that are deployed across large geographical areas. In such circumstances, fog gateway devices will move through a path that will allow them to make temporary connections with edge ICOs in order to perform different types of management tasks, such as data acquisition, ICO configuration, evaluating device health statuses and so on. Mobility also highlights the importance of dynamic discovery and configuration. Each time a gateway moves, the discoverer including security procedure needs to be executed repeatedly to make sure ICOs are well maintained \cite{Perera2014Book}. 

\textcolor{black}{Discovery in fog computing need to be done efficiently and effectively. The reason is that each ICO will have very limited amount of time to connected to a gateway and perform some data communication before either the ICO itself or the corresponding gateway moves away \cite{Ishino2015}. The challenge is to find a common ground as soon as possible so hey can initiate thee communication \cite{PereraC011}. First, they need to find out which communication technology to use. Specially, in circumstance where a gateway may have multiple communication capabilities (e.g. WiFi, Bluetooth, Zigbee) and the ICO may only have one. For energy reasons, gateway may not want to keep all its communication technologies `ON' all the time. So the challenge is to efficiently and effectively find out, which technology to use on a given location and time period while on the move. Next, they need to find out which application level protocol they would want to use by consider the capabilities of both ICOs and gateway devices. As discussed in Section \ref{Multi_Protocol_Support_Application_Level}, there are many frameworks that has been proposed to make discovery in IoT much easier and efficient (e.g. HyperCat, AllJoyn, IoTivity). Predictive and opportunistic models will be useful in handling these challenges \cite{Pozza2015,Higuchi2014}.}

Fog gateways can also be in the form of drones \cite{Cloudrone}. Such drone based fog gateways are useful in many smart city applications including disaster recovery, wild fire monitoring and so on. \textcolor{black}{In Phenonet \cite{P412}, mobile robots are used as gateways.}

\subsection{General Data Considerations} 
\label{sec:General_Data_Considerations}
\textcolor{black}{Cosnidering the exlpoitations so far in the paper, it should be very clear that data is a key aspect of Fog computing systems; the next few subsections will be concerned with different aspects of data analytics and gathering but it is sensible to consider some generic Data issues first. In this section we will briefly explore management of data and the different kinds of data that can exist based on their origin.}

\textcolor{black}{Managing the vast amount of data, which can exist in different parts of the system is a very important issue, and one that requires further research. However, Fog computing presents a unique opportunity to move from data to information very quickly, with information that is of long-term value then being send and stored in the cloud (a problem that is reasonably well understood) and information that is of short-term value being used to act immediately and then being discarded. Information of the latter kind could be room temperature readings that are used to activate cooloing or heating systems. The challenge  which requires further research in this space lies in determining , ideally automatically, the value proposition of specific data items and making the decision to not keep certain data. The section on context awareness discusses these aspects, and data analytics does of course have the purpose to convert data to information.} 

\textcolor{black}{A further aspect of data management is focused around the idea that we have different data in the system: data can be sensed or meassured directly; data could exist because it was communicated from a nother part of the system or data could be derived by some analytics or processing approach. This data is often combined with stored data to make decisions or enrich analytics. Ultimately the question of managament is less a question of where the data stems from, but rather on of how realiable and current the data is. For example a temperature reading gathered from a trusted high quality sensor is likely very good, while a derived conclusion of the wereabouts of a person based on the last registration to a cell phone tower from a while ago is probably not too reliable \cite{QoD}. A whole field of study has emerged that is considering quality of data (QoD) and is generally cosnidered as `Data Science'.}

\subsection{Context Discovery and Awareness} 
\label{sec:Context_Discovery}
\begin{figure}[h!]
 \centering
 \includegraphics[scale=0.40]{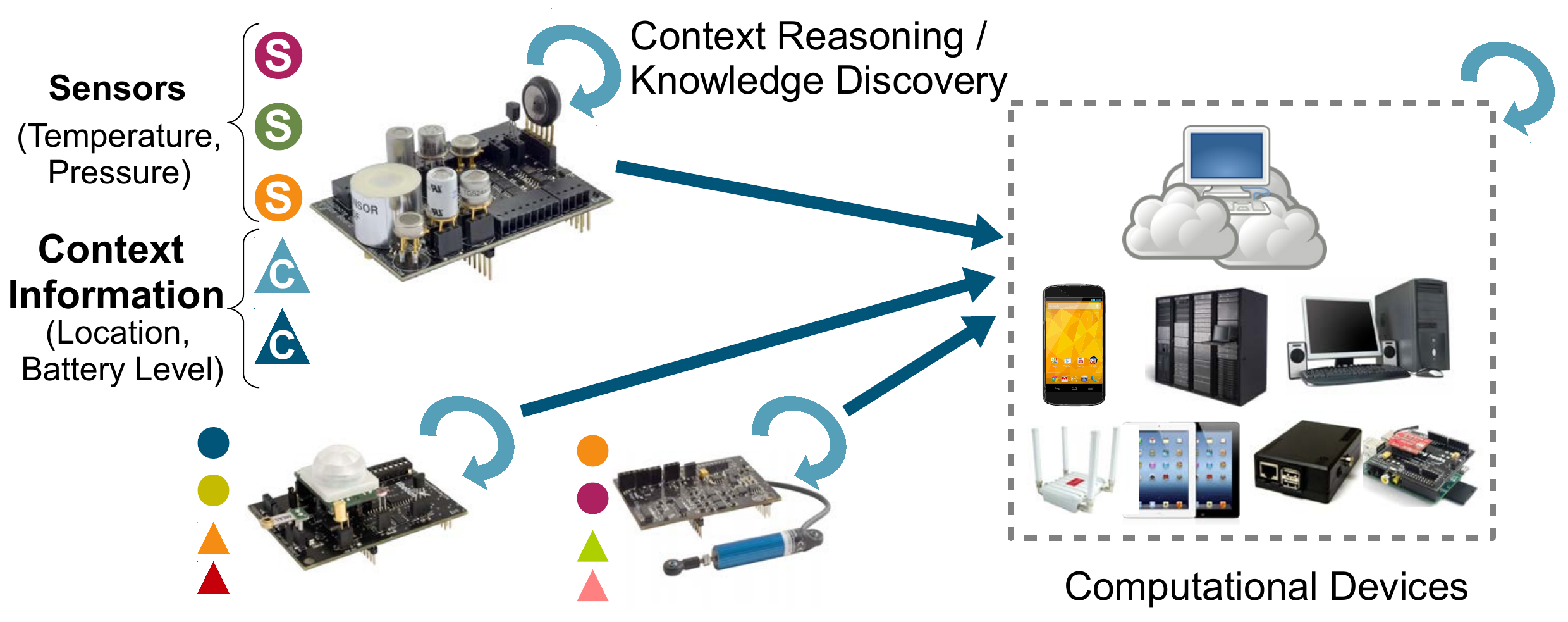}
 \caption{Context Discovery}
 \label{Figure:ContextDiscovery}
\end{figure}

This is an important feature that needs to be supported in fog computing platforms as shown in Figure \ref{Figure:ContextDiscovery}. Specially, the ability to discover context is one of the primary advantages in fog computing over cloud computing. Being close to the edge nodes (ICOs), fog gateway devices have more chances and ability to infer context information such as location, environmental conditions, nearby devices and their capabilities, comparing different ICOs and identifying any malfunctioning ICOs, and so on.
\textit{``Context is any information that can be used to characterise the situation of an entity. An entity is a person, place, or object that is considered relevant to the interaction between a user and an application, including the user and applications themselves''} \cite{P132}. \textcolor{black}{Further, context information is \cite{ZMP007} typically useful in inferring knowledge about what is exactly happening in the field. For example, identifying a malfunctioning ICO by comparing data from nearby ICOs is critical, so the cloud can ignore the data items captured by the malfunctioning ICOs.  Context data can  provide information about data quality which is also has a direct impact on the fused results \cite{ZMP007}. Context data can be used to develop efficient and effective data collection plans specially, when multiple ICOs available near by that offer similar information.}

In data analytics, inaccurate input data is likely to produce inaccurate results. \textit{``A system is context-aware if it uses context to provide relevant information and/or services to the user, where relevancy depends on the user's task''} \cite{P132}. Ideally, fog gateways should be context-aware and intelligent enough so it automatically configures itself to semantically annotate the sensor data appropriately based on the location it is being deployed and capabilities it has. Another context-aware requirement is co-operative and opportunistic sensing. ICOs and fog gateways should be able to balance their workload with neighbouring gateways and ICOs in order to make sure no resources are wasted due to redundant sensing \cite{FC17}. Context-aware data communication is an important part of connected vehicles applications where data resources are constrained \cite{FC05}. Policy driven resources management in fog computing is discussed in detail in \cite{FC16}. \textcolor{black}{Context data also has a significant impact on data collection and fusion strategies \cite{DePaola2016,Jiang2014}. For example, as explained in Section \ref{sec:SmartAgriculture}, IoT application may decide to change the sampling rates of the data collection depending on the context information (e.g., temperature). Further, data aggregation time frames may also get altered based on context data. }

\subsection{Semantic Annotation}
\label{sec:SemanticAnnotation}
The process of \textit{``associating metadata with resources (audio, video, structured text, unstructured text, web pages, images, etc.) is called annotation''} \cite{Cardoso2006}. Further, semantic annotation can be identified as method of using semantic metadata to annotate a given resource \cite{Cardoso2006}. Metadata is \textit{``data that provides information about other data''} \cite{NISO2004}. There are three main types of metadata, namely 1) structural, 2) descriptive, and  3) administrative metadata. Structural metadata provides information about who the data is being structure and stored (e.g. whether the data is  structure based on the location they are being generated or whether data is structured based on the type of the sensor). Descriptive metadata provides a way to annotate data to help identify and discover a given resource (e.g. when is the data generated, who is the originator, and so on). Administrative metadata provides information about managing a given resource (e.g., how a particular resources being created, managed, who have access to a particular resource and so on) \cite{NISO2004}.

Semantic annotation is strongly linked to the context discovery we discussed in Section \ref{sec:Context_Discovery}. Typically, in smart city applications, context information is semantically annotated to the data. We have discussed what context information is in \cite{ZMP007} in detail. Any piece of context information becomes an potential candidate to be used to annotate sensor data \cite{Wei2009}. semantic annotation could be inferred from the sensor data or entered by users. Typically, ICOs are resource constrained devices and sometimes they may not have sufficient resources or knowledge to annotate data with semantics. Therefore, responsibility of semantic annotation could be assigned to fog gateways \cite{Things2015}

\subsection{Data Analytics} 
\label{sec:DataAnalytics}
Fog gateway devices are expected to have limited computational capabilities. In such circumstances, one of requirements is to have module based data analytical components that can be remotely pushed into fog devices on demand. As these fog gateway devices need to be designed to support a large variety of smart city applications, it is wasteful to install a large number of data analytical modules. Ideally, these devices should have only installed with the most common data analytical capabilities, such as average, ignoring outliers, etc. Other types of data analytics should be installed on-demand.  It is also important to provide necessary tools that can be used as templates to build such data analytical components. It is also ideal to support different runtime paradigms that are commonly used to build data analytics components such as Java, Python, R, Matlab, and so on \cite{Bordawekar2015}. Each of these modules would be black boxes running on an individual sandboxes where they intake certain types of inputs and generate certain types of outputs. Some type of permission and validation procedures will be required to ensure privacy and security. It is important to note that fog gateway devices are also resource constrained devices (better off compared to edge ICOs) \cite{FC12FC61}. Therefore, some types of data analytics may not be able to perform on them due to computational complexities and resource limitations \cite{Bordawekar2015,FC17}. \textcolor{black}{Dimension reduction techniques can be used to reduce the  amount data that need to be communicated to the cloud for further analysis \cite{Fodor2002}.}

 Aazam and Huh \cite{FC17} have presented an approach to estimate resource consumption in a fog computing IoT application. For example, data analytics could vary from identifying user activity by fusing accelerometer data to identify sentiment by analysing an image captured by a fog gateway located in a bus station. It is important to note that most the data in IoT are captured as data streaming. Therefore, it is critical to support data stream analytics capabilities in both go gateways and cloud companion platforms \cite{FC03,P022}. 
 
 \textcolor{black}{Data stream analytics is an important aspect in IoT domain as well as in fog computing. In IoT, ICOs are more like to stream data into IoT application than sending data as batches. Though an IoT application may received large number of data streams from large number of ICOs, each gateway will receive only few data stream. Therefore, each gateway is only need to deal with few data streams \cite{Puschmann2016}. Data stream analytics are discussed in detail in \cite{Liu2014,Singh2016,Singh2015a}. The challenge is to find out how to bring data stream analytics towards fog gateway from the cloud, in order to reduce network communication and energy consumptions.}

\subsection{Security and Privacy}
\label{sec:SecurityandPrivacy}
This is an important non-functional requirement in smart city applications \cite{FC15}. We can identify four broader categories under this theme namely, 1) privacy, 2) authenticity, 3) confidentiality, and 4) integrity. Privacy safeguards that only the approved ICOs and gateways are a part of the network. Authentication is a major challenge in fog computing \cite{FC02}. Authenticity aims to verify genuineness of the data sender. For example,  public-Key crypto systems \cite{Rivest1978} are used to  achieve this. Confidentiality make sure  that only the proposed destination can read the data. AES 128 and AES 256 \cite{Daemen:2002} or similar techniques are used to assure the confidentiality.  Integrity safeguards that  the original data is not being  altered during the transmission and the information received by the destination is same as the information send by the originator.   For example,  hash algorithms such as MD5 and SHA \cite{Delfs2007} can be used to generate checksum of the message and to ensure that the integrity of the message.   In fog computing paradigms, we can identify several different data communication patterns that need to be secure: 1) ICO to ICO , 2) ICO to Gateway, 3) ICO to Cloud, 4) Gateway to Cloud, 5) Cloud to ICO/Gateway \cite{LibeliumSecurity}. Let us briefly introduce each of these patterns, which can be secured using different techniques.

\textbf{1) ICO to ICO:} In order to avoid external ICOs (including malicious ICOs) being connected to a network and observing the data communicated back and forth, ICOs can use link layer based encryption techniques. AES 128 \cite{Daemen:2002} (Symmetric Encryption) is the most widely used algorithm to address this issue. Specialized hardware can be used to manage the encryptions. In this communication pattern, each network will have a shared common  key pre-shared by all the ICOs  \cite{LibeliumSecurity}.

\textbf{2) ICO to Gateway:}  Networking  and application layer based symmetric encryption can be used to secure this type of communication. In ICO to ICO communication, each node has its own private key to ensure authenticity. Gateways are expected to have all the encryption key corresponds to each ICO. Gateways need them to decrypt the data sent by ICOs. Data generated by one ICO may hop through multiple neighboring ICOs to reach the gateway. However, intermediary ICOs will not be able to read the packet as it does not have the private key of the generator. With the help of random internal seed and  a sequence number, fog gateways can verify the integrity of data using checksums \cite{LibeliumSecurity}.

AES 256 \cite{Daemen:2002} (Symmetric Encryption) is the most widely used algorithm to secure this type of communication pattern. Man-in-the-middle attacks are the most common type of security challenges in fog networks \cite{FC02}. Such attacks are hard to detect specially in the scenarios we discussed in Section \ref{sec:UseCase}.

\textbf{3) ICO to Cloud:} At times, it is necessary for ICOs to directly connect to the cloud platform. In this pattern, without storing the keys  in the gateway, the keys are stored in the cloud, so the cloud can decry the data sent by a known ICO. AES 256 \cite{Daemen:2002} (Symmetric Encryption) is the most widely used algorithm to secure this type of communication pattern and typically software libraries help to handle the encryption.

\textbf{ 4) Gateway to Cloud:}. HyperText Transfer Protocol over SSL/TLS (HTTPS) is commonly used to secure this type of communication. Typically, IoT cloud platforms are Web servers and fog gateways may use HTTPS to communicate with the server.

\textbf{5) Cloud to ICO/Gateway:} This type of communication pattern is primarily required to initiate secure connections between ICOs, gateways, and the cloud. To initiate AES 256, it is essential to use a shared key between each origin node (i.e. sensor node) and the gateway or the cloud destination server. Public-Key cryptosystems \cite{Rivest1978} are used to  achieve this.

Most IoT platforms have the implicit assumption that \textit{``You feed your data to our platform, we support the data analytics, event detection, actuation and other functionalities''}. However, as these platforms receive more and more personal data and once these platforms start sharing data between different parties, privacy becomes a serious concern and a great challenge to address \cite{Perera2015}. 
 
 Sending all the collected data all the time to the cloud consumes more data communication bandwidth and energy. Another major issue is privacy \cite{FC15}.  Raw data collection can always lead to privacy violations at later stages during the data life cycle. Therefore, an important principle in protecting user privacy is to collect only the amount of data that is sufficient to achieve the task at hand \cite{Gaura2013}. Data collection and analytics is a double edged sword. Data gathering and analytics help us to derive useful new knowledge. However, such discovery of new knowledge could also lead to violation of someone's privacy. Therefore, data analytics in the IoT domain need to be carefully and ethically monitored to ensure no privacy violations would occur during the process. Privacy issues become more critical in IoT applications in smart homes and smart wearable domains.

\subsection{Cloud Companion Support}
\label{sec:CloudCompanionSupport} As  we mentioned earlier, fog computing is not a solution fitting for all scenarios, neither can it be used as a standalone solution. Fog computing generally goes hand-in-hand with cloud computing \cite{FC14}. As a result, it is important for fog gateways to have the built in ability to communicate with cloud IoT platforms. Ideally, fog computing should have a pluggable architecture so adding support to a new cloud  platform should not be difficult. Standardized interfaces opened-up through HTTP REST APIs (or similar) would enable interoperability. Use of semantic annotation or adoption of techniques such as HyperCat \cite{Hypercat} would also enable such interoperability. Cloud platforms and fog gateways should be able to exchange commands (i.e., the control plane) as well as data (i.e., the data plane). Some common cloud computing platforms are discussed in \cite{Perera2015a}. For example, industrial IoT cloud platform IBM Blumix provides a way to covert \textit{Raspberry Pi} devices into fog gateways through an open source framework called \textit{Node-RED} (nodered.org).

\section{Existing Research Efforts and Trends}
\label{sec:Comparison}

Based on the identified characteristics and common features in the previous section, in this section, we select and compare more than 30 representative existing research efforts in Fog computing. All these research efforts focus on at least two of the common features of Fog computing, as shown in Table \ref{TbL:LiteratureReview}. 

It is easy to observe that Cloud Companion Support and Data Analytics are the most popular features in these Fog computing research efforts. This indicates that many applications employ Fog computing to do small data analytics tasks but meanwhile, still rely on Cloud computing platforms to perform large tasks, which normally cannot be done on Fog devices. There also exist a few research efforts where Fog devices will take care of almost all the data analytics tasks, such as Gazis et al. \cite{FC36}, Kulkarni et al. \cite{FC15}, Preden et al. \cite{FC18}, and Oueis et al. \cite{FC22}. The reasons for these include real-time local decision making \cite{FC36,FC18}, privacy \cite{FC15}, and load balancing \cite{FC36,FC22}. 

Following the above two most popular features, Multi-Protocol Support at Communication Level, Mobility, and Security and Privacy are also popular features in Fog computing applications. Regarding Multi-Protocol Support at Communication Level, since computing happens at the edge of a network, supporting communication to a variety of devices is of great importance. Further, to make Fog computing more useful, mobility support is also critical. Some research effort even emphasizes that, to this end, a distributed directory system is necessary and mobility techniques support, such as LISP protocol support, should be provided \cite{FC09}. Finally, Security and Privacy attract much attention. The main reason for this is that in Fog computing, the generated data is very sensitive to locations, identities, and times, which all pose great risk of information and privacy disclosure.

\begin{table}[t!]
\begin{center}

\renewcommand{\arraystretch}{1.3}
\tbl{Evaluation of Existing Research Efforts}{
\begin{tabular}{p{3.3cm}ccccccclcl}
\multicolumn{1}{l}{\textbf{}} & \multicolumn{1}{l}{\textbf{DDIO}} & \multicolumn{1}{l}{\textbf{DCDM}} & \multicolumn{1}{l}{\textbf{MPS-C}} & \multicolumn{1}{l}{\textbf{MPS-A}} & \multicolumn{1}{l}{\textbf{Mob}} & \multicolumn{1}{l}{\textbf{CDA}} & \multicolumn{1}{l}{\textbf{SA}} & \textbf{DA} & \multicolumn{1}{l}{\textbf{SP}} & \textbf{CCS} \\ 
\hline
\cite{FC47} & \multicolumn{1}{l}{$\checkmark$} &  & \multicolumn{1}{l}{} &  & \multicolumn{1}{l}{$\checkmark$} &  &  & \multicolumn{1}{c}{} &  & $\checkmark$ \\ 
\cite{FC53} &  & \multicolumn{1}{l}{} &  &  &  &  &  & $\checkmark$ &  & $\checkmark$ \\ 
\cite{FC03} &  & \multicolumn{1}{l}{$\checkmark$} &  & \multicolumn{1}{l}{$\checkmark$} &  &  &  & $\checkmark$ &  & $\checkmark$ \\ 
\cite{FC36} &  &  & \multicolumn{1}{l}{$\checkmark$} &  &  & \multicolumn{1}{l}{} &  & $\checkmark$ &  &  \\ 
\cite{FC57} & \multicolumn{1}{l}{$\checkmark$} &  & \multicolumn{1}{l}{$\checkmark$} &  &  &  &  & \multicolumn{1}{c}{} &  & $\checkmark$ \\ 
\cite{FC39} & \multicolumn{1}{l}{$\checkmark$} & \multicolumn{1}{l}{$\checkmark$} &  &  & \multicolumn{1}{l}{$\checkmark$} &  &  & \multicolumn{1}{c}{} &  & \multicolumn{1}{c}{} \\ 
\cite{FC05} &  & \multicolumn{1}{l}{$\checkmark$} & \multicolumn{1}{l}{$\checkmark$} &  &  &  &  & \multicolumn{1}{c}{} & \multicolumn{1}{l}{} & $\checkmark$ \\ 
\cite{FC08} &  &  &  &  & \multicolumn{1}{l}{$\checkmark$} &  &  & \multicolumn{1}{c}{} & \multicolumn{1}{l}{$\checkmark$} & $\checkmark$ \\ 
\cite{FC09} &  & \multicolumn{1}{l}{$\checkmark$} & \multicolumn{1}{l}{$\checkmark$} &  & \multicolumn{1}{l}{$\checkmark$} & \multicolumn{1}{l}{} &  & $\checkmark$ &  & $\checkmark$ \\ 
\cite{FC11} &  &  &  & \multicolumn{1}{l}{$\checkmark$} &  &  & \multicolumn{1}{l}{} & $\checkmark$ & \multicolumn{1}{l}{$\checkmark$} & $\checkmark$ \\ 
\cite{FC12FC61} &  &  &  &  &  &  &  & $\checkmark$ & \multicolumn{1}{l}{$\checkmark$} & $\checkmark$ \\ 
 \cite{FC15} &  &  &  &  &  &  &  & $\checkmark$ & \multicolumn{1}{l}{$\checkmark$} & \multicolumn{1}{c}{} \\ 
\cite{FC16} &  &  &  &  & \multicolumn{1}{l}{$\checkmark$} &  & \multicolumn{1}{l}{$\checkmark$} & $\checkmark$ & \multicolumn{1}{l}{$\checkmark$} & $\checkmark$ \\ 
\multicolumn{1}{l}{\cite{FC17}} & \multicolumn{1}{l}{$\checkmark$} & \multicolumn{1}{l}{$\checkmark$} & \multicolumn{1}{l}{$\checkmark$} & \multicolumn{1}{l}{} & \multicolumn{1}{l}{} & \multicolumn{1}{l}{} & \multicolumn{1}{l}{} & $\checkmark$ & \multicolumn{1}{l}{} & $\checkmark$ \\ 
 \cite{FC18} & \multicolumn{1}{l}{$\checkmark$} & \multicolumn{1}{l}{$\checkmark$} &  &  & \multicolumn{1}{l}{$\checkmark$} & \multicolumn{1}{l}{$\checkmark$} &  & $\checkmark$ &  & \multicolumn{1}{c}{} \\ 
\cite{FC22} &  & \multicolumn{1}{l}{$\checkmark$} &  &  &  &  &  & $\checkmark$ &  & \multicolumn{1}{c}{} \\ 
\cite{FC25} &  &  &  &  &  &  &  & $\checkmark$ &  & $\checkmark$ \\ 
\cite{FC26} & \multicolumn{1}{l}{} & \multicolumn{1}{l}{$\checkmark$} &  &  &  &  &  & \multicolumn{1}{c}{} &  & $\checkmark$ \\ 
\cite{FC27} & \multicolumn{1}{l}{$\checkmark$} & \multicolumn{1}{l}{$\checkmark$} & \multicolumn{1}{l}{$\checkmark$} & \multicolumn{1}{l}{$\checkmark$} &  &  &  & \multicolumn{1}{c}{} & \multicolumn{1}{l}{$\checkmark$} & $\checkmark$ \\ 
 \cite{FC28} &  &  &  &  & \multicolumn{1}{l}{$\checkmark$} &  &  & $\checkmark$ & \multicolumn{1}{l}{$\checkmark$} & $\checkmark$ \\ 
\cite{FC38} &  &  &  &  & \multicolumn{1}{l}{$\checkmark$} &  &  & $\checkmark$ &  & $\checkmark$ \\ 
 \cite{FC41} & \multicolumn{1}{l}{} & \multicolumn{1}{l}{$\checkmark$} &  &  &  &  &  & \multicolumn{1}{c}{} &  & $\checkmark$ \\ 
\cite{FC43} &  &  & \multicolumn{1}{l}{$\checkmark$} &  &  & \multicolumn{1}{l}{$\checkmark$} &  & $\checkmark$ &  & $\checkmark$ \\ 
\cite{FC44} &  &  & \multicolumn{1}{l}{$\checkmark$} &  &  &  &  & $\checkmark$ &  & $\checkmark$ \\ 
\cite{FC49} &  & \multicolumn{1}{l}{$\checkmark$} &  &  & \multicolumn{1}{l}{$\checkmark$} &  &  & $\checkmark$ & \multicolumn{1}{l}{$\checkmark$} & $\checkmark$ \\ 
\cite{FC52} &  &  & \multicolumn{1}{l}{$\checkmark$} &  &  &  &  & $\checkmark$ & \multicolumn{1}{l}{$\checkmark$} & $\checkmark$ \\ 
\cite{FC54} &  &  & \multicolumn{1}{l}{$\checkmark$} &  & \multicolumn{1}{l}{$\checkmark$} &  &  & $\checkmark$ &  & $\checkmark$ \\ 
 \cite{FC67} &  &  & \multicolumn{1}{l}{$\checkmark$} &  &  &  &  & \multicolumn{1}{c}{} & \multicolumn{1}{l}{$\checkmark$} & $\checkmark$ \\ 
\cite{FC78} & \multicolumn{1}{l}{$\checkmark$} &  & \multicolumn{1}{l}{$\checkmark$} &  &  &  &  & $\checkmark$ & \multicolumn{1}{l}{$\checkmark$} & $\checkmark$ \\ 
\cite{FC84} &  &  &  &  & \multicolumn{1}{l}{$\checkmark$} & \multicolumn{1}{l}{$\checkmark$} & \multicolumn{1}{l}{$\checkmark$} & $\checkmark$ &  & $\checkmark$ \\ 
\cite{SecureIoT} &  &  & $\checkmark$ &  & \multicolumn{1}{l}{$\checkmark$} &  &  & $\checkmark$ & $\checkmark$ & $\checkmark$ \\ 
\cite{Stack4Things} &  & $\checkmark$ &  &  & \multicolumn{1}{l}{$\checkmark$} & \multicolumn{1}{l}{$\checkmark$} & & $\checkmark$ &  & $\checkmark$ \\ 
\hline

\multicolumn{11}{p{14cm}}{\vspace{2pt} \textbf{DDIO:} Dynamic Discovery of Internet Objects, \textbf{DCDM:} Dynamic Configuration and Device Management, \textbf{MPS-C:} Multi-Protocol Support: Communication Level, \textbf{MPS-A:} Multi-Protocol Support: Application Level, \textbf{Mob}: Mobility, \textbf{CDA:} Context Discovery and Awareness, \textbf{SA}: Semantic Annotation, \textbf{DA:} Data Analytics, \textbf{SP:} Security and Privacy, \textbf{CCS:} Cloud Companion Support.}  
\end{tabular}}
\end{center}
\label{TbL:LiteratureReview}
\end{table}

According to the table, the least implemented common features of Fog computing are Multi-Protocol Support at Application Level, Context Discovery and Awareness, and Semantic Annotation. It seems that many research efforts are happy with supporting only one application protocol, which is sufficient for most of the time. There are also not many applications requiring Context Discovery and Awareness, where the main reason could be that the application scenarios are very clear and contexts are quite certain. In terms of Semantic Annotation, for now it receives least attention in the listed research efforts shown in Table III. This may be due to the fact that Fog computing is still at its early stage and more efforts are devoted to other seemingly more important features.

Among these research efforts, Gia et al. \cite{FC43} provide an interesting case study on Fog computing in Healthcare Internet of Things. When applying the Fog computing technology, smart gateways are deployed with embedded data mining, distributed storage, and notification services. As a case study, Electrocardiogram (ECG) feature extraction is examined. Specifically, ECG signals are transmitted to smart gateways, where feature extractions are performed. In this way, it is possible to provide real-time analytics and offer low-latency response. It is demonstrated that such Fog computing approach can help to save more than 90\% bandwidth compared with traditional Cloud computing approaches with lower latency.

Regarding security, Stolfo et al. \cite{FC08} propose an cloud-based approach that can be used to mitigate attacks, that are focused on data theft,   by exploiting the advantages of Fog computing. The main idea is to implement two additional security features, including \textit{User Behavior Profiling} and \textit{Decoys}, on fog gateway devices  \cite{FC08}. Firstly, User Behavior Profiling can be achieved by training one-class support vector machines, where building a classifier without sharing data from different users is feasible  \cite{FC08}. Meanwhile, Decoys feature refers to placing different types of important-looking documents in some highly conspicuous locations by the legitimate user, in order to detect suspicious access. Since these two new features are built on top of existing cloud security features, better security can be achieved. 

Similarly, in another scenario, Dsouza et al. \cite{FC16} investigates Fog computing in supporting secure, private and safe real-time smart transportation systems, which should accommodate dynamic traffic changes and alert potential conflicts and safety issues to travelers. To achieve this, a robust policy management framework is proposed, which can ensure both interoperability and secure communication in a fog ecosystem.

\section{Lessons Learned}
\label{sec:Lessons_Learned}

In this section, we take each of the major features identified in Section \ref{sec:Characteristics} and explain how they would contribute towards building sustainable sensing infrastructures for smart cities. Due to high demand and public pressure towards building smarter living spaces for people, both government and private sector entities  are forced to initiated smart cite programs. Smart cities investments are designed to focus of finding ICT based sustainable solutions to address growing issues \cite{Perera2014}. By definition, Smart Cities \cite{P532} have \textit{``six characteristics: smart economy,  smart people, smart governance, smart mobility, smart  environment and smart living''} \cite{P528}. Sustainability is defined as \textit{``...able to be used without being completely used up or destroyed..''} and \textit{``...able to last or continue for a long time''} (merriam-webster.com/dictionary/sustainability). Sustainability has the same meaning in smart cities context as well. 
 
 Smart Cities require to accommodate growing population and their needs with limited resources. Smart Cities also need to make sure the limited resources do not run out \cite{FC17}. The best strategy to achieve this is to use resources in the most efficient and optimum ways. To do this, it is essential to understand how cities and its citizens behave and consume resources (e.g., patterns, practices, norms and so on). All the information that is required to understand cities and their citizens is hidden in IoT data. To discover knowledge and insights from IoT data, we need to collect and analyses them in large scale. Sensing infrastructure is a critical element in collecting and analyzing IoT data. Fog computing brings efficiency and sustainability to the sensing infrastructure as follows.

 Dynamic discovery of ICOs help towards building sustainable sensing infrastructure in multiple ways. First, it reduces the deployment time (therefore cost) of sensing infrastructure. Secondly, it allows ICOs to be moved at runtime. For example, a new ICO can be deployed in the field at any time without changing existing infrastructure. Dynamic configuration plays a critical role in sustainability as it allows to reconfigure sensing parameters at run time (e.g., sampling rate, communication frequency, which sensors to activate). These parameters have direct impact on the energy consumption and the life time of the infrastructure, especially when ICOs are battery/solar powered. Efficient and optimum management of these parameters can make sure that a given infrastructure can sustain for a long period of time without recharging or replacement.
 
 Choosing the most appropriate communication protocols is paramount in building sustainable sensing infrastructure. As we have presented in Section \ref{Multi_Protocol_Support_Communication_Level}, each protocol has its own strengthens and weaknesses.   Some protocols use a far less amount of energy as they are specifically designed to operate using batteries for years. Further, fog gateways may also change the communication protocols they use at run time opportunistically to get the advantage of their surrounding resources. For example, cellular based mobile fog gateways may opportunistically use a Wi-Fi zone to upload the data quickly to the cloud so it can save energy and costs compared to cellular communication. Similarly, selecting the right application level protocols also has a significant impact of he sustainability of the infrastructure. For example, we have presented a comparison between HTTP and MQTT in Table I. It clearly shows how much energy can be saved by using MQTT over HTTP. Therefore, it is critical to think about which application protocol is sufficient to accomplish the needs of a smart city application.
 
 Mobility helps to build smart city applications with less resources. For example, in Section \ref{sec:UseCase}, we discussed a number of use cases that involve mobility of ICOs and fog gateways. Mobility helps to cover a large geographical area with fewer ICOs and gateways by reducing the deployment costs significantly. Such reductions can help to build sustainable sensing infrastructure for smart cities. 
 
\textcolor{black}{Basic management of data and assuring quality of data has challenges that are fundamental to Fog computing, but Fog computing also provides great opportunities to allow for storage of valuable information rather than data which might quickly become irrelevant.} Semantic annotation allows  data to be more meaningful. It reduces the amount of efforts that need to be put in when recovering the meaning out of data. When sensor data is annotated with context information, it is easy for algorithms to discover knowledge later in the data analytical pipeline. This reduces overall computation costs (i.e., energy requirement throughout the data flow) significantly, which is better for sustainability. 
 
 It is vital to apply analytics over data at the right time and in the right place. Data in smart cities flows from ICOs to the cloud. It is always better to apply the analytics to the data at early stages of the pipeline. This reduces the amount of data that needs to be communicated to the cloud and save data communication costs, storage costs, and computational costs. Data analytics typical analyzes comparatively large amounts of data and produces summarized results. This reduction leads to less amounts of data being transferred to the cloud and saving significant costs (e.g., bandwidth, energy, storage).
 
 Security and privacy are two of the most important factors that impact towards sustainability. No sensing infrastructure can survive without them. In order to function without disruption, security and privacy need to be guaranteed. Otherwise, it would affect sustainability in two ways; 1) citizens will oppose the deployments and the usage of sensing infrastructure  as it affects their personal security and privacy, and 2) applications that are built around sensing infrastructure will get disrupted and users who depend on those applications will get affected significantly and could lead to losses, disappointments, frustrations, and chaoses.

Context-awareness helps to reduce energy consumption in most of the use case scenarios. It can decide 1) when to start sensing and when to stop, 2) when to use which protocol, 3) when to transmit data, and 4)when to increase or decrease sampling rates using context information. We discussed these aspects earlier in Section \ref{sec:Context_Discovery}. Further, load balancing and opportunistic sensing will help fog gateways and ICOs to efficiently balance their workload with neighbors and avoid redundant sensing and resource wastage. Intelligent and efficient planning can be used to make sure all neighboring ICOs and gateways do not go off-line at the same time due to energy running out. This allows recharging and replenishment to be taken place without significant disruption. Having said all these advantages in fog computing,  almost no sensing infrastructure can survive without cloud platforms. Cloud platforms are  scalable and can accommodate the growing needs in smart cities. Cloud based large scale data analytics and knowledge discovery (e.g., prediction) can make all the sensing efforts useful and create values. Such value creation directly contributes towards the long term sustainability of sensing infrastructure.

\section{Challenges and Opportunities}
\label{sec:Challenges_and_Opportunities}

So far we have discussed ten common features that are important in a fog computing platform in order to support building sustainable smart cities. It is important to note that these features are research areas that need to be explored further and significantly. We have provided a number of references that readers can use to look further into challenges on each of these areas.

In this section, we highlight the needs of fog platforms that support all the features discussed in this paper. Both IoT and fog computing are comparatively immature fields. Therefore, a large amount of experimentation needs to be done. An ideal fog computing platform should be able to provide a framework that others can use it to test  different approaches, techniques, and algorithms. For example, there are many ways to autonomously annotate data with semantics within a fog gateway. It is not possible to develop one universal approach or algorithm to annotate data. Therefore, a fog platform should be built in such a away that anyone can write exertions to support new ways of annotating data. This is similar to what we see today in mobile app markets. In a  mobile app  market, there are many different apps that are designed to  perform the same task at high level (e.g., TODO List apps). However, each of these applications is designed in a unique way and offers slightly different capabilities. Users can easily uninstall one app and install a new one depending on their requirements at a given time. However, it is important to note that  a mobile app does need to follow certain guidelines in order for them to work in a given platform. Similarly, there should be some standardizations so these plug-ins  (or extensions) would work seamlessly in the fog platforms, but the implementation should be open for creativity.

Further, such platforms should provide built-in supports for different types of communication and application level protocols. This does not mean that all these protocols should be supported at all time. However, a fog platform should be able to  download plug-ins to provide support for these different protocols without requiring much effort. Further, providing support for existing data analytics frameworks is also important. For example, some analytical plug-ins may be written in R, Python, Matlab, Java, C/C++ and so on. All these different approaches need to be supported so the existing code bases can be easily reused to perform data analytics. At high level, it is useful to build an ecosystem with interchangeable plug-ins that support different implementations of the features we discussed.

 Due to low computational resources of fog gateways, it is important that these plug-ins can be easily removed to avoid resource wastage when not required in a given fog gateway. Despite we see a large number IoT cloud platforms in the market in both academia (e.g., OpenIoT) and industry (e.g., IBM Bluemix, Microsoft Azure IoT), a fog computing platform that supports all the features we listed in this article are yet to be researched and developed. We believe such fog platforms would be greatly beneficial to the research and industrial communities once available as by then people can easily test new fog computing related approaches, techniques and algorithms.

According to the literature comparison presented in Table III, context discovery and awareness \cite{ZMP007} and semantic annotation is fairly ignored. As we discussed in Section \ref{sec:Lessons_Learned}, these two functionalities has the potential to improve the sustainability of a sensing infrastructure significantly. Challenges in semantic annotation and interoperability in IoT domain are discussed in detail in \cite{Things2015}.

\textcolor{black}{Distributed intelligence will be very critical in making fog computing platforms successful in building future smart cities. The main reason is that prompt reactions and best possible decisions should be achieved in a timely manner to make future cities smarter, safer and more living-enjoyable. This requires a large amount of research efforts in putting distributed intelligence in place properly across smart things, buildings, fog devices/gateways, and cloud computing infrastructure in a city. This process will involve many important aspects, such as available domain-dependent knowledge/intelligence, combination of business logic, engineering processes and government policies, cost-efficient and computation-efficient and context-/semantics-aware computing models for real-time decision making, etc.}

\textcolor{black}{Security is also a critical issue in building future smart cities. This mainly refers to security of fog computing platforms, including potential cyber attacks to smart things, fog devices/gateways, and trust and authentication, network security, and data security, etc. For example, cyber attacks to smart things, fog devices/gateways can dysfunction smart things, fog devices/gateways and  pose risks in failure of providing proper services to the city and making wrong decisions in reaction to emergencies and disasters. Failure of ensuring trust and authentication will also put any large-scale fog computing platforms at risk, potentially leading to intentional and accidental misbehavior, criminal activities and so on. Network security is also of great importance since  network attacks such as jamming attacks, sniffer attacks and so on can create huge risks in fog computing systems, potentially leading to chaos of the whole fog computing systems. Data security will also be critical. Sensitive and/or valuable data generated from any fog computing platforms should be kept secure.}

\section{Conclusions}
\label{sec:Conclusions}

Due to the improvements in sensing technology and reduction in costs, sensing capabilities are expected to be integrated into  everyday objects around us. There is a natural tendency that smart city applications are being built in a centralized manner. That means all the data collected by sensors are  transferred to a cloud node for knowledge discovery. However, this is a very inefficient approach from both computational and communication perspectives. To address this issue, the fog computing paradigm has been proposed. Specially, more and more industrial IoT platform developers, such as Microsoft, IBM, and Intel, are now  moving towards utilizing fog gateway devices to perform edge analytics. Fog computing brings sustainability to the smart city applications. In this survey paper, we have analysed and evaluated different types of fog computing and edge analytics research efforts to understand what are the most important functionalities of a fog computing platform.  We have also discussed the major trends in this field that were identified during the survey. Our finding clearly highlight the importance of fog computing platforms in order to build sustainable IoT infrastructure for smart cities. 



%
%

\begin{acks}
We acknowledge the financial support of European Research Council Advanced Grant 291652 (ASAP).
\end{acks}

\bibliographystyle{ACM-Reference-Format-Journals}
\bibliography{library}


\begin{thebibliography}{00}


\ifx \showCODEN    \undefined \def \showCODEN     #1{\unskip}     \fi
\ifx \showDOI      \undefined \def \showDOI       #1{{\tt DOI:}\penalty0{#1}\ }
  \fi
\ifx \showISBNx    \undefined \def \showISBNx     #1{\unskip}     \fi
\ifx \showISBNxiii \undefined \def \showISBNxiii  #1{\unskip}     \fi
\ifx \showISSN     \undefined \def \showISSN      #1{\unskip}     \fi
\ifx \showLCCN     \undefined \def \showLCCN      #1{\unskip}     \fi
\ifx \shownote     \undefined \def \shownote      #1{#1}          \fi
\ifx \showarticletitle \undefined \def \showarticletitle #1{#1}   \fi
\ifx \showURL      \undefined \def \showURL       #1{#1}          \fi

\bibitem[\protect\citeauthoryear{Aazam and Huh}{Aazam and Huh}{2014}]%
        {FC12FC61}
{Mohammad Aazam} {and} {Eui~Nam Huh}. 2014.
\newblock \showarticletitle{{Fog computing and smart gateway based
  communication for cloud of things}}. In {\em Proceedings - 2014 International
  Conference on Future Internet of Things and Cloud, FiCloud 2014}. IEEE,
  464--470.
\newblock
\showISBNx{9781479943586}
\showDOI{%
\url{http://dx.doi.org/10.1109/FiCloud.2014.83}}


\bibitem[\protect\citeauthoryear{Aazam and Huh}{Aazam and Huh}{2015a}]%
        {FC84}
{Mohammad Aazam} {and} {Eui~Nam Huh}. 2015a.
\newblock \showarticletitle{{E-HAMC: Leveraging Fog computing for emergency
  alert service}}. In {\em 2015 IEEE International Conference on Pervasive
  Computing and Communication Workshops, PerCom Workshops 2015}. IEEE,
  518--523.
\newblock
\showISBNx{9781479984251}
\showDOI{%
\url{http://dx.doi.org/10.1109/PERCOMW.2015.7134091}}


\bibitem[\protect\citeauthoryear{Aazam and Huh}{Aazam and Huh}{2015b}]%
        {FC17}
{Mohammad Aazam} {and} {Eui~Nam Huh}. 2015b.
\newblock \showarticletitle{{Fog computing micro datacenter based dynamic
  resource estimation and pricing model for IoT}}. In {\em Proceedings -
  International Conference on Advanced Information Networking and Applications,
  AINA}, Vol. 2015-April. IEEE, 687--694.
\newblock
\showISBNx{9781479979042}
\showISSN{1550445X}
\showDOI{%
\url{http://dx.doi.org/10.1109/AINA.2015.254}}


\bibitem[\protect\citeauthoryear{Abdullahi, Arif, and Hassan}{Abdullahi
  et~al\mbox{.}}{2015}]%
        {FC57}
{Ibrahim Abdullahi}, {Suki Arif}, {and} {Suhaidi Hassan}. 2015.
\newblock \showarticletitle{{Ubiquitous shift with information centric network
  caching using fog computing}}. In {\em Advances in Intelligent Systems and
  Computing}, Vol. 331. Springer International Publishing, 327--335.
\newblock
\showISBNx{9783319131528}
\showISSN{21945357}
\showDOI{%
\url{http://dx.doi.org/10.1007/978-3-319-13153-5_32}}


\bibitem[\protect\citeauthoryear{Aberer, Hauswirth, and Salehi}{Aberer
  et~al\mbox{.}}{2007}]%
        {P022}
{Karl Aberer}, {Manfred Hauswirth}, {and} {Ali Salehi}. 2007.
\newblock \showarticletitle{{Infrastructure for Data Processing in Large-Scale
  Interconnected Sensor Networks}}. In {\em International Conference on Mobile
  Data Management}. 198--205.
\newblock
\showDOI{%
\url{http://dx.doi.org/10.1109/MDM.2007.36}}


\bibitem[\protect\citeauthoryear{Akyildiz and Jornet}{Akyildiz and
  Jornet}{2010}]%
        {nanoThings}
{I~F Akyildiz} {and} {J~M Jornet}. 2010.
\newblock \showarticletitle{{The Internet of Nano-Things}}.
\newblock {\em IEEE Wireless Communications\/} {17}, 6 (dec 2010), 58--63.
\newblock
\showISSN{1536-1284}
\showDOI{%
\url{http://dx.doi.org/10.1109/MWC.2010.5675779}}


\bibitem[\protect\citeauthoryear{{Al Faruque} and Vatanparvar}{{Al Faruque} and
  Vatanparvar}{2016}]%
        {FC27}
{Mohammad~Abdullah {Al Faruque}} {and} {Korosh Vatanparvar}. 2016.
\newblock \showarticletitle{{Energy Management-as-a-Service over Fog Computing
  Platform}}.
\newblock {\em IEEE Internet of Things Journal\/} {3}, 2 (apr 2016), 161--169.
\newblock
\showISSN{23274662}
\showDOI{%
\url{http://dx.doi.org/10.1109/JIOT.2015.2471260}}


\bibitem[\protect\citeauthoryear{Alliance}{Alliance}{2008}]%
        {Alliance2008}
{Zigbee Alliance}. 2008.
\newblock \showarticletitle{{Zigbee Specification}}.
\newblock {\em Zigbee Alliance website\/} (2008), 1--604.
\newblock
\showISBNx{053474r17}


\bibitem[\protect\citeauthoryear{Alvarado, Juanicorena, Adin, Sedano, Gutirrez,
  and de~N}{Alvarado et~al\mbox{.}}{2012}]%
        {P633}
{U Alvarado}, {A Juanicorena}, {I Adin}, {B Sedano}, {I Gutirrez}, {and} {J de
  N}. 2012.
\newblock \showarticletitle{{Energy harvesting technologies for low-power
  electronics}}.
\newblock {\em Transactions on Emerging Tele communications Technologies\/}
  {23}, 8 (2012), 728--741.
\newblock
\showISSN{2161-3915}
\showDOI{%
\url{http://dx.doi.org/10.1002/ett.2529}}


\bibitem[\protect\citeauthoryear{Asin and Gascon}{Asin and Gascon}{2012}]%
        {P416}
{Alicia Asin} {and} {David Gascon}. 2012.
\newblock {\em {50 Sensor Applications for a Smarter World}}.
\newblock {T}echnical {R}eport. Libelium Comunicaciones Distribuidas.
\newblock


\bibitem[\protect\citeauthoryear{Belli, Cirani, Ferrari, Melegari, and
  Picone}{Belli et~al\mbox{.}}{2015}]%
        {FC03}
{Laura Belli}, {Simone Cirani}, {Gianluigi Ferrari}, {Lorenzo Melegari}, {and}
  {Marco Picone}. 2015.
\newblock \showarticletitle{{A graph-based cloud architecture for big stream
  real-time applications in the internet of things}}. In {\em Communications in
  Computer and Information Science}, Vol. 508. Springer International
  Publishing, 91--105.
\newblock
\showISBNx{9783319148854}
\showISSN{18650929}
\showDOI{%
\url{http://dx.doi.org/10.1007/978-3-319-14886-1_10}}


\bibitem[\protect\citeauthoryear{Bluetooth}{Bluetooth}{2005}]%
        {Bluetooth2005}
{SIG Bluetooth}. 2005.
\newblock \showarticletitle{{Specification of the Bluetooth system}}.
\newblock {\em Core, version\/}  {1} (2005), 2005--10.
\newblock
\showISBNx{Version 1.0 B}


\bibitem[\protect\citeauthoryear{Bonomi, Milito, Zhu, and Addepalli}{Bonomi
  et~al\mbox{.}}{2012}]%
        {FC01}
{Flavio Bonomi}, {Rodolfo Milito}, {Jiang Zhu}, {and} {Sateesh Addepalli}.
  2012.
\newblock \showarticletitle{{Fog computing and its role in the internet of
  things}}. In {\em Proceedings of the first edition of the MCC workshop on
  Mobile cloud computing - MCC '12}. ACM Press, New York, New York, USA, 13.
\newblock
\showISBNx{9781450315197}
\showDOI{%
\url{http://dx.doi.org/10.1145/2342509.2342513}}


\bibitem[\protect\citeauthoryear{Bordawekar, Blainey, and Puri}{Bordawekar
  et~al\mbox{.}}{2015}]%
        {Bordawekar2015}
{Rajesh Bordawekar}, {Bob Blainey}, {and} {Ruchir Puri}. 2015.
\newblock \showarticletitle{{Analyzing Analytics}}.
\newblock {\em Synthesis Lectures on Computer Architecture\/} {10}, 4 (nov
  2015), 1--124.
\newblock
\showISBNx{9781627058353}
\showISSN{1935-3235}
\showDOI{%
\url{http://dx.doi.org/10.2200/S00678ED1V01Y201511CAC035}}


\bibitem[\protect\citeauthoryear{Bormann, Castellani, and Shelby}{Bormann
  et~al\mbox{.}}{2012}]%
        {Bormann2012}
{Carsten Bormann}, {Angelo~P. Castellani}, {and} {Zach Shelby}. 2012.
\newblock \showarticletitle{{CoAP: An application protocol for billions of tiny
  internet nodes}}.
\newblock {\em IEEE Internet Computing\/} {16}, 2 (2012), 62--67.
\newblock
\showISBNx{1089-7801 VO - 16}
\showISSN{10897801}
\showDOI{%
\url{http://dx.doi.org/10.1109/MIC.2012.29}}


\bibitem[\protect\citeauthoryear{Botts and Robin}{Botts and Robin}{2007}]%
        {P256}
{Mike Botts} {and} {Alexandre Robin}. 2007.
\newblock {\em {OpenGIS Sensor Model Language (SensorML) Implementation
  Specification}}.
\newblock {T}echnical {R}eport. Open Geospatial Consortium Inc.
\newblock


\bibitem[\protect\citeauthoryear{Bruneo, Distefano, Longo, Merlino, Puliafito,
  D'Amico, Sapienza, and Torrisi}{Bruneo et~al\mbox{.}}{2016}]%
        {Stack4Things}
{D Bruneo}, {S Distefano}, {F Longo}, {G Merlino}, {A Puliafito}, {V D'Amico},
  {M Sapienza}, {and} {G Torrisi}. 2016.
\newblock \showarticletitle{{Stack4Things as a fog computing platform for Smart
  City applications}}. In {\em 2016 IEEE Conference on Computer Communications
  Workshops (INFOCOM Workshops)}. 848--853.
\newblock
\showDOI{%
\url{http://dx.doi.org/10.1109/INFCOMW.2016.7562195}}


\bibitem[\protect\citeauthoryear{Buratti, Conti, Dardari, and Verdone}{Buratti
  et~al\mbox{.}}{2009}]%
        {Buratti2009}
{Chiara Buratti}, {Andrea Conti}, {Davide Dardari}, {and} {Roberto Verdone}.
  2009.
\newblock {An overview on wireless sensor networks technology and evolution}.
\newblock   (2009).
\newblock
\showISSN{14248220}
\showDOI{%
\url{http://dx.doi.org/10.3390/s90906869}}


\bibitem[\protect\citeauthoryear{Buyya and {Vahid Dastjerdi}}{Buyya and {Vahid
  Dastjerdi}}{2016}]%
        {Buyya2016}
{Rajkumar Buyya} {and} {Amir {Vahid Dastjerdi}}. 2016.
\newblock {\em {Internet of things : principles and paradigms}}.
\newblock Morgan Kaufmann. 354 pages.
\newblock
\showISBNx{9780128053959}


\bibitem[\protect\citeauthoryear{Cai and Zhu}{Cai and Zhu}{2015}]%
        {QoD}
{Li Cai} {and} {Yangyong Zhu}. 2015.
\newblock \showarticletitle{The Challenges of Data Quality and Data Quality
  Assessment in the Big Data Era}.
\newblock {\em Data Science Journal\/} {14}, 2 (2015).
\newblock
\showDOI{%
\url{http://dx.doi.org/10.5334/dsj-2015-002}}


\bibitem[\protect\citeauthoryear{Cao, Chen, Hou, and Brown}{Cao
  et~al\mbox{.}}{2015}]%
        {FC25}
{Yu Cao}, {Songqing Chen}, {Peng Hou}, {and} {Donald Brown}. 2015.
\newblock \showarticletitle{{FAST: A fog computing assisted distributed
  analytics system to monitor fall for stroke mitigation}}. In {\em Proceedings
  of the 2015 IEEE International Conference on Networking, Architecture and
  Storage, NAS 2015}. IEEE, 2--11.
\newblock
\showISBNx{9781467378918}
\showDOI{%
\url{http://dx.doi.org/10.1109/NAS.2015.7255196}}


\bibitem[\protect\citeauthoryear{Caragliu, Bo, and Nijkamp}{Caragliu
  et~al\mbox{.}}{2009}]%
        {P532}
{Andrea Caragliu}, {Chiara~Del Bo}, {and} {Peter Nijkamp}. 2009.
\newblock \showarticletitle{{Smart cities in Europe}}. In {\em 3rd Central
  European Conference in Regional Science-CERS}. 45--59.
\newblock


\bibitem[\protect\citeauthoryear{Cardoso and Sheth}{Cardoso and Sheth}{2006}]%
        {Cardoso2006}
{Jorge Cardoso} {and} {Amit Sheth}. 2006.
\newblock \showarticletitle{{The Semantic Web and its Applications}}.
\newblock In {\em Semantic Web Services Processes and Applications}. Vol.~3.
  Springer US, Boston, MA, 3--33.
\newblock
\showISBNx{0387302395}
\showISSN{0264-410X}
\showDOI{%
\url{http://dx.doi.org/10.1007/978-0-387-34685-4_1}}


\bibitem[\protect\citeauthoryear{Castellani, Rahman, Dijk, Fossati, and
  Loreto}{Castellani et~al\mbox{.}}{2011}]%
        {Castellani2011}
{Angelo~P. Castellani}, {Akbar Rahman}, {Esko Dijk}, {Thomas Fossati}, {and}
  {Salvatore Loreto}. 2011.
\newblock {Best practices for HTTP-CoAP mapping implementation}.
\newblock   (2011).
\newblock
\showURL{%
\url{http://tools.ietf.org/html/draft-castellani-core-http-mapping-02}}


\bibitem[\protect\citeauthoryear{Chaves and Decker}{Chaves and Decker}{2010}]%
        {P017}
{Leonardo Weiss~Ferreira Chaves} {and} {Christian Decker}. 2010.
\newblock \showarticletitle{{A survey on organic smart labels for the
  Internet-of-Things}}. In {\em Networked Sensing Systems (INSS), 2010 Seventh
  International Conference on}. 161--164.
\newblock
\showDOI{%
\url{http://dx.doi.org/10.1109/INSS.2010.5573467}}


\bibitem[\protect\citeauthoryear{Chiang}{Chiang}{2015}]%
        {Chiang2015}
{Mung Chiang}. 2015.
\newblock {\em {Fog Networking:An Overview on Research Opportunities}}.
\newblock {T}echnical {R}eport.
\newblock
\showURL{%
\url{http://www.princeton.edu/}}


\bibitem[\protect\citeauthoryear{Chirila, Lemnaru, and Dinsoreanu}{Chirila
  et~al\mbox{.}}{2016}]%
        {Chirila2016}
{Stefana Chirila}, {Camelia Lemnaru}, {and} {Mihaela Dinsoreanu}. 2016.
\newblock \showarticletitle{{Semantic-based IoT device discovery and
  recommendation mechanism}}. In {\em 2016 IEEE 12th International Conference
  on Intelligent Computer Communication and Processing (ICCP)}. IEEE, 111--116.
\newblock
\showISBNx{978-1-5090-3899-2}
\showDOI{%
\url{http://dx.doi.org/10.1109/ICCP.2016.7737131}}


\bibitem[\protect\citeauthoryear{{Commonwealth Scientific and Industrial
  Research Organisation (CSIRO), Australia}}{{Commonwealth Scientific and
  Industrial Research Organisation (CSIRO), Australia}}{2011}]%
        {P412}
{{Commonwealth Scientific and Industrial Research Organisation (CSIRO),
  Australia}}. 2011.
\newblock {Phenonet: Distributed Sensor Network for Phenomics supported by High
  Resolution Plant Phenomics Centre, CSIRO ICT Centre, and CSIRO Sensor and
  Sensor Networks TCP.}
\newblock   (2011).
\newblock


\bibitem[\protect\citeauthoryear{Compton, Henson, Neuhaus, Lefort, and
  Sheth}{Compton et~al\mbox{.}}{2009}]%
        {P103}
{Michael Compton}, {Corey Henson}, {Holger Neuhaus}, {Laurent Lefort}, {and}
  {Amit Sheth}. 2009.
\newblock \showarticletitle{{A Survey of the Semantic Specification of
  Sensors}}. In {\em 2nd International Workshop on Semantic Sensor Networks, at
  8th International Semantic Web Conference,}.
\newblock


\bibitem[\protect\citeauthoryear{Coskun, Ozdenizci, and Ok}{Coskun
  et~al\mbox{.}}{2013}]%
        {Coskun2013}
{Vedat Coskun}, {Busra Ozdenizci}, {and} {Kerem Ok}. 2013.
\newblock {A survey on near field communication (NFC) technology}.
\newblock   (2013).
\newblock
\showISBNx{9783642228759}
\showISSN{09296212}
\showDOI{%
\url{http://dx.doi.org/10.1007/s11277-012-0935-5}}


\bibitem[\protect\citeauthoryear{Daemen and Rijmen}{Daemen and Rijmen}{2002}]%
        {Daemen:2002}
{Joan Daemen} {and} {Vincent Rijmen}. 2002.
\newblock {\em {The design of AES- the Advanced Encryption Standard}}.
\newblock Spring{\{}$\backslash$-{\}}er-Ver{\{}$\backslash$-{\}}lag. 238 pages.
\newblock
\showISBNx{3-540-42580-2}


\bibitem[\protect\citeauthoryear{Datta, Bonnet, and Haerri}{Datta
  et~al\mbox{.}}{2015}]%
        {Datta2015}
{Soumya~Kanti Datta}, {Christian Bonnet}, {and} {Jerome Haerri}. 2015.
\newblock \showarticletitle{{Fog Computing architecture to enable consumer
  centric Internet of Things services}}. In {\em Proceedings of the
  International Symposium on Consumer Electronics, ISCE}, Vol. 2015-Augus.
\newblock
\showISBNx{9781467373654}
\showISSN{0747-668X}
\showDOI{%
\url{http://dx.doi.org/10.1109/ISCE.2015.7177778}}


\bibitem[\protect\citeauthoryear{{De Paola}, Ferraro, Gaglio, {Lo Re}, and
  Das}{{De Paola} et~al\mbox{.}}{2016}]%
        {DePaola2016}
{Alessandra {De Paola}}, {Pierluca Ferraro}, {Salvatore Gaglio}, {Giuseppe {Lo
  Re}}, {and} {Sajal Das}. 2016.
\newblock \showarticletitle{{An Adaptive Bayesian System for Context-Aware Data
  Fusion in Smart Environments}}.
\newblock {\em IEEE Transactions on Mobile Computing\/} (2016), 1--1.
\newblock
\showISSN{1536-1233}
\showDOI{%
\url{http://dx.doi.org/10.1109/TMC.2016.2599158}}


\bibitem[\protect\citeauthoryear{Delfs and Knebl}{Delfs and Knebl}{2015}]%
        {Delfs2007}
{Hans. Delfs} {and} {Helmut. Knebl}. 2015.
\newblock {\em {Introduction to cryptography: Principles and applications:
  Third edition}}.
\newblock Springer. 1--508 pages.
\newblock
\showISBNx{9783662479742}
\showISSN{1619-7100}
\showDOI{%
\url{http://dx.doi.org/10.1007/978-3-662-47974-2}}


\bibitem[\protect\citeauthoryear{Dementyev, Hodges, Taylor, and
  Smith}{Dementyev et~al\mbox{.}}{2013}]%
        {Dementyev2013}
{Artem Dementyev}, {Steve Hodges}, {Stuart Taylor}, {and} {Josh Smith}. 2013.
\newblock \showarticletitle{{Power Consumption Analysis of Bluetooth Low
  Energy, ZigBee, and ANT Sensor Nodes in a Cyclic Sleep Scenario}}. IEEE.
\newblock


\bibitem[\protect\citeauthoryear{Dey}{Dey}{2001}]%
        {P132}
{Anind~K Dey}. 2001.
\newblock \showarticletitle{{Understanding and Using Context}}.
\newblock {\em Personal Ubiquitous Comput.\/} {5}, 1 (jan 2001), 4--7.
\newblock
\showISSN{1617-4909}
\showDOI{%
\url{http://dx.doi.org/10.1007/s007790170019}}


\bibitem[\protect\citeauthoryear{Dsouza, Ahn, and Taguinod}{Dsouza
  et~al\mbox{.}}{2014}]%
        {FC16}
{Clinton Dsouza}, {Gail~Joon Ahn}, {and} {Marthony Taguinod}. 2014.
\newblock \showarticletitle{{Policy-driven security management for fog
  computing: Preliminary framework and a case study}}. In {\em Proceedings of
  the 2014 IEEE 15th International Conference on Information Reuse and
  Integration, IEEE IRI 2014}. IEEE, 16--23.
\newblock
\showISBNx{9781479958801}
\showDOI{%
\url{http://dx.doi.org/10.1109/IRI.2014.7051866}}


\bibitem[\protect\citeauthoryear{Dubey, Yang, Constant, Amiri, Yang, and
  Makodiya}{Dubey et~al\mbox{.}}{2015}]%
        {FC52}
{Harishchandra Dubey}, {Jing Yang}, {Nick Constant}, {Amir~Mohammad Amiri},
  {Qing Yang}, {and} {Kunal Makodiya}. 2015.
\newblock \showarticletitle{{Fog Data: Enhancing Telehealth Big Data Through
  Fog Computing}}.
\newblock {\em Proceedings of the ASE BigData {\&} SocialInformatics 2015\/}
  (2015), 14:1----14:6.
\newblock
\showISBNx{978-1-4503-3735-9}
\showDOI{%
\url{http://dx.doi.org/10.1145/2818869.2818889}}


\bibitem[\protect\citeauthoryear{Dutta}{Dutta}{2013}]%
        {Karasiewicz2013}
{Aditya Dutta}. 2013.
\newblock {Why HTTP is not enough for the Internet of Things}.
\newblock   (2013).
\newblock


\bibitem[\protect\citeauthoryear{{European Commission}}{{European
  Commission}}{2008}]%
        {P006}
{{European Commission}}. 2008.
\newblock {\em {Internet of Things in 2020 Road Map For The Future}}.
\newblock {T}echnical {R}eport. Working Group RFID of the ETP EPOSS.
\newblock


\bibitem[\protect\citeauthoryear{{European Research Cluster on the Internet of
  Things}}{{European Research Cluster on the Internet of Things}}{2015}]%
        {Things2015}
{{European Research Cluster on the Internet of Things}}. 2015.
\newblock {\em {Internet of Things - IoT Semantic Interoperability: research
  challeges, best practices, recommendations and next steps}}.
\newblock {T}echnical {R}eport. 48 pages.
\newblock


\bibitem[\protect\citeauthoryear{Farris, Girau, Militano, Nitti, Atzori, Iera,
  and Morabito}{Farris et~al\mbox{.}}{2015}]%
        {FC49}
{Ivan Farris}, {Roberto Girau}, {Leonardo Militano}, {Michele Nitti}, {Luigi
  Atzori}, {Antonio Iera}, {and} {Giacomo Morabito}. 2015.
\newblock \showarticletitle{{Social Virtual Objects in the Edge Cloud}}.
\newblock {\em IEEE Cloud Computing\/} {2}, 6 (nov 2015), 20--28.
\newblock
\showISSN{2325-6095}
\showDOI{%
\url{http://dx.doi.org/10.1109/MCC.2015.116}}


\bibitem[\protect\citeauthoryear{Farris, Militano, Nitti, Atzori, and
  Iera}{Farris et~al\mbox{.}}{2016}]%
        {FC47}
{I. Farris}, {L. Militano}, {M. Nitti}, {L. Atzori}, {and} {A. Iera}. 2016.
\newblock \showarticletitle{{Federated edge-assisted mobile clouds for service
  provisioning in heterogeneous IoT environments}}. In {\em IEEE World Forum on
  Internet of Things, WF-IoT 2015 - Proceedings}. IEEE, 591--596.
\newblock
\showISBNx{9781509003655}
\showDOI{%
\url{http://dx.doi.org/10.1109/WF-IoT.2015.7389120}}


\bibitem[\protect\citeauthoryear{Fielding, Gettys, Mogul, Frystyk, Masinter,
  Leach, and Berners-Lee}{Fielding et~al\mbox{.}}{1999}]%
        {Fielding1999}
{R Fielding}, {J Gettys}, {J Mogul}, {H Frystyk}, {L Masinter}, {P Leach},
  {and} {T Berners-Lee}. 1999.
\newblock {RFC 2616 - Hypertext Transfer Protocol - HTTP/1.1}.
\newblock   (1999).
\newblock
\showISBNx{2616}
\showISSN{20701721}
\showDOI{%
\url{http://dx.doi.org/rfc/rfc2616.txt}}


\bibitem[\protect\citeauthoryear{Fodor}{Fodor}{2002}]%
        {Fodor2002}
{Imola~K Fodor}. 2002.
\newblock \showarticletitle{{A survey of dimension reduction techniques}}.
\newblock {\em Library\/} {18}, 1 (2002), 1--18.
\newblock
\showDOI{%
\url{http://dx.doi.org/10.2172/15002155}}


\bibitem[\protect\citeauthoryear{{Forrest Stroud}}{{Forrest Stroud}}{}]%
        {FogDef}
{{Forrest Stroud}}.
\newblock {Fog Computing}.
\newblock   (????).
\newblock
\showURL{%
\url{http://www.webopedia.com/TERM/F/fog-computing.html}}


\bibitem[\protect\citeauthoryear{Garcia-Martinez}{Garcia-Martinez}{}]%
        {iotGoesNano}
{Javier Garcia-Martinez}.
\newblock \showarticletitle{{The Internet of Things Goes Nano}}.
\newblock {\em
  https://www.scientificamerican.com/article/the-internet-of-things-goes-nano/,
  Retrieved November 2016\/} (????).
\newblock


\bibitem[\protect\citeauthoryear{Gascon}{Gascon}{2015}]%
        {LibeliumSecurity}
{David Gascon}. 2015.
\newblock {\em {IoT Security Infographic – Privacy, Authenticity,
  Confidentiality and Integrity of the Sensor Data. “The Invisible Asset}}.
\newblock {T}echnical {R}eport. Libelium.
\newblock


\bibitem[\protect\citeauthoryear{Gaura, Brusey, Allen, Wilkins, Goldsmith, and
  Rednic}{Gaura et~al\mbox{.}}{2013}]%
        {Gaura2013}
{Elena~I. Gaura}, {James Brusey}, {Michael Allen}, {Ross Wilkins}, {Dan
  Goldsmith}, {and} {Ramona Rednic}. 2013.
\newblock \showarticletitle{{Edge mining the internet of things}}.
\newblock {\em IEEE Sensors Journal\/} {13}, 10 (oct 2013), 3816--3825.
\newblock
\showISSN{1530437X}


\bibitem[\protect\citeauthoryear{Gazis, Leonardi, Mathioudakis, Sasloglou,
  Kikiras, and Sudhaakar}{Gazis et~al\mbox{.}}{2015}]%
        {FC36}
{Vangelis Gazis}, {Alessandro Leonardi}, {Kostas Mathioudakis}, {Konstantinos
  Sasloglou}, {Panayotis Kikiras}, {and} {Raghuram Sudhaakar}. 2015.
\newblock \showarticletitle{{Components of fog computing in an industrial
  internet of things context}}. In {\em 2015 12th Annual IEEE International
  Conference on Sensing, Communication, and Networking - Workshops, SECON
  Workshops 2015}. IEEE, 37--42.
\newblock
\showISBNx{9781467373920}
\showDOI{%
\url{http://dx.doi.org/10.1109/SECONW.2015.7328144}}


\bibitem[\protect\citeauthoryear{Gia, Jiang, Rahmani, Westerlund, Liljeberg,
  and Tenhunen}{Gia et~al\mbox{.}}{2015}]%
        {FC43}
{Tuan~Nguyen Gia}, {Mingzhe Jiang}, {Amir~Mohammad Rahmani}, {Tomi Westerlund},
  {Pasi Liljeberg}, {and} {Hannu Tenhunen}. 2015.
\newblock \showarticletitle{{Fog computing in healthcare Internet of Things: A
  case study on ECG feature extraction}}. In {\em Proceedings - 15th IEEE
  International Conference on Computer and Information Technology, CIT 2015,
  14th IEEE International Conference on Ubiquitous Computing and
  Communications, IUCC 2015, 13th IEEE International Conference on Dependable,
  Autonomic and Se}. IEEE, 356--363.
\newblock
\showISBNx{9781509001545}
\showDOI{%
\url{http://dx.doi.org/10.1109/CIT/IUCC/DASC/PICOM.2015.51}}


\bibitem[\protect\citeauthoryear{Giang, Blackstock, Lea, and Leung}{Giang
  et~al\mbox{.}}{2015}]%
        {FC28}
{Nam~Ky Giang}, {Michael Blackstock}, {Rodger Lea}, {and} {Victor C~M Leung}.
  2015.
\newblock \showarticletitle{{Developing IoT applications in the Fog: A
  Distributed Dataflow approach}}. In {\em Proceedings - 2015 5th International
  Conference on the Internet of Things, IoT 2015}. IEEE, 155--162.
\newblock
\showISBNx{9781467380584}
\showDOI{%
\url{http://dx.doi.org/10.1109/IOT.2015.7356560}}


\bibitem[\protect\citeauthoryear{Giffinger, Fertner, Kramar, Kalasek,
  Pichler-Milanovic, and Meijers}{Giffinger et~al\mbox{.}}{2007}]%
        {P528}
{Rudolf Giffinger}, {Christian Fertner}, {Hans Kramar}, {Robert Kalasek},
  {Natasa Pichler-Milanovic}, {and} {Evert Meijers}. 2007.
\newblock {\em {Smart cities Ranking of European medium-sized cities}}.
\newblock Research project report. Centre of Regional Science, Vienna UT.
\newblock


\bibitem[\protect\citeauthoryear{Gomez, Rasheed, Riggio, Miorandi, Sengul, and
  Bayer}{Gomez et~al\mbox{.}}{2013}]%
        {Gomez2013}
{Karina Gomez}, {Tinku Rasheed}, {Roberto Riggio}, {Daniele Miorandi}, {Cigdem
  Sengul}, {and} {Nico Bayer}. 2013.
\newblock \showarticletitle{{Achilles and the tortoise: Power consumption in
  IEEE 802.11n and IEEE 802.11g networks}}. In {\em 2013 IEEE Online Conference
  on Green Communications, OnlineGreenComm 2013}. IEEE, 20--26.
\newblock
\showISBNx{9781479903542}
\showDOI{%
\url{http://dx.doi.org/10.1109/OnlineGreenCom.2013.6731023}}


\bibitem[\protect\citeauthoryear{Gonzalez}{Gonzalez}{2016}]%
        {diffOfActuators}
{Carlos Gonzalez}. 2016.
\newblock {What is the Difference between Pneumatic, Hydraulic, and Electrical
  Actuators}.
\newblock   (2016).
\newblock
\newblock
\shownote{\url{http://machinedesign.com/linear-motion/what-s-difference-between-pneumatic-hydraulic-and-electrical-actuators}
  [Retrieved November 2016].}


\bibitem[\protect\citeauthoryear{Gu, Zeng, Guo, Barnawi, and Xiang}{Gu
  et~al\mbox{.}}{2015}]%
        {FC39}
{Lin Gu}, {Deze Zeng}, {Song Guo}, {Ahmed Barnawi}, {and} {Yong Xiang}. 2015.
\newblock \showarticletitle{{Cost-Efficient Resource Management in Fog
  Computing Supported Medical CPS}}.
\newblock {\em IEEE Transactions on Emerging Topics in Computing\/} {6750}, c
  (2015), 1--1.
\newblock
\showISSN{2168-6750}
\showDOI{%
\url{http://dx.doi.org/10.1109/TETC.2015.2508382}}


\bibitem[\protect\citeauthoryear{Halperin, Greenstein, Sheth, and
  Wetherall}{Halperin et~al\mbox{.}}{2010}]%
        {Halperin2010}
{Daniel Halperin}, {Ben Greenstein}, {Anmol Sheth}, {and} {David Wetherall}.
  2010.
\newblock {Demystifying 802.11n power consumption}.
\newblock   (2010).
\newblock


\bibitem[\protect\citeauthoryear{Harmening}{Harmening}{2013}]%
        {Harmening2013}
{James~T. Harmening}. 2013.
\newblock \showarticletitle{{Virtual Private Networks}}.
\newblock In {\em Computer and Information Security Handbook}. 855--867.
\newblock
\showISBNx{9780123943972}
\showISSN{1060-5487}
\showDOI{%
\url{http://dx.doi.org/10.1016/B978-0-12-394397-2.00048-9}}


\bibitem[\protect\citeauthoryear{Hassan, Xiao, Wei, and Chen}{Hassan
  et~al\mbox{.}}{2015}]%
        {FC38}
{Mohammed~A. Hassan}, {Mengbai Xiao}, {Qi Wei}, {and} {Songqing Chen}. 2015.
\newblock \showarticletitle{{Help your mobile applications with fog
  computing}}. In {\em 2015 12th Annual IEEE International Conference on
  Sensing, Communication, and Networking - Workshops, SECON Workshops 2015}.
  IEEE, 49--54.
\newblock
\showISBNx{9781467373920}
\showDOI{%
\url{http://dx.doi.org/10.1109/SECONW.2015.7328146}}


\bibitem[\protect\citeauthoryear{Henry and Luo}{Henry and Luo}{2002}]%
        {Henry2002}
{Paul~S. Henry} {and} {Hui Luo}. 2002.
\newblock \showarticletitle{{WiFi: What's next?}}
\newblock {\em IEEE Communications Magazine\/} {40}, 12 (dec 2002), 66--72.
\newblock
\showISSN{01636804}
\showDOI{%
\url{http://dx.doi.org/10.1109/MCOM.2002.1106162}}


\bibitem[\protect\citeauthoryear{Heydon and Hunn}{Heydon and Hunn}{2012}]%
        {Heydon2012}
{R Heydon} {and} {N Hunn}. 2012.
\newblock {\em {Bluetooth Low Energy : The Developer's Handbook}}.
\newblock 368 pages.
\newblock
\showISBNx{9780132888363}
\showURL{%
\url{https://www.bluetooth.org/DocMan/handlers/DownloadDoc.ashx}}


\bibitem[\protect\citeauthoryear{Higuchi, Yamaguchi, Higashino, and
  Takai}{Higuchi et~al\mbox{.}}{2014}]%
        {Higuchi2014}
{Takamasa Higuchi}, {Hirozumi Yamaguchi}, {Teruo Higashino}, {and} {Mineo
  Takai}. 2014.
\newblock \showarticletitle{{A neighbor collaboration mechanism for mobile
  crowd sensing in opportunistic networks}}. In {\em 2014 IEEE International
  Conference on Communications (ICC)}. IEEE, 42--47.
\newblock
\showISBNx{978-1-4799-2003-7}
\showDOI{%
\url{http://dx.doi.org/10.1109/ICC.2014.6883292}}


\bibitem[\protect\citeauthoryear{Hunkeler, Truong, and Stanford-Clark}{Hunkeler
  et~al\mbox{.}}{2008}]%
        {Hunkeler2008}
{Urs Hunkeler}, {Hong~Linh Truong}, {and} {Andy Stanford-Clark}. 2008.
\newblock \showarticletitle{{MQTT-S — A publish/subscribe protocol for
  Wireless Sensor Networks}}.
\newblock {\em 2008 3rd International Conference on Communication Systems
  Software and Middleware and Workshops (COMSWARE '08)\/} (2008), 791--798.
\newblock
\showISBNx{978-1-4244-1796-4}
\showDOI{%
\url{http://dx.doi.org/10.1109/COMSWA.2008.4554519}}


\bibitem[\protect\citeauthoryear{{Hypercat Consortium}}{{Hypercat
  Consortium}}{2016}]%
        {Hypercat}
{{Hypercat Consortium}}. 2016.
\newblock {Hypercat}.
\newblock   (2016).
\newblock
\showURL{%
\url{http://www.hypercat.io/}}


\bibitem[\protect\citeauthoryear{{IEEE Instrumentation and Measurement
  Society}}{{IEEE Instrumentation and Measurement Society}}{2007}]%
        {P258}
{{IEEE Instrumentation and Measurement Society}}. 2007.
\newblock \showarticletitle{{IEEE Standard for a Smart Transducer Interface for
  Sensors and Actuators Wireless Communication Protocols and Transducer
  Electronic Data Sheet (TEDS) Formats}}.
\newblock {\em IEEE Std 1451.5-2007\/} (2007), C1 --236.
\newblock
\showDOI{%
\url{http://dx.doi.org/10.1109/IEEESTD.2007.4346346}}


\bibitem[\protect\citeauthoryear{Ishino, Koizumi, and Hasegawa}{Ishino
  et~al\mbox{.}}{2015}]%
        {Ishino2015}
{Masanori Ishino}, {Yuki Koizumi}, {and} {Toru Hasegawa}. 2015.
\newblock \showarticletitle{{Leveraging proximity services for relay device
  discovery in user-provided IoT networks}}. In {\em 2015 IEEE 2nd World Forum
  on Internet of Things (WF-IoT)}. IEEE, 553--558.
\newblock
\showISBNx{978-1-5090-0366-2}
\showDOI{%
\url{http://dx.doi.org/10.1109/WF-IoT.2015.7389114}}


\bibitem[\protect\citeauthoryear{Ismail, {Mostajeran Goortani}, {Ab Karim},
  {Ming Tat}, Setapa, Luke, and {Hong Hoe}}{Ismail et~al\mbox{.}}{2015}]%
        {FC41}
{Bukhary~Ikhwan Ismail}, {Ehsan {Mostajeran Goortani}}, {Mohd~Bazli {Ab
  Karim}}, {Wong {Ming Tat}}, {Sharipah Setapa}, {Jing~Yuan Luke}, {and} {Ong
  {Hong Hoe}}. 2015.
\newblock \showarticletitle{{Evaluation of Docker as Edge computing platform}}.
  In {\em 2015 IEEE Conference on Open Systems (ICOS)}. IEEE, 130--135.
\newblock
\showISBNx{978-1-4673-9434-5}
\showDOI{%
\url{http://dx.doi.org/10.1109/ICOS.2015.7377291}}


\bibitem[\protect\citeauthoryear{Jia and Xu}{Jia and Xu}{2013}]%
        {rpme/JiaX2013}
{Yukun Jia} {and} {Qingsong Xu}. 2013.
\newblock \showarticletitle{{MEMS Microgripper Actuators and Sensors: The
  State-of-the-Art Survey}}.
\newblock {\em Recent Patents on Mechanical Engineering\/} {6}, 2 (2013),
  132--142.
\newblock
\showURL{%
\url{http://www.eurekaselect.com/108839}}


\bibitem[\protect\citeauthoryear{Jiang, Tang, Liu, Qiu, and Xu}{Jiang
  et~al\mbox{.}}{2014}]%
        {Jiang2014}
{Qingyun Jiang}, {Ruichun Tang}, {Peishun Liu}, {Yue Qiu}, {and} {Huimin Xu}.
  2014.
\newblock \showarticletitle{{Research on dynamic data fusion algorithm based on
  context awareness}}. In {\em 2014 IEEE International Conference on Progress
  in Informatics and Computing}. IEEE, 529--534.
\newblock
\showISBNx{978-1-4799-2030-3}
\showDOI{%
\url{http://dx.doi.org/10.1109/PIC.2014.6972391}}


\bibitem[\protect\citeauthoryear{{Jiang Zhu}, Chan, Prabhu, Natarajan, {Hao
  Hu}, and Bonomi}{{Jiang Zhu} et~al\mbox{.}}{2013}]%
        {FC09}
{Jiang {Jiang Zhu}}, {D.~S. Chan}, {M.~S. Prabhu}, {P. Natarajan}, {Hao {Hao
  Hu}}, {and} {F. Bonomi}. 2013.
\newblock \showarticletitle{{Improving Web Sites Performance Using Edge Servers
  in Fog Computing Architecture}}.
\newblock {\em 2013 IEEE Seventh International Symposium on Service-Oriented
  System Engineering\/} (mar 2013), 320--323.
\newblock
\showISBNx{978-0-7695-4944-6}
\showDOI{%
\url{http://dx.doi.org/10.1109/SOSE.2013.73}}


\bibitem[\protect\citeauthoryear{Kulkarni, Saha, and Hockenbury}{Kulkarni
  et~al\mbox{.}}{2012}]%
        {FC15}
{Saurabh Kulkarni}, {Shayan Saha}, {and} {Ryler Hockenbury}. 2012.
\newblock \showarticletitle{{Preserving privacy in sensor-fog networks}}. In
  {\em 2014 9th International Conference for Internet Technology and Secured
  Transactions, ICITST 2014}. IEEE, 96--99.
\newblock
\showISBNx{9781908320391}
\showDOI{%
\url{http://dx.doi.org/10.1109/ICITST.2014.7038785}}


\bibitem[\protect\citeauthoryear{Lampkin, Leong, Olivera, Rawat, Subrahmanyam,
  and Xiang}{Lampkin et~al\mbox{.}}{2012}]%
        {Lampkin2012}
{Valerie Lampkin}, {Weng~Tat Leong}, {Leonardo Olivera}, {Sweta Rawat}, {Nagesh
  Subrahmanyam}, {and} {Rong Xiang}. 2012.
\newblock \showarticletitle{{Building Smarter Planet Solutions with MQTT and
  IBM WebSphere MQ Telemetry}}.
\newblock {\em IBM Redbooks\/} (2012), 270.
\newblock
\showISBNx{0738437085}


\bibitem[\protect\citeauthoryear{Lashkari, Mansoori, and Danesh}{Lashkari
  et~al\mbox{.}}{2009}]%
        {Lashkari2009}
{Arash~Habibi Lashkari}, {Masood Mansoori}, {and} {Amir~Seyed Danesh}. 2009.
\newblock \showarticletitle{{Wired equivalent privacy (WEP) versus Wi-Fi
  protected access (WPA)}}. In {\em 2009 International Conference on Signal
  Processing Systems, ICSPS 2009}. 445--449.
\newblock
\showISBNx{9780769536545}
\showDOI{%
\url{http://dx.doi.org/10.1109/ICSPS.2009.87}}


\bibitem[\protect\citeauthoryear{Li, Jin, Yuan, Palaniswami, and Moessner}{Li
  et~al\mbox{.}}{2015}]%
        {FC44}
{Jianhua Li}, {Jiong Jin}, {Dong Yuan}, {Marimuthu Palaniswami}, {and} {Klaus
  Moessner}. 2015.
\newblock \showarticletitle{{EHOPES: Data-centered Fog platform for smart
  living}}. In {\em 25th International Telecommunication Networks and
  Applications Conference, ITNAC 2015}. IEEE, 308--313.
\newblock
\showISBNx{9781467393485}
\showDOI{%
\url{http://dx.doi.org/10.1109/ATNAC.2015.7366831}}


\bibitem[\protect\citeauthoryear{LinkLabs}{LinkLabs}{2016}]%
        {LinkLabs2016}
{LinkLabs}. 2016.
\newblock {\em {Low Power, Wide Area Networks}}.
\newblock {T}echnical {R}eport.
\newblock


\bibitem[\protect\citeauthoryear{{Linux Foundation}}{{Linux
  Foundation}}{2016a}]%
        {AllJoyn}
{{Linux Foundation}}. 2016a.
\newblock {AllJoyn Framework}.
\newblock   (2016).
\newblock
\showURL{%
\url{https://allseenalliance.org/framework}}


\bibitem[\protect\citeauthoryear{{Linux Foundation}}{{Linux
  Foundation}}{2016b}]%
        {IoTivity}
{{Linux Foundation}}. 2016b.
\newblock {IoTivity}.
\newblock   (2016).
\newblock
\showURL{%
\url{https://www.iotivity.org/}}


\bibitem[\protect\citeauthoryear{Liu, Iftikhar, and Xie}{Liu
  et~al\mbox{.}}{2014}]%
        {Liu2014}
{Xiufeng Liu}, {Nadeem Iftikhar}, {and} {Xike Xie}. 2014.
\newblock \showarticletitle{{Survey of real-time processing systems for big
  data}}. In {\em Proceedings of the 18th International Database Engineering
  {\&} Applications Symposium on - IDEAS '14}. ACM Press, New York, New York,
  USA, 356--361.
\newblock
\showISBNx{9781450326278}
\showDOI{%
\url{http://dx.doi.org/10.1145/2628194.2628251}}


\bibitem[\protect\citeauthoryear{Ma and Luo}{Ma and Luo}{2008}]%
        {Ma2008}
{Xin Ma} {and} {Wei Luo}. 2008.
\newblock \showarticletitle{{The analysis of 6LowPAN technology}}. In {\em
  Proceedings - 2008 Pacific-Asia Workshop on Computational Intelligence and
  Industrial Application, PACIIA 2008}, Vol.~1. IEEE, 963--966.
\newblock
\showISBNx{9780769534909}
\showDOI{%
\url{http://dx.doi.org/10.1109/PACIIA.2008.72}}


\bibitem[\protect\citeauthoryear{Madsen, Albeanu, Burtschy, and
  Popentiu-Vladicescu}{Madsen et~al\mbox{.}}{2013}]%
        {FC10}
{H. Madsen}, {G. Albeanu}, {Bernard Burtschy}, {and} {Fl. Popentiu-Vladicescu}.
  2013.
\newblock \showarticletitle{{Reliability in the utility computing era: Towards
  reliable fog computing}}. In {\em International Conference on Systems,
  Signals, and Image Processing}. IEEE, 43--46.
\newblock
\showISBNx{9781479909414}
\showISSN{21578702}
\showDOI{%
\url{http://dx.doi.org/10.1109/IWSSIP.2013.6623445}}


\bibitem[\protect\citeauthoryear{Markkanen}{Markkanen}{2015}]%
        {ABIResearch2015}
{Aapo Markkanen}. 2015.
\newblock {Competitive Edge from Edge Intelligence IoT Analytics Today and in
  2020}.
\newblock   (2015).
\newblock
\showURL{%
\url{http://www.4cad.fr/content/files/Competitive-Edge-from-Edge-Intelligence-IoT-Whitepaper.pdf}}


\bibitem[\protect\citeauthoryear{Mohandas, Doctor, Jayawant, Pattni, and
  Johri}{Mohandas et~al\mbox{.}}{1060}]%
        {Mohandas1060}
{Ajay Mohandas}, {Khoshrav Doctor}, {Shubham Jayawant}, {Mohit Pattni}, {and}
  {Era Johri}. 1060.
\newblock \showarticletitle{{NFC vs Bluetooth}}.
\newblock {\em International Journal of Multidisciplinary and Scientific
  Emerging Research\/} {4}, 1 (1060), 2349--6037.
\newblock


\bibitem[\protect\citeauthoryear{{National Information Standards
  Organization}}{{National Information Standards Organization}}{2004}]%
        {NISO2004}
{{National Information Standards Organization}}. 2004.
\newblock \showarticletitle{{Understanding Metadata}}.
\newblock {\em National Information Standards\/} MD:NISO Press (2004), 20.
\newblock
\showISBNx{1880124629}
\showISSN{00943061}
\showDOI{%
\url{http://dx.doi.org/10.1017/S0003055403000534}}


\bibitem[\protect\citeauthoryear{Nilsson, Hogsden, Perera, Aghaee, Scruton,
  Lund, and Blackwell}{Nilsson et~al\mbox{.}}{2016}]%
        {Tommy2016}
{Tommy Nilsson}, {Carl Hogsden}, {Charith Perera}, {Saeed Aghaee}, {David
  Scruton}, {Andreas Lund}, {and} {Alan~F. Blackwell}. 2016.
\newblock \showarticletitle{{Applying Seamful Design in Location-based Mobile
  Museum Applications}}.
\newblock {\em ACM Transactions on Multimedia Computing Communications and
  Applications (TOMM)\/} (2016).
\newblock


\bibitem[\protect\citeauthoryear{Nishio, Shinkuma, Takahashi, and
  Mandayam}{Nishio et~al\mbox{.}}{2013}]%
        {FC54}
{Takayuki Nishio}, {Ryoichi Shinkuma}, {Tatsuro Takahashi}, {and} {Narayan~B.
  Mandayam}. 2013.
\newblock \showarticletitle{{Service-oriented heterogeneous resource sharing
  for optimizing service latency in mobile cloud}}. In {\em Proceedings of the
  first international workshop on Mobile cloud computing {\&} networking -
  MobileCloud '13}. ACM Press, New York, New York, USA, 19.
\newblock
\showISBNx{9781450322065}
\showDOI{%
\url{http://dx.doi.org/10.1145/2492348.2492354}}


\bibitem[\protect\citeauthoryear{Nuaymi}{Nuaymi}{2007}]%
        {Nuaymi2007}
{Loutfi Nuaymi}. 2007.
\newblock {\em {WiMAX: Technology for Broadband Wireless Access}}.
\newblock 1--283 pages.
\newblock
\showISBNx{9780470319055}
\showDOI{%
\url{http://dx.doi.org/10.1002/9780470319055}}


\bibitem[\protect\citeauthoryear{OASIS}{OASIS}{2014}]%
        {OASIS2014}
{OASIS}. 2014.
\newblock \showarticletitle{{MQTT Version 3.1.1}}.
\newblock {\em OASIS Standard\/} October (2014), 81.
\newblock
\showURL{%
\url{http://docs.oasis-open.org/mqtt/mqtt/v3.1.1/os/mqtt-v3.1.1-os.html}}


\bibitem[\protect\citeauthoryear{O'Hara}{O'Hara}{2007}]%
        {OHara2007}
{John O'Hara}. 2007.
\newblock \showarticletitle{{Toward a commodity enterprise middleware}}.
\newblock {\em Queue\/} {5}, 4 (may 2007), 48--55.
\newblock
\showISBNx{0001-0782}
\showISSN{15427730}
\showDOI{%
\url{http://dx.doi.org/10.1145/1255421.1255424}}


\bibitem[\protect\citeauthoryear{{OpenIoT Consortium}}{{OpenIoT
  Consortium}}{2012}]%
        {P377}
{{OpenIoT Consortium}}. 2012.
\newblock {Open Source Solution for the Internet of Things into the Cloud}.
\newblock   (jan 2012).
\newblock


\bibitem[\protect\citeauthoryear{Oueis, Strinati, and Barbarossa}{Oueis
  et~al\mbox{.}}{2015}]%
        {FC22}
{Jessica Oueis}, {Emilio~Calvanese Strinati}, {and} {Sergio Barbarossa}. 2015.
\newblock \showarticletitle{{The Fog Balancing: Load Distribution for Small
  Cell Cloud Computing}}.
\newblock {\em 2015 IEEE 81st Vehicular Technology Conference (VTC Spring)\/}
  (may 2015), 1--6.
\newblock
\showISBNx{978-1-4799-8088-8}
\showISSN{15502252}
\showDOI{%
\url{http://dx.doi.org/10.1109/VTCSpring.2015.7146129}}


\bibitem[\protect\citeauthoryear{Perera, Jayaraman, Zaslavsky, Christen, and
  Georgakopoulos}{Perera et~al\mbox{.}}{2014a}]%
        {Perera2014Book}
{Charith Perera}, {Prem~Prakash Jayaraman}, {Arkady Zaslavsky}, {Peter
  Christen}, {and} {Dimitrios Georgakopoulos}. 2014a.
\newblock {\em {Context-aware Dynamic Discovery and Configuration of Things in
  Smart Environments}}.
\newblock Springer International Publishing, Cham, Chapter Context-Aw,
  215--241.
\newblock
\showISBNx{978-3-319-05029-4}
\showDOI{%
\url{http://dx.doi.org/10.1007/978-3-319-05029-4_9}}


\bibitem[\protect\citeauthoryear{Perera, Jayaraman, Zaslavsky, Georgakopoulos,
  and Christen}{Perera et~al\mbox{.}}{2014b}]%
        {PereraC011}
{C Perera}, {P~P Jayaraman}, {A Zaslavsky}, {D Georgakopoulos}, {and} {P
  Christen}. 2014b.
\newblock \showarticletitle{{Sensor discovery and configuration framework for
  the Internet of Things paradigm}}. In {\em Internet of Things (WF-IoT), 2014
  IEEE World Forum on}. 94--99.
\newblock
\showDOI{%
\url{http://dx.doi.org/10.1109/WF-IoT.2014.6803127}}


\bibitem[\protect\citeauthoryear{Perera, Liu, and Jayawardena}{Perera
  et~al\mbox{.}}{2014c}]%
        {PereraIEEEAccess}
{Charith Perera}, {Chi~Harold Liu}, {and} {Srimal Jayawardena}. 2014c.
\newblock \showarticletitle{{A Survey on Internet of Things From Industrial
  Market Perspective}}.
\newblock {\em IEEE Access\/}  {2} (2014), 1660--1679.
\newblock
\showISSN{2169-3536}
\showDOI{%
\url{http://dx.doi.org/10.1109/ACCESS.2015.2389854}}


\bibitem[\protect\citeauthoryear{Perera, Liu, and Jayawardena}{Perera
  et~al\mbox{.}}{2015a}]%
        {Perera2015a}
{Charith Perera}, {Chi~Harold Liu}, {and} {Srimal Jayawardena}. 2015a.
\newblock \showarticletitle{{The Emerging Internet of Things Marketplace from
  an Industrial Perspective: A Survey}}.
\newblock {\em IEEE Transactions on Emerging Topics in Computing\/} {3}, 4
  (2015), 585--598.
\newblock
\showISBNx{2168-6750 VO - 3}
\showISSN{21686750}


\bibitem[\protect\citeauthoryear{Perera, Ranjan, Wang, Khan, and Zomaya}{Perera
  et~al\mbox{.}}{2015b}]%
        {Perera2015}
{Charith Perera}, {Rajiv Ranjan}, {Lizhe Wang}, {Samee~U. Khan}, {and}
  {Albert~Y. Zomaya}. 2015b.
\newblock \showarticletitle{{Big data privacy in the internet of things era}}.
\newblock {\em IT Professional\/} {17}, 3 (2015), 32--39.
\newblock
\showISSN{15209202}


\bibitem[\protect\citeauthoryear{Perera, Talagala, Liu, and Estrella}{Perera
  et~al\mbox{.}}{2015c}]%
        {Perera2016}
{Charith Perera}, {Dumidu Talagala}, {Chi~Harold Liu}, {and} {Julio~C.
  Estrella}. 2015c.
\newblock \showarticletitle{{Energy-Efficient Location and Activity-Aware
  On-Demand Mobile Distributed Sensing Platform for Sensing as a Service in IoT
  Clouds}}.
\newblock {\em IEEE Transactions on Computational Social Systems\/} {2}, 4 (dec
  2015), 171--181.
\newblock
\showISSN{2329-924X}


\bibitem[\protect\citeauthoryear{Perera, Zaslavsky, Christen, and
  Georgakopoulos}{Perera et~al\mbox{.}}{2014a}]%
        {ZMP007}
{Charith Perera}, {Arkady Zaslavsky}, {Peter Christen}, {and} {Dimitrios
  Georgakopoulos}. 2014a.
\newblock \showarticletitle{{Context Aware Computing for The Internet of
  Things: A Survey}}.
\newblock {\em Communications Surveys Tutorials, IEEE\/} {16}, 1 (2014), 414 --
  454.
\newblock
\showISSN{1553-877X}


\bibitem[\protect\citeauthoryear{Perera, Zaslavsky, Christen, and
  Georgakopoulos}{Perera et~al\mbox{.}}{2014b}]%
        {Perera2014}
{Charith Perera}, {Arkady Zaslavsky}, {Peter Christen}, {and} {Dimitrios
  Georgakopoulos}. 2014b.
\newblock \showarticletitle{{Sensing as a service model for smart cities
  supported by Internet of Things}}.
\newblock {\em European Transactions on Telecommunications\/} {25}, 1 (2014),
  81--93.
\newblock
\showISBNx{1424403553}
\showISSN{1124318X}
\showDOI{%
\url{http://dx.doi.org/10.1002/ett.2704}}


\bibitem[\protect\citeauthoryear{Perera, Zaslavsky, Christen, Salehi, and
  Georgakopoulos}{Perera et~al\mbox{.}}{2012}]%
        {ZMP002}
{Charith Perera}, {Arkady Zaslavsky}, {Peter Christen}, {Ali Salehi}, {and}
  {Dimitrios Georgakopoulos}. 2012.
\newblock \showarticletitle{{Connecting Mobile Things to Global Sensor Network
  Middleware using System-generated Wrappers}}. In {\em Eleventh ACM
  International Workshop on Data Engineering for Wireless and Mobile Access
  (ACM SIGMOD/PODS 2012-Workshop-MobiDE)}. Scottsdale, Arizona, USA, 23--30.
\newblock
\showDOI{%
\url{http://dx.doi.org/10.1145/2258056.2258062}}


\bibitem[\protect\citeauthoryear{Pietschmann, Mitschick, Winkler, and
  Meissner}{Pietschmann et~al\mbox{.}}{2008}]%
        {P334}
{S Pietschmann}, {A Mitschick}, {R Winkler}, {and} {K Meissner}. 2008.
\newblock \showarticletitle{{CroCo: Ontology-Based, Cross-Application Context
  Management}}. In {\em Semantic Media Adaptation and Personalization, 2008.
  SMAP '08. Third International Workshop on}. 88--93.
\newblock
\showDOI{%
\url{http://dx.doi.org/10.1109/SMAP.2008.10}}


\bibitem[\protect\citeauthoryear{Pozza, Nati, Georgoulas, Moessner, and
  Gluhak}{Pozza et~al\mbox{.}}{2015}]%
        {Pozza2015}
{R. Pozza}, {M. Nati}, {S. Georgoulas}, {K. Moessner}, {and} {A. Gluhak}. 2015.
\newblock \showarticletitle{{Neighbor Discovery for Opportunistic Networking in
  Internet of Things Scenarios: A Survey}}.
\newblock {\em IEEE Access\/}  {3} (2015), 1101--1131.
\newblock
\showISSN{2169-3536}
\showDOI{%
\url{http://dx.doi.org/10.1109/ACCESS.2015.2457031}}


\bibitem[\protect\citeauthoryear{Preden, Kaugerand, Suurjaak, Astapov, Motus,
  and Pahtma}{Preden et~al\mbox{.}}{2015}]%
        {FC18}
{Jurgo Preden}, {Jaanus Kaugerand}, {Erki Suurjaak}, {Sergei Astapov}, {Leo
  Motus}, {and} {Raido Pahtma}. 2015.
\newblock \showarticletitle{{Data to decision: pushing situational information
  needs to the edge of the network}}. In {\em 2015 IEEE International
  Multi-Disciplinary Conference on Cognitive Methods in Situation Awareness and
  Decision}. IEEE, 158--164.
\newblock
\showISBNx{978-1-4799-8015-4}
\showDOI{%
\url{http://dx.doi.org/10.1109/COGSIMA.2015.7108192}}


\bibitem[\protect\citeauthoryear{Puschmann, Barnaghi, and Tafazolli}{Puschmann
  et~al\mbox{.}}{2016}]%
        {Puschmann2016}
{Daniel Puschmann}, {Payam Barnaghi}, {and} {Rahim Tafazolli}. 2016.
\newblock \showarticletitle{{Adaptive Clustering for Dynamic IoT Data
  Streams}}.
\newblock {\em IEEE Internet of Things Journal\/} (2016), 1--1.
\newblock
\showISSN{2327-4662}
\showDOI{%
\url{http://dx.doi.org/10.1109/JIOT.2016.2618909}}


\bibitem[\protect\citeauthoryear{Rahman and Dijk}{Rahman and Dijk}{2014}]%
        {Rahman2014}
{A. Rahman} {and} {E. Dijk}. 2014.
\newblock {Group Communication for the Constrained Application Protocol
  (CoAP)}.
\newblock   (2014).
\newblock
\showISSN{2070-1721}
\showURL{%
\url{https://tools.ietf.org/pdf/rfc7390.pdf}}


\bibitem[\protect\citeauthoryear{Raza, Shafagh, Hewage, Hummen, and Voigt}{Raza
  et~al\mbox{.}}{2013}]%
        {Raza2013}
{Shahid Raza}, {Hossein Shafagh}, {Kasun Hewage}, {Rene Hummen}, {and} {Thiemo
  Voigt}. 2013.
\newblock \showarticletitle{{Lithe: Lightweight secure CoAP for the internet of
  things}}.
\newblock {\em IEEE Sensors Journal\/} {13}, 10 (2013), 3711--3720.
\newblock
\showISSN{1530437X}


\bibitem[\protect\citeauthoryear{Raza, Trabalza, and Voigt}{Raza
  et~al\mbox{.}}{2012}]%
        {Raza2012}
{Shahid Raza}, {Daniele Trabalza}, {and} {Thiemo Voigt}. 2012.
\newblock \showarticletitle{{6LoWPAN compressed DTLS for CoAP}}. In {\em
  Proceedings - IEEE International Conference on Distributed Computing in
  Sensor Systems, DCOSS 2012}. 287--289.
\newblock
\showISBNx{9780769547077}
\showDOI{%
\url{http://dx.doi.org/10.1109/DCOSS.2012.55}}


\bibitem[\protect\citeauthoryear{{RCUK Digital Economy HAT Project}}{{RCUK
  Digital Economy HAT Project}}{2014}]%
        {HAT1}
{{RCUK Digital Economy HAT Project}}. 2014.
\newblock {\em {Engineering a Market for Personal Data: The Hub-of-all-Things
  (HAT) A Briefing Paper}}.
\newblock {T}echnical {R}eport. RCUK Digital Economy.
\newblock


\bibitem[\protect\citeauthoryear{Reiter}{Reiter}{2014}]%
        {Texas2014}
{Gil Reiter}. 2014.
\newblock {\em {Wireless connectivity for the Internet of Things}}.
\newblock {T}echnical {R}eport. Texas Instruments Incorporated, Texas.
\newblock
\showURL{%
\url{http://www.ti.com.cn/cn/lit/wp/swry010/swry010.pdf}}


\bibitem[\protect\citeauthoryear{Rivest, Shamir, and Adleman}{Rivest
  et~al\mbox{.}}{1978}]%
        {Rivest1978}
{R.~L. Rivest}, {A. Shamir}, {and} {L. Adleman}. 1978.
\newblock \showarticletitle{{A method for obtaining digital signatures and
  public-key cryptosystems}}.
\newblock {\it Commun. ACM} {21}, 2 (feb 1978), 120--126.
\newblock
\showISBNx{0001-0782}
\showISSN{00010782}
\showDOI{%
\url{http://dx.doi.org/10.1145/359340.359342}}


\bibitem[\protect\citeauthoryear{{RS Components}}{{RS Components}}{2015}]%
        {RSComponents}
{{RS Components}}. 2015.
\newblock {11 Internet of Things (IoT) Protocols You Need to Know About »
  DesignSpark}.
\newblock   (2015).
\newblock
\showURL{%
\url{http://www.rs-online.com/designspark/electronics/knowledge-item/eleven-internet-of-things-iot-protocols-you-need-to-know-about}}


\bibitem[\protect\citeauthoryear{Saint-Andre}{Saint-Andre}{2011}]%
        {SaintAndre2011}
{P Saint-Andre}. 2011.
\newblock {Extensible Messaging and Presence Protocol (XMPP) : Core}.
\newblock   (2011).
\newblock
\showISSN{2070-1721}


\bibitem[\protect\citeauthoryear{Sathiaseelan, Lertsinsrubtavee, Jagan,
  Baskaran, and Crowcroft}{Sathiaseelan et~al\mbox{.}}{2016}]%
        {Cloudrone}
{Arjuna Sathiaseelan}, {Adisorn Lertsinsrubtavee}, {Adarsh Jagan}, {Prakash
  Baskaran}, {and} {Jon Crowcroft}. 2016.
\newblock \showarticletitle{{Cloudrone: Micro Clouds in the Sky}}. In {\em ACM
  Mobisys Dronet}.
\newblock


\bibitem[\protect\citeauthoryear{Schneider}{Schneider}{2013}]%
        {Schneider2013}
{Stan Schneider}. 2013.
\newblock {Understanding The Protocols Behind The Internet Of Things}.
\newblock   (2013).
\newblock
\showURL{%
\url{http://electronicdesign.com/iot/understanding-protocols-behind-internet-things}}


\bibitem[\protect\citeauthoryear{Sehgal, Patrick, Soni, and Rajput}{Sehgal
  et~al\mbox{.}}{2015}]%
        {FC78}
{Vivek~Kumar Sehgal}, {Anubhav Patrick}, {Ashutosh Soni}, {and} {Lucky Rajput}.
  2015.
\newblock \showarticletitle{{Smart human security framework using internet of
  things, Cloud and fog computing}}.
\newblock {\em Advances in Intelligent Systems and Computing\/}  {321} (2015),
  251--263.
\newblock
\showDOI{%
\url{http://dx.doi.org/10.1007/978-3-319-11227-5_22}}


\bibitem[\protect\citeauthoryear{Shelby and Bormann}{Shelby and
  Bormann}{2009}]%
        {Shelby2009}
{Zach Shelby} {and} {Carsten Bormann}. 2009.
\newblock {\em {6LoWPAN: The Wireless Embedded Internet}}.
\newblock 1--223 pages.
\newblock
\showISBNx{9780470747995}
\showDOI{%
\url{http://dx.doi.org/10.1002/9780470686218}}


\bibitem[\protect\citeauthoryear{Shelby, Hartke, and Bormann}{Shelby
  et~al\mbox{.}}{2014}]%
        {Shelby2014}
{Z Shelby}, {K Hartke}, {and} {C Bormann}. 2014.
\newblock \showarticletitle{{The Constrained Application Protocol (CoAP)}}.
\newblock {\em Rfc 7252\/} (2014), 112.
\newblock
\showISBNx{2070-1721}
\showISSN{2070-1721}
\showDOI{%
\url{http://dx.doi.org/10.1007/s13398-014-0173-7.2}}


\bibitem[\protect\citeauthoryear{Singh and Reddy}{Singh and Reddy}{2015}]%
        {Singh2015a}
{Dilpreet Singh} {and} {Chandan~K Reddy}. 2015.
\newblock \showarticletitle{{A survey on platforms for big data analytics}}.
\newblock {\em Journal of Big Data\/} {2}, 1 (dec 2015), 8.
\newblock
\showISSN{2196-1115}
\showDOI{%
\url{http://dx.doi.org/10.1186/s40537-014-0008-6}}


\bibitem[\protect\citeauthoryear{Singh, Rajan, Shivraj, and
  Balamuralidhar}{Singh et~al\mbox{.}}{2015}]%
        {Singh2015}
{Meena Singh}, {M.~A. Rajan}, {V.~L. Shivraj}, {and} {P. Balamuralidhar}. 2015.
\newblock \showarticletitle{{Secure MQTT for Internet of Things (IoT)}}. In
  {\em Proceedings - 2015 5th International Conference on Communication Systems
  and Network Technologies, CSNT 2015}. 746--751.
\newblock
\showISBNx{9781479917976}
\showDOI{%
\url{http://dx.doi.org/10.1109/CSNT.2015.16}}


\bibitem[\protect\citeauthoryear{Singh, Hoque, and Tarkoma}{Singh
  et~al\mbox{.}}{2016}]%
        {Singh2016}
{Maninder~Pal Singh}, {Mohammad~A. Hoque}, {and} {Sasu Tarkoma}. 2016.
\newblock \showarticletitle{{A survey of systems for massive stream
  analytics}}.
\newblock  (may 2016).
\newblock
\showURL{%
\url{http://arxiv.org/abs/1605.09021}}


\bibitem[\protect\citeauthoryear{Stanford-clark and Truong}{Stanford-clark and
  Truong}{2008}]%
        {Stanfordclark2008}
{Andy Stanford-clark} {and} {Hong~Linh Truong}. 2008.
\newblock \showarticletitle{{MQTT For Sensor Networks ( MQTT-S ) Protocol
  Specification}}.
\newblock {\em mqttorg\/} (2008), 1--27.
\newblock


\bibitem[\protect\citeauthoryear{Stantchev, Barnawi, Ghulam, Schubert, and
  Tamm}{Stantchev et~al\mbox{.}}{2015}]%
        {FC67}
{Vladimir Stantchev}, {Ahmed Barnawi}, {Sarfaraz Ghulam}, {Johannes Schubert},
  {and} {Gerrit Tamm}. 2015.
\newblock \showarticletitle{{Smart Items, Fog and Cloud Computing as Enablers
  of Servitization in Healthcare}}.
\newblock {\em Sensors {\&} Transducers\/} {185}, 2 (2015), 121--128.
\newblock


\bibitem[\protect\citeauthoryear{Stojmenovic and Wen}{Stojmenovic and
  Wen}{2014}]%
        {FC02}
{Ivan Stojmenovic} {and} {Sheng Wen}. 2014.
\newblock \showarticletitle{{The Fog Computing Paradigm: Scenarios and Security
  Issues}}. In {\em Computer Science and Information Systems (FedCSIS), 2014
  Federated Conference on}. IEEE, 1--8.
\newblock
\showDOI{%
\url{http://dx.doi.org/10.15439/2014F503}}


\bibitem[\protect\citeauthoryear{Stolfo, Salem, and Keromytis}{Stolfo
  et~al\mbox{.}}{2012}]%
        {FC08}
{Salvatore~J. Stolfo}, {Malek~Ben Salem}, {and} {Angelos~D. Keromytis}. 2012.
\newblock \showarticletitle{{Fog computing: Mitigating insider data theft
  attacks in the cloud}}. In {\em Proceedings - IEEE CS Security and Privacy
  Workshops, SPW 2012}. IEEE, 125--128.
\newblock
\showISBNx{9780769547404}
\showDOI{%
\url{http://dx.doi.org/10.1109/SPW.2012.19}}


\bibitem[\protect\citeauthoryear{Su, Lin, Zhou, and Lu}{Su
  et~al\mbox{.}}{2015}]%
        {FC26}
{Jingtao Su}, {Fuhong Lin}, {Xianwei Zhou}, {and} {Xing Lu}. 2015.
\newblock \showarticletitle{{Steiner tree based optimal resource caching scheme
  in fog computing}}.
\newblock {\em China Communications\/} {12}, 8 (aug 2015), 161--168.
\newblock
\showISSN{16735447}
\showDOI{%
\url{http://dx.doi.org/10.1109/CC.2015.7224698}}


\bibitem[\protect\citeauthoryear{Suciu, Vulpe, Halunga, Fratu, Todoran, and
  Suciu}{Suciu et~al\mbox{.}}{2013}]%
        {SecureIoT}
{G Suciu}, {A Vulpe}, {S Halunga}, {O Fratu}, {G Todoran}, {and} {V Suciu}.
  2013.
\newblock \showarticletitle{{Smart Cities Built on Resilient Cloud Computing
  and Secure Internet of Things}}. In {\em 2013 19th International Conference
  on Control Systems and Computer Science}. 513--518.
\newblock
\showDOI{%
\url{http://dx.doi.org/10.1109/CSCS.2013.58}}


\bibitem[\protect\citeauthoryear{Tang, Chen, Hefferman, Wei, He, and Yang}{Tang
  et~al\mbox{.}}{2015}]%
        {FC53}
{Bo Tang}, {Zhen Chen}, {Gerald Hefferman}, {Tao Wei}, {Haibo He}, {and} {Qing
  Yang}. 2015.
\newblock \showarticletitle{{A Hierarchical Distributed Fog Computing
  Architecture for Big Data Analysis in Smart Cities}}.
\newblock {\em Proceedings of the ASE BigData {\&} SocialInformatics 2015\/}
  (2015), 28:1----28:6.
\newblock
\showISBNx{978-1-4503-3735-9}
\showDOI{%
\url{http://dx.doi.org/10.1145/2818869.2818898}}


\bibitem[\protect\citeauthoryear{Team}{Team}{2013}]%
        {Team2013}
{VFabric Team}. 2013.
\newblock {Choosing Your Messaging Protocol: AMQP, MQTT, or STOMP}.
\newblock   (2013).
\newblock
\showURL{%
\url{http://blogs.vmware.com/vfabric/2013/02/choosing-your-messaging-protocol-amqp-mqtt-or-stomp.html}}


\bibitem[\protect\citeauthoryear{{Texas Instruments}}{{Texas
  Instruments}}{2014}]%
        {TexasInstruments}
{{Texas Instruments}}. 2014.
\newblock {\em {Wireless Connectivity}}.
\newblock {T}echnical {R}eport. Texas.
\newblock


\bibitem[\protect\citeauthoryear{Thangavel, Ma, Valera, Tan, and Tan}{Thangavel
  et~al\mbox{.}}{2014}]%
        {Thangavel2014}
{Dinesh Thangavel}, {Xiaoping Ma}, {Alvin Valera}, {Hwee~Xian Tan}, {and}
  {Colin Keng~Yan Tan}. 2014.
\newblock \showarticletitle{{Performance evaluation of MQTT and CoAP via a
  common middleware}}. In {\em IEEE ISSNIP 2014 - 2014 IEEE 9th International
  Conference on Intelligent Sensors, Sensor Networks and Information
  Processing, Conference Proceedings}.
\newblock
\showISBNx{9781479928439}
\showDOI{%
\url{http://dx.doi.org/10.1109/ISSNIP.2014.6827678}}


\bibitem[\protect\citeauthoryear{Thomas}{Thomas}{2000}]%
        {Thomas2000}
{Stephen Thomas}. 2000.
\newblock {\em {SSL and TLS essentials}}.
\newblock
\showISBNx{9780471383543}


\bibitem[\protect\citeauthoryear{{Thread Group}}{{Thread Group}}{2015}]%
        {ThreadGroup2015}
{{Thread Group}}. 2015.
\newblock {\em {Thread Stack Fundamentals}}.
\newblock {T}echnical {R}eport.
\newblock


\bibitem[\protect\citeauthoryear{Tracy}{Tracy}{2016}]%
        {iotAtNanoScale}
{Phillip Tracy}. 2016.
\newblock {What is the internet of things at nanoscale}.
\newblock   (2016).
\newblock
\newblock
\shownote{\url{http://www.rcrwireless.com/20160912/big-data-analytics/nano-scale-iot-tag31-tag99}
  [Retrieved November 2016].}


\bibitem[\protect\citeauthoryear{Truong, Lee, and Ghamri-Doudane}{Truong
  et~al\mbox{.}}{2015}]%
        {FC05}
{Nguyen~B. Truong}, {Gyu~Myoung Lee}, {and} {Yacine Ghamri-Doudane}. 2015.
\newblock \showarticletitle{{Software defined networking-based vehicular Adhoc
  Network with Fog Computing}}. In {\em Proceedings of the 2015 IFIP/IEEE
  International Symposium on Integrated Network Management, IM 2015}. IEEE,
  1202--1207.
\newblock
\showISBNx{9783901882760}
\showDOI{%
\url{http://dx.doi.org/10.1109/INM.2015.7140467}}


\bibitem[\protect\citeauthoryear{Vangelista, Zanella, and Zorzi}{Vangelista
  et~al\mbox{.}}{2015}]%
        {Vangelista2015}
{Lorenzo Vangelista}, {Andrea Zanella}, {and} {Michele Zorzi}. 2015.
\newblock \showarticletitle{{Long-range IoT technologies: The dawn of LoRaTM}}.
  In {\em Lecture Notes of the Institute for Computer Sciences,
  Social-Informatics and Telecommunications Engineering, LNICST}, Vol. 159.
  51--58.
\newblock
\showISBNx{9783319270715}
\showISSN{18678211}
\showDOI{%
\url{http://dx.doi.org/10.1007/978-3-319-27072-2_7}}


\bibitem[\protect\citeauthoryear{Vinoski}{Vinoski}{2006}]%
        {Vinoski2006}
{Steve Vinoski}. 2006.
\newblock \showarticletitle{{Advanced message queuing protocol}}.
\newblock {\em IEEE Internet Computing\/} {10}, 6 (2006), 87--89.
\newblock
\showISBNx{978-1-4673-5990-0}
\showISSN{10897801}
\showDOI{%
\url{http://dx.doi.org/10.1109/MIC.2006.116}}


\bibitem[\protect\citeauthoryear{Want}{Want}{2006}]%
        {Want2006}
{Roy Want}. 2006.
\newblock {An introduction to RFID technology}.
\newblock   (2006).
\newblock
\showISBNx{1536-1268}
\showISSN{15361268}
\showDOI{%
\url{http://dx.doi.org/10.1109/MPRV.2006.2}}


\bibitem[\protect\citeauthoryear{Want}{Want}{2011}]%
        {Want2011}
{Roy Want}. 2011.
\newblock \showarticletitle{{Near Field Communication}}.
\newblock {\em Pervasive Computing, IEEE, 10(3)\/} (2011), 4--7.
\newblock
\showISBNx{978-0-7695-3577-7}
\showISSN{1932-6203}
\showDOI{%
\url{http://dx.doi.org/10.1109/MPRV.2011.55}}


\bibitem[\protect\citeauthoryear{Watteyne, Vilajosana, Kerkez, Chraim, Weekly,
  Wang, Glaser, and Pister}{Watteyne et~al\mbox{.}}{2012}]%
        {P634}
{Thomas Watteyne}, {Xavier Vilajosana}, {Branko Kerkez}, {Fabien Chraim},
  {Kevin Weekly}, {Qin Wang}, {Steven Glaser}, {and} {Kris Pister}. 2012.
\newblock \showarticletitle{{OpenWSN: a standards-based low-power wireless
  development environment}}.
\newblock {\em Transactions on Emerging Telecommunications Technologies\/}
  {23}, 5 (2012), 480--493.
\newblock
\showISSN{2161-3915}
\showDOI{%
\url{http://dx.doi.org/10.1002/ett.2558}}


\bibitem[\protect\citeauthoryear{Wei and Barnaghi}{Wei and Barnaghi}{2009}]%
        {Wei2009}
{Wang Wei} {and} {Payam Barnaghi}. 2009.
\newblock \showarticletitle{{Semantic annotation and reasoning for sensor
  data}}. In {\em Lecture Notes in Computer Science (including subseries
  Lecture Notes in Artificial Intelligence and Lecture Notes in
  Bioinformatics)}, Vol. 5741 LNCS. Springer Berlin Heidelberg, 66--76.
\newblock
\showISBNx{3642044700}
\showISSN{03029743}
\showDOI{%
\url{http://dx.doi.org/10.1007/978-3-642-04471-7_6}}


\bibitem[\protect\citeauthoryear{Yannuzzi, Milito, Serral-Gracia, Montero, and
  Nemirovsky}{Yannuzzi et~al\mbox{.}}{2014}]%
        {FC14}
{M. Yannuzzi}, {R. Milito}, {R. Serral-Gracia}, {D. Montero}, {and} {M.
  Nemirovsky}. 2014.
\newblock \showarticletitle{{Key ingredients in an IoT recipe: Fog Computing,
  Cloud computing, and more Fog Computing}}.
\newblock {\em 2014 IEEE 19th International Workshop on Computer Aided Modeling
  and Design of Communication Links and Networks (CAMAD)\/} (dec 2014),
  325--329.
\newblock
\showISBNx{978-1-4799-5725-5}
\showISSN{2378-4865}
\showDOI{%
\url{http://dx.doi.org/10.1109/CAMAD.2014.7033259}}


\bibitem[\protect\citeauthoryear{Yeh, Chen, and Lee}{Yeh et~al\mbox{.}}{2003}]%
        {Yeh2003}
{Jui-Hung Yeh}, {Jyh-Cheng Chen}, {and} {Chi-Chen Lee}. 2003.
\newblock \showarticletitle{{WLAN standards}}.
\newblock {\em IEEE Potentials\/} {22}, 4 (2003), 16--22.
\newblock
\showISSN{0278-6648}
\showDOI{%
\url{http://dx.doi.org/10.1109/MP.2003.1238688}}


\bibitem[\protect\citeauthoryear{Yi, Li, and Li}{Yi et~al\mbox{.}}{2015a}]%
        {FC66}
{Shanhe Yi}, {Cheng Li}, {and} {Qun Li}. 2015a.
\newblock \showarticletitle{{A Survey of Fog Computing}}. In {\em Proceedings
  of the 2015 Workshop on Mobile Big Data - Mobidata '15}. ACM Press, New York,
  New York, USA, 37--42.
\newblock
\showISBNx{9781450335249}
\showDOI{%
\url{http://dx.doi.org/10.1145/2757384.2757397}}


\bibitem[\protect\citeauthoryear{Yi, Qin, and Li}{Yi et~al\mbox{.}}{2015b}]%
        {FC77}
{Shanhe Yi}, {Zhengrui Qin}, {and} {Qun Li}. 2015b.
\newblock \showarticletitle{{Security and Privacy Issues of Fog Computing: A
  Survey}}.
\newblock Springer International Publishing, 685--695.
\newblock
\showDOI{%
\url{http://dx.doi.org/10.1007/978-3-319-21837-3_67}}


\bibitem[\protect\citeauthoryear{Z-Wave}{Z-Wave}{2015}]%
        {ZWave2015}
{Z-Wave}. 2015.
\newblock {About Z-Wave}.
\newblock   (2015).
\newblock
\showURL{%
\url{http://www.z-wave.com/z-wave}}


\bibitem[\protect\citeauthoryear{Zao, Gan, You, Mendez, Chung, Wang, Mullen,
  and Jung}{Zao et~al\mbox{.}}{2014}]%
        {FC11}
{John~K. Zao}, {Tchin~Tze Gan}, {Chun~Kai You}, {Sergio Jose~Rodriguez Mendez},
  {Cheng~En Chung}, {Yu~Te Wang}, {Tim Mullen}, {and} {Tzyy~Ping Jung}. 2014.
\newblock \showarticletitle{{Augmented brain computer interaction based on fog
  computing and linked data}}. In {\em Proceedings - 2014 International
  Conference on Intelligent Environments, IE 2014}. IEEE, 374--377.
\newblock
\showISBNx{9781479929474}
\showDOI{%
\url{http://dx.doi.org/10.1109/IE.2014.54}}


\bibitem[\protect\citeauthoryear{Zaslavsky, Perera, and
  Georgakopoulos}{Zaslavsky et~al\mbox{.}}{2012}]%
        {ZMP003}
{Arkady Zaslavsky}, {Charith Perera}, {and} {Dimitrios Georgakopoulos}. 2012.
\newblock \showarticletitle{{Sensing as a Service and Big Data}}. In {\em
  International Conference on Advances in Cloud Computing (ACC)}. Bangalore,
  India, 21--29.
\newblock


\bibitem[\protect\citeauthoryear{Zikopoulos}{Zikopoulos}{2012}]%
        {IBMBigData}
{Paul. Zikopoulos}. 2012.
\newblock {\em {Understanding big data : analytics for enterprise class Hadoop
  and streaming data}}.
\newblock McGraw-Hill. 141 pages.
\newblock
\showISBNx{0071790535}


\bibitem[\protect\citeauthoryear{Zimmerman}{Zimmerman}{1996}]%
        {Zimmerman1996}
{T.~G. Zimmerman}. 1996.
\newblock \showarticletitle{{Personal Area Networks: Near-field intrabody
  communication}}.
\newblock {\em IBM Systems Journal\/} {35}, 3.4 (1996), 609--617.
\newblock
\showISBNx{1755879415}
\showISSN{0018-8670}
\showDOI{%
\url{http://dx.doi.org/10.1147/sj.353.0609}}


\end{thebibliography}

\received{Month Year}{Month Year}{Month Year}


\medskip

%
%
%
%

\end{document}